\documentclass[11pt]{article}
\pdfoutput=1
\usepackage{amsfonts,amsmath,amssymb,epsf}
\usepackage{graphicx}
\usepackage{color}
\usepackage{wrapfig}
\usepackage{cancel}

\usepackage{xcolor}

\newcommand{\Slash}[1]{{\ooalign{\hfil$#1$\hfil\crcr\raise.167ex\hbox{/}}}}
\newcommand{\be}{\begin{equation}}
\newcommand{\ba}{\begin{eqnarray}}
\newcommand{\ea}{\end{eqnarray}}
\newcommand{\ee}{\end{equation}}

\def\abs#1{\mid \! #1 \! \mid}

\begin{document}

\begin{flushright}

\end{flushright}

\vspace{0.1cm}

\begin{center}

{\LARGE Effective Hopping in Holographic Bose and Fermi-Hubbard Models}
\end{center}
\vspace{0.2cm}
\begin{center}
 Mitsutoshi F{\sc ujita}$^{a}$\footnote{e-mail: {\tt fujita@mail.sysu.edu.cn}}, 
                  Ren\'e M{\sc eyer}$^{b}$\footnote{e-mail: {\tt rene.meyer@physik.uni-wuerzburg.de}}, 
	Sumiran P{\sc ujari}$^{c,d}$\footnote{e-mail: {\tt sumiran.pujari@gmail.com}}, and Masaki T{\sc ezuka}$^{e}$\footnote{e-mail: {\tt tezuka@scphys.kyoto-u.ac.jp}}, \\

\vspace{0.5cm}$^a$ {\it School of Physics and Astronomy, Sun Yat-Sen University, Guangzhou 510275, China}\\

\vspace{0.1cm}$^b$ {\it Institute for Theoretical Physics and Astrophysics, University of W\"urzburg, 97074 W\"urzburg, Germany}\\

\vspace{0.1cm}$^c$ {\it Department of Physics and Astronomy, University of Kentucky, Lexington, KY 40506, USA}\\

\vspace{0.1cm}$^d$ {\it 
    Department of Physics, IIT Bombay, Powai, Mumbai, Maharastra 400076, India}\\

\vspace{0.1cm}$^e$ {\it 
    Department of Physics, Kyoto University, Kyoto 606-8502, Japan}\\

\end{center}

\begin{center}
  {\bf Abstract}
\end{center}
In this paper, we analyze a proposed gravity dual to a $SU(N)$ Bose-Hubbard model, as well as construct a holographic dual of a $SU(N)$ Fermi-Hubbard model from D-branes in string theory. In both cases, the $SU(N)$ is dynamical, i.e. the hopping degrees of freedom are strongly coupled to $SU(N)$ gauge bosons which themselves are strongly interacting.  The vacuum expectation value (VEV) of the hopping term (i.e. the hopping energy) is analyzed in the gravity dual as a function of the bulk mass of the field dual to the hopping term, as well as of the coupling constants of the model. The bulk mass controls the anomalous dimension (i.e. the critical exponent) of the hopping term in the $SU(N)$ Bose-Hubbard model. We compare the hopping energy to the corresponding result in a numerical simulation of the ungauged $SU(N)$ Bose-Hubbard model. We find agreement when the hopping parameter is smaller than the other couplings. Our analysis shows that the kinetic energy increases as the bulk mass increases, due to increased contributions from the IR. The holographic Bose-Hubbard model is then compared with the string theory construction of a $SU(N)$ Fermi-Hubbard model. The string theory construction makes it possible to describe {fluctuations around a half-filled state in the supergravity limit, which map to ${\cal O}(1)$ occupation number fluctuations in the Fermi-Hubbard model at half filling.} Finally, the VEV of the Bose-Hubbard model is shown to agree with the one of the fermionic Hubbard model with the help of a two-site version of the Jordan-Wigner transformation. 

\newpage

\tableofcontents

\newpage

\section{Introduction}

The Bose-Hubbard model is an effective lattice theory of {bosons (as e.g. realized in cold atoms experiments \cite{Jaksch:1998zz})} that includes hopping or kinetic energy terms and short-range interactions. The hopping term, with the coefficient being equal to the hopping integral, is especially important to describe the motion of particles. It is known~\cite{Fisher:1989zza} that a) there are only two phases in the Bose-Hubbard model {in the absence of disorder or impurities}, namely, the Mott insulator phase and the coherent superfluid (SF) phase, b) the hopping term has long-range correlations in the coherent superfluid phase, and c) the condensate of the hopping term is of the same order as the occupation number.

A generalization of the Bose-Hubbard model to many species of bosons is also of interest. 
If all bosons are of the same type an additional global $SU(N)$ symmetry exists, and the relevant model is the $SU(N)$ Bose-Hubbard model.\footnote{There is an additional $U(1)$ rotating all colors with the same phase and hence enhancing $SU(N)$ to $U(N)$, which after rescaling by $N$ is the baryonic $U(1)$.} The Hamiltonian of the $SU(N)$ Bose-Hubbard model is given by\footnote{\label{BHNscaling}The Coulomb repulsion U should scale as ${\cal O}(1/N)$, and the hopping term should scale as ${\cal O}(N)$. In this way, all the terms in \protect\eqref{SUNBoseHubbard} have the same large N scaling if $t_\mathrm{hop}$ and $\mu$ are ${\cal O}(1)$, which corresponds to color independent chemical potential and hopping. The energy then scales as ${\cal O}(N).$} 
\ba\label{SUNBoseHubbard}
H=-t_\mathrm{hop}\sum_{\langle ij \rangle}(b_{i}^{a\dagger}b_{ja}+c.c.)+\dfrac{U}{2}\sum_{j}n_{j}(n_{j}-1)-\mu \sum_{j}n_{j},\label{BHmodel}
\ea
where the sum over the repeated index $a=1,2,\ldots,N$ is taken in the first term, the coefficient of the hopping term $t_\mathrm{hop}$ is the hopping integral and $n_{j}=b_{j}^{a\dagger}b_{ja}$ is the occupation number operator for site $j$. The second term and the third term are the on-site Hubbard interactions and the chemical potential, respectively. 

In this paper, we analyze {a cousin of the $SU(N)$ Bose-Hubbard model in which the $SU(N)$ is gauged and {strongly} coupled to a sector of itself strongly interacting gluonic degrees of freedoms} with the help of the gauge/gravity correspondence~\cite{Maldacena:1997re}. This correspondence, also called AdS/CFT duality, is a duality between strongly coupled gauge theories and weakly coupled gravity theories. 
Recently, a gravity dual to the large $N$ Bose-Hubbard model has been proposed as 2-dimensional gravity on $AdS_{2}$ with a hard wall~\cite{Fujita:2014mqa}.~\footnote{Note that the large $N$ Bose-Hubbard model is not a theory on a single site but a lattice theory on multiple sites, while the dual gravity theory lives in two dimensions.} Since the large $N$ limit is assumed in the gravity dual, the field theory side {correspondingly is the large $N$} $SU(N)$ Bose-Hubbard model \cite{Fujita:2014mqa}. The holographic model contains gauge fields, bi-fundamental scalars, and an IR potential. The number of gauge fields is equal to the number of sites in the large $N$ Bose-Hubbard model. An IR potential is needed to derive the phase structure of the Bose-Hubbard model and is an additional input {which we have  not yet succeeded to determine from a top-down construction.}~\footnote{The Mott insulator/non-homogeneous phase transition exists even in the finite size Bose-Hubbard model due to the large $N$ limit involved. Moreover, the spontaneous symmetry breaking superfluid-like phase transition exists in the $AdS_2$ geometry of the holographic Bose-Hubbard model.}

{A motivation to use a bottom-up model is to obtain non-perturbative aspects of the actual Bose-Hubbard model and to generalize to an interesting higher dimensional model (See section 5 of \cite{Fujita:2014mqa} for generalization to the higher dimensional model). Especially, the Bose-Hubbard model in higher dimensions is difficult to analyze only by using field theory techniques unworthy of one spatial dimension. Possible numerics also tend to give only small perspective of non-perturbative physics. Higher dimensional models with numerous lattices (triangular, Kagome, honeycomb etc.) can describe separate physics of the frustration and spin liquids. In the gravity dual, 3-site is required at least to understand these physics of numerous lattices. Moreover, making the holographic Bose-Hubbard model is a first step to make a top-down model of the Fermi-Hubbard model. Unworthy of the Bose-Hubbard model, Monte Carlo simulations suffer from a sign problem in the fermionic case. The top-down model will give non-perturbative perspective of the Fermi-Hubbard model.}

{The gauge/gravity correspondence has already been successfully applied to a range of holographic defect lattices. Holographic lattice models using probe branes~\cite{Karch:2003nh} have been proposed as a model of dimerization transition~\cite{Kachru:2009xf,Kachru:2010dk}. 
More recently, a holographic $AdS_2$ superconductor without spatial directions coupled to a $AdS_3$ metallic state has been used to describe the screening of impurities in a holographic Kondo model~\cite{Erdmenger:2013dpa}. 
A holographic Kondo model with two impurities constructed in~\cite{OBannon:2015cqy} is rather similar to our lattice constructions. For example, like the holographic Bose-Hubbard model, the gauge field on $AdS_2$ has strong leading divergences due to the additional strongly interacting gluonic sector, which in turn affect the asymptotic behavior of the matter fields.  } 

The purpose of this paper is to analyze the vacuum expectation value (VEV) of the hopping term on both sides of the gauge/gravity correspondence and to compare their behaviors, focusing on {the two-site holographic $SU(N)$ Bose-Hubbard model.} The definition of the VEV of the hopping term is the derivative of the free energy with respect to $t_\mathrm{hop}$.  Note that this VEV is qualitatively different from off-diagonal long-range order of the superfluid phase (long-range correlations), since the holographic Bose-Hubbard model is defined at finite volume. Off-diagonal long range order  is the superfluid order parameter in the Bose-Hubbard model at infinite volume and can be  decomposed in terms of the condensed order parameter in the large hopping integral limit.  The Gross-Pitaevskii equations of motion are more useful to describe such a condensate~\cite{PethickS}. The hopping VEV on two sites is rather the nearest-neighbor correlations and thus representative of short-range correlations. 

In~\cite{Fujita:2014mqa}, this two-site correlator was shown to become the order parameter of the Mott insulator/non-homogeneous phase transition in a holographic bottom-up Bose-Hubbard model.  {In the present work we in particular derive the $t_\mathrm{hop}/U$ fall off behavior {of the hopping kinetic energy} for large $U$ in the Mott insulator phase.} In this phase, the particles are hopping with effectively small amplitudes. {As we will show in sec.~\ref{sec4}, the same behavior can also be derived in second order perturbation theory in a two-site Bose-Hubbard model with an even number of particles.} We furthermore compare our result with the numerical simulation of the effective hopping in the $SU(N)$ Bose-Hubbard model at a fixed number of particles. In particular, we change the bulk mass parameter in the gravity dual for the purpose of comparison. {In all cases, we find qualitative agreement for a large range of $t_\mathrm{hop}/U$. This is the first main result of this work.}

The second main result, presented in sec.~\ref{holohubbard}, is a top-down construction of an $n_F$-site $SU(N)$ Fermi-Hubbard model by means of a D3-D5-D7 configuration.\footnote{See also the previous approach in~\cite{Kachru:2010dk}), which describes a holographic dimerization transition from a bound state of D5 and anti-D5 to a disconnected D5 and anti-D5 system. In contrast, our top-down model does not contain anti-D5 branes. There are no phase transitions making a bound state of D5 and anti-D5, or of two D5 branes. } We qualitatively analyze its phase structure and compare it with the bottom-up construction of \cite{Fujita:2014mqa}. Our string theoretic construction introduces $n_F$ non-Abelian D5-branes into the D3-D7 system of~\cite{Fujita:2009kw}. These D5-branes stretch between the asymptotic AdS boundary and the D7 brane at the bottom of the soliton cigar. We then separate the D5-branes along the boundary directions to become the lattice impurities, with fundamental strings attached between them describing the hopping dynamics. For two sites, quantization of the relative charge density between the two sites in terms of the fundamental charge of the F1 string is equivalent to quantization of the angular transverse direction of the embedding of the D5 brane wrapping a $S^4$ inside the $S^5$. If the two branes are not separated in the angular transverse direction, a phase corresponds to a homogeneous phase of the Hubbard model at half filling. If the D5 branes are separated in the angular direction as well, the system is in the non-homogeneous superfluid phase. In this way, the top-down construction has the same phase structure as the bottom-up model of \cite{Fujita:2014mqa}. The mapping of the matter content, which is the same as the bottom-up holographic Bose-Hubbard model of \cite{Fujita:2014mqa}, is summarized in table~\ref{TU2}.

{This paper is organized as follows: In section \ref{sec2} we review the bottom-up holographic Bose-Hubbard model of \cite{Fujita:2014mqa}, and in particular the lobe-shaped phase structure well-known from the mean-field treatment of the Bose-Hubbard model \cite{Fisher:1989zza}. We also derive the $1/\rho$ behavior of the values of $t_\mathrm{hop}$ at the lobe tips by a special choice of boundary conditions in the IR potential. In section \ref{sec3} we introduce a bulk mass for the field dual to the hopping operator in the holographic Bose-Hubbard model and calculate the VEV of the hopping term as a function of that mass. We find that the qualitative behavior of the VEV is comparable with the $SU(N)$ Bose-Hubbard model at small $t_\mathrm{hop}$. In section \ref{holohubbard}, we then present our top-down construction based on the D3-D5-D7 brane configuration and compare it with the bottom-up model of \cite{Fujita:2014mqa}. We in particular map the model to lowest order in the string tension to the Fermi-Hubbard hamiltonian at half filling. In section \ref{sec4} we compute the effective hopping kinetic energy by numerically simulating the $SU(N)$ Bose-Hubbard model. We  then analyze the effective hopping kinetic VEV for the single species $SU(N)$ Bose-Hubbard model for all $N$. Finally, we show that the hopping VEV agrees with the one of the fermionic Hubbard model with the help of a two-site version of the Jordan-Wigner transformation. We conclude by discussing our results in section ~\ref{sec:discussion}. Several technical details of the calculations are relegated to the appendices. In particular, the variation principle in $AdS_2$ and the holographic renormalization procedure is discussed in relation with~\cite{Cvetic:2016eiv} in app.~\ref{FMNEQ}. 
}

\section{{Phase structure of the holographic Bose-Hubbard model}}\label{sec2} 

{In this section, we review the holographic Bose-Hubbard model proposed in~\cite{Fujita:2014mqa}. We in particular explain the lobe-shape of the Mott insulating phases in the $t_\mathrm{hop}-\mu$ phase diagram and derive the $1/\rho$ behavior of the values of $t_\mathrm{hop}$ at the lobe tips by a special choice of boundary conditions in the IR potential. }

\begin{table}[t]
\caption{The AdS/CFT dictionary of \protect\cite{Fujita:2014mqa}}
  \label{TU1}
  \begin{center}
    \begin{tabular}{|c|c|} \hline
Dual Gravity Side   & Large $N$   Bose-Hubbard model            \\ \hline
$A_{t,i}$  &  $\mu$ (chemical potential) \ \& $b_{i}^{a\dagger}b_{i a}$ (occupation number) \\ \hline
$\phi_{i,j}$ &  $t_\mathrm{hop}$ (hopping amplitude) \ \&  $b_{i}^{a\dagger}b_{j a}$ (hopping operator) \\ \hline
 hard wall cut-off $u_\mathrm{h}$ & $U$ (on-site Coulomb interaction) \\ \hline
    \end{tabular}
  \end{center}
\end{table}

The matter content of the holographic Bose-Hubbard model of \cite{Fujita:2014mqa} is summarized in table~\ref{TU1}. It consists of $n$ $U(1)$ gauge fields $A_{\mu,i}$, one on each lattice site, and bi-fundamental scalars\footnote{A bi-fundamental scalar field is one charged under two $U(1)$ gauge symmetries with a priori different charges. In the model considered here the charge will be of equal magnitude but opposite sign.} $\phi_{i,j}$ linking two different sites. Indices $i,\ j$ label the lattice sites in the field theory, and run from $1$ to $n$. Gauge fields $A_{\mu,i}$ and bi-fundamentals $\phi_{i,j}$ are dual to the occupation numbers $\langle n_{i}\rangle =\langle b_{i}^{a\dagger}b_{i a}\rangle$ for each site and the bi-local hopping condensates $\langle b_{i}^{a\dagger}b_{ja} \rangle$, respectively.  The $U(1)^{n}$ gauge symmetry of the bulk theory corresponds to $U(1)^{n}$ global charge symmetry in the large $N$ Bose-Hubbard model, which rotates bosons $b_{ia}$ independent of the indices $a$. This $U(1)^{n}$ symmetry is broken to a charge $U(1)$ symmetry in the presence of the hopping term $t_\mathrm{hop}\neq 0$. Besides this global symmetry, the gravity dual also describes a single gauged $SU(N)$ acting on the index $a$ of the $SU(N)$ Bose-Hubbard model (c.f. eq.~\eqref{SUNBoseHubbard}) which, as usual, is hidden in the gravity dual which only describes gauge-invariant observables.
  
In the rest of this paper we focus on a two-site model, i.e. we restrict our discussion to the case of $n=2$. Under the assumption of the hopping amplitude on each link and the charges on each site being the same, it is straightforward to generalize to any number of sites $n$.\footnote{If the hopping is different on different sites or if the charges on either side of the bifundamental are not the same, translational symmetry will be broken and persistent currents introduced. Care also needs to be taken for chains of sites that are closed, such as e.g. a triangle. Since the condensing hopping scalars want to imbalance the charge density of the sites they are attached to, the boundary conditions on closed chains may lead to charge frustration.} The relevant gravitational background for the model of \cite{Fujita:2014mqa} is the $AdS_{2}$ hard wall geometry
\ba
ds^{2}=g_{\mu\nu}dx^\mu dx^\nu = -u^{2}dt^{2}+\dfrac{du^{2}}{u^{2}},
\ea
where the hard wall is located at $u=u_\mathrm{h}$ and we have set the AdS radius $L=1$.\footnote{For more information on this background c.f. \cite{Erlich:2005qh}. The hard wall cutoff $u_\mathrm{h}$ should be large compared to the other scales (e.g. temperature, chemical potential, AdS radius) in order to prevent possible instabilities to appear at energy scales below $u_\mathrm{h} $~\cite{Maldacena:1998uz,Sen:2011cn}.} 
This background is confining (has a discrete spectrum of excitations) due to the hard wall, and it was shown in \cite{Fujita:2014mqa} that the cutoff $u_\mathrm{h}$ plays the role of the on-site Coulomb interaction energy. The action of the holographic Bose-Hubbard model of \cite{Fujita:2014mqa} is\footnote{Here $\Lambda$ was absorbed into the parameters of the IR potential. It could also be absorbed into scalar fields via rescaling. These scaling symmetries will become important when analyzing the solutions later in this paper.}
\ba\label{bottomupmodel}
&I&=I_\mathrm{gauge}+I_\mathrm{matter}+I^\mathrm{IR}_\mathrm{mixed}, \label{ACT11} \\
&I_\mathrm{gauge}&=\sum_{n=1}^{2}\int d^{2}x\sqrt{-g}\Big(-\dfrac{1}{4}F_{(n)\mu\nu}F^{\mu\nu}_{(n)}\Big),\\
&I_\mathrm{matter}&=-\int d^{2}x\sqrt{-g}\dfrac{1}{\Lambda}(\abs{\vec{D}\phi}^{2}+M^{2}\abs{\phi}^{2}),  \\
&I^\mathrm{IR}_\mathrm{mixed}&\equiv - \int dt \mathcal{I}^\mathrm{IR}_\mathrm{mixed} =-\int_{u=u_\mathrm{h}}dt{u_{h}}( 2w^{2}\abs{\phi}^{2}+\lambda \abs{\phi}^{4}+ \nonumber \\\label{IRpotential}
& & +\sum_{p,r\ge 1}\Lambda_{(p,r)}\abs{\phi}^{2p}\sum_{n}(F^{(n)}_{\mu}F^{(n)\mu})^{r}+\dots,
\ea
with the covariant derivative $D_{\mu}=\partial_{\mu}-iq A_{\mu}^{(1)}+iq A_{\mu}^{(2)}$ ($\mu=u, t$).  $F_{\mu}^{(i)}$ is the field strength projected onto the outward pointing unit normal $n_{\mu}$ to the hard wall, {$F_\mu^{(i)}\equiv F_{\mu\nu}^{(i)}n^{\nu}$}. {The last two lines in \eqref{IRpotential} are a general Ansatz for an IR potential parametrizing the boundary conditions at the hard wall.} The first term in $I_\mathrm{mixed}^\mathrm{IR}$ is an IR mass for fields~\cite{Csaki:2003dt,Csaki:2003zu}. {Dots represent couplings with IR localized (Higgs) fields of~\cite{Kachru:2010dk}, which are  ignored in this paper since we do not need IR Higgs fields to derive the qualitative phase structure of the Bose-Hubbard model. Following \cite{Basu:2012gg}, we in particular included a tension-like coefficient $\Lambda$.~\footnote{$\Lambda$ can be {set to one} via  rescaling $(\phi,w^2,\lambda,\Lambda_{(p,r)})\to (\sqrt{\Lambda} \phi,w^2/\Lambda , \lambda/\Lambda^2,\Lambda_{(p,r)}/\Lambda^{p})$.} In the region where $\Lambda,\ 1/w^2,\ 1/\lambda$, and  $1/\Lambda_{(p,q)}$ are much larger than the gravitational coupling constant, the bi-fundamental scalar can be considered as a probe field w.r.t. the AdS hard wall background. Furthermore, if $\Lambda \gg q^2$, the backreaction of the scalars to the gauge field is expected to be small.\footnote{This can be seen by noting that the stress tensor of the scalars and gauge fields is proportional to $1/\Lambda$, and $1/q^2$, respectively. So the scalar?s energy momentum contribution is relatively small if $q^2/\Lambda$ is small.}} The two-site model \eqref{bottomupmodel} is invariant under a vector $U(1)=U(1)_1+U(1)_2$ that decouples from the bifundamental $\phi$, as well as an axial $U(1)=U(1)_1-U(1)_2$ symmetry rotating the phase of $\phi$. 

In this section we follow \cite{Fujita:2014mqa} and first consider vanishing bifundamental bulk mass $M=0$. By choosing the radial gauge $A_{u}^{(n)}=0$ and considering an ansatz for the background field $A_{t}^{(n)}=A_{t}^{(n)}(u)$ and $\phi=\phi (u)$, the equations of motion (EOM) derived from \eqref{ACT11}  are
\ba\label{EOL27}
&(u^{2}\phi')'+\dfrac{q^{2}}{u^{2}}(A_{t}^{(1)}-A_{t}^{(2)})^{2}\phi=0, \\\label{EOL28}
&A_{t}^{(m)\prime\prime}-\dfrac{2q^{2}\abs{\phi}^{2}}{\Lambda u^{2}}(A_{t}^{(m)}-A_{t}^{(m+1)})=0, \quad m=1,2,3,
\ea
where  primes denote differentiation with respect to $u$ and $A_{t}^{(3)}=A_{t}^{(1)}$. {Splitting  $\phi$ into absolute value and phase, one finds from \eqref{EOL27} that the phase part is constant. The IR potential \eqref{IRpotential} does not affect the EOM, but only the boundary conditions on the hard wall and the free energy. The latter fact will enable us to derive the position of the maxima of the lobe-shaped Mott insulator phases in the phase diagram analytically later in this section.}

\subsection{Homogeneous Mott Insulating Phase}
 
The solutions to \eqref{EOL27}-\eqref{EOL28} can be classified into two phases, namely, (i) a homogeneous phase and (ii) a non-homogeneous phase. In the homogeneous phase the gauge fields on both sites are equal to each other, $A_{t}^{(1)}=A_{t}^{(2)}$, while in the non-homogeneous phase they differ, $A_{t}^{(1)}\neq A_{t}^{(2)}$. In the homogeneous phase the interactions in \eqref{EOL27} between the gauge field and bifundamental vanish, and analytic solutions of \eqref{EOL27} can be obtained. We specify a generalized Dirichlet boundary condition for the fields on the hard wall,\footnote{{Other boundary conditions lead to quantitatively slightly different but qualitatively similar results. We will see an example of this in sec.~\ref{sec3}.}}
\ba\label{IRBChom}
A_{t}^{(1,2)}|_{u=u_\mathrm{h}}=-\mu +\rho u_\mathrm{h},\quad \phi|_{u=u_\mathrm{h}}=t_\mathrm{hop}\,.
\ea 
This boundary condition corresponds to the choice where the UV parameters $\mu, \ t_\mathrm{hop}$ become {sources}~\cite{Erlich:2005qh}. {The effect of \eqref{IRBChom} is to switch of the subleading VEV term $\varphi$ in the solution $\phi = t_\mathrm{hop} + \varphi\,u^{-1}$.} The solutions satisfying \eqref{IRBChom}  are 
\ba\label{FIR29}
A_{t}^{(1)}=A_{t}^{(2)}=-\mu+\rho u,\quad \phi=t_\mathrm{hop},
\ea
with the chemical potential chosen to be negative. By the standard AdS/CFT dictionary, $t_\mathrm{hop}$ is identified as the source for a hopping kinetic energy operator with scaling dimension $\Delta=1$. {Similarly, $\rho$ is identified as the charge density at each site dual to the chemical potential $\mu$. Switching off the VEV piece $\varphi$ via \eqref{IRBChom} allows to analytically obtain the free energy of the homogeneous phase, defined in \eqref{FreeEnergy}, as a function of $(\mu,t_\mathrm{hop})$ within the grand canonical ensemble.} 
 The on-shell action obtained by substituting \eqref{FIR29} into the action \eqref{ACT11} is linearly divergent in the UV. Its divergence may be canceled by adding a counterterm~\cite{Henningson:1998gx,de Haro:2000xn,Karch:2005ms}
\ba\label{homCT}
I_\mathrm{cut}=\sum_{k}\dfrac{1}{2}\int_{u=u_\mathrm{max}}dt\sqrt{-h}A_{(k)t}A^{t}_{(k)},
\ea
where $u_\mathrm{max}$ is the UV cutoff and $\sqrt{-h}$ is the induced metric at $u=u_\mathrm{max}$. On the field theory side of the AdS/CFT correspondence, $u_\mathrm{max}$ corresponds to the UV cutoff and $u_{h}$ is the IR cutoff yielding a mass gap~\cite{Erlich:2005qh,Witten:1998zw}.

Gauge invariance is not manifest in the counterterm \eqref{homCT}. As we show now, this enforces charge quantization: $\rho$ must be an integer due to a Dirac quantization condition. Gauge transformations $A_{(k)\mu}\to A_{(k)\mu}+\partial_{\mu}\Lambda^{(k)}$ which leave the action invariant should vanish after integration by parts in the bulk AdS$_2$ and should not change the leading coefficient (charges) in the solution for the gauge fields. Moreover,  large and discontinuous gauge transformations of the form $\Lambda^{(k)} =2\pi Q_M^{(k)}\theta (t-t_{0})$ can be considered. In this case $Q_{M}^{(k)}$ is the monopole charge and $\theta(t)$ is the Heaviside step function.  
  Requiring that such {large gauge transformations do not change the action leads to the Dirac quantization of the charge $\rho_{(k)}$ in the presence of the monopole charge: $\rho_{(k)} Q_{M}^{(k)}\in \mathbf{Z}$~\cite{Castro:2008ms}, implying $\rho_{(k)}\in \mathbb{Z}$ for the charges on the site $k$. For $k=1$, this requirement can be shown to arise from the worldvolume theory of a fundamental string coupled to an $NSNS$ $B$-field~\cite{Camino:2001at}.
  
The free energy is then evaluated by adding \eqref{homCT} to the on-shell action of \eqref{ACT11} in Euclidean signature~\cite{Hartnoll:2008kx,Donos:2012yu}, 
\ba\label{FreeEnergy}
F^\mathrm{Hom}=-(I+I_\mathrm{cut})/\beta=-2\mu\rho +U\rho^{2}+\mathcal{I}^\mathrm{IR}_\mathrm{mixed}.
\ea
Note that $\mathcal{I}^\mathrm{IR}_\mathrm{mixed}$, defined in \eqref{IRpotential}, vanishes when $t_\mathrm{hop}=0$ because all interactions include $\phi$ (see eq.~\eqref{FIR29}). 
 Following~\cite{Fujita:2014mqa}, the parameters $(\mu,u_\mathrm{h})$ are matched with the parameters in the Bose-Hubbard model $(\mu_{b},U)$ by comparing the result with $t_\mathrm{hop}=0$, 
\ba
u_\mathrm{h}=U,\quad \mu =\mu_{b}-\dfrac{U}{2}.
\ea 
At zero hopping level-crossing phase transitions are then observed by varying the chemical potential $\mu_b$. This transition is of first order, and the quantized charge density (occupation number) jumps by unity between the different Mott insulating ground states. The phase transition points are drawn on the $\mu_{b}$-axis of Fig. \ref{fig:mix} for $\mu_{b}/U=1,2,3$.

\begin{figure}[tb]
     \begin{center}
          \includegraphics[height=4.5cm,clip]{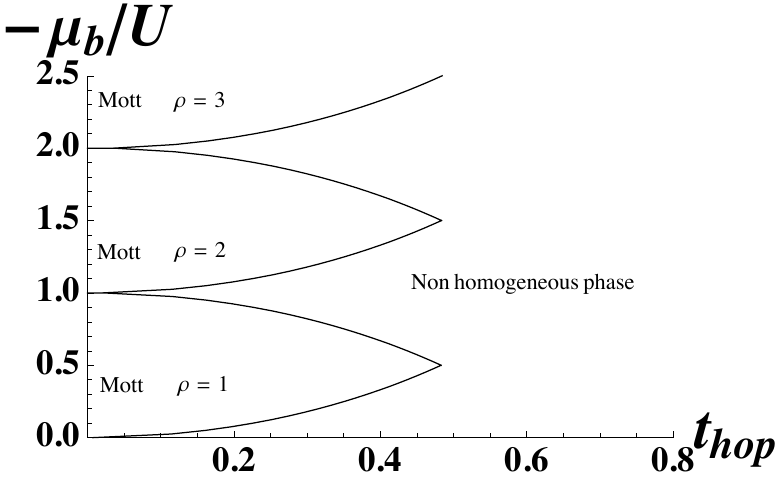} 
             \includegraphics[height=4.5cm,clip]{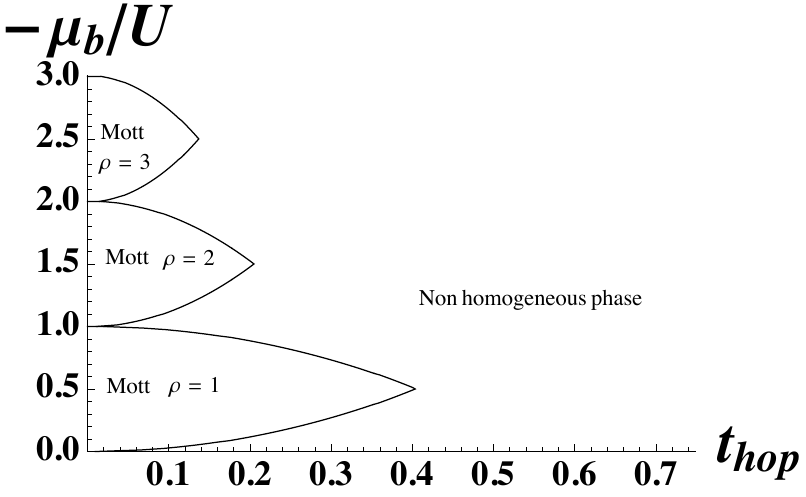}  
          \hspace{1.6cm}
           \caption{The lobe-shaped phase structure of the holographic two-site model in the $(\mu_{b},t_\mathrm{hop})$-plane ($\Lambda=1$, $u_\mathrm{h}=U=40$, and $q=\sqrt{6}/5$). Left: for $\lambda=w^{2}=1$, and all $\Lambda_{(p,r)}=0$. In the absence of IR interactions among the gauge fields and $\phi$, the amplitude of the lobes is periodic under the shift $\mu_{b}\to \mu_{b}+1$. Right: for $\lambda=1$, $w=0$, $\Lambda_{(1,1)}=-3/2$ and other $\Lambda_{(p,r)}=0$. We see the $1/\rho$ behavior of the phase structure.} 
    \label{fig:mix}
    \end{center}
\end{figure}

\begin{figure}[htbp]
     \begin{center}
        \includegraphics[height=4.5cm,clip]{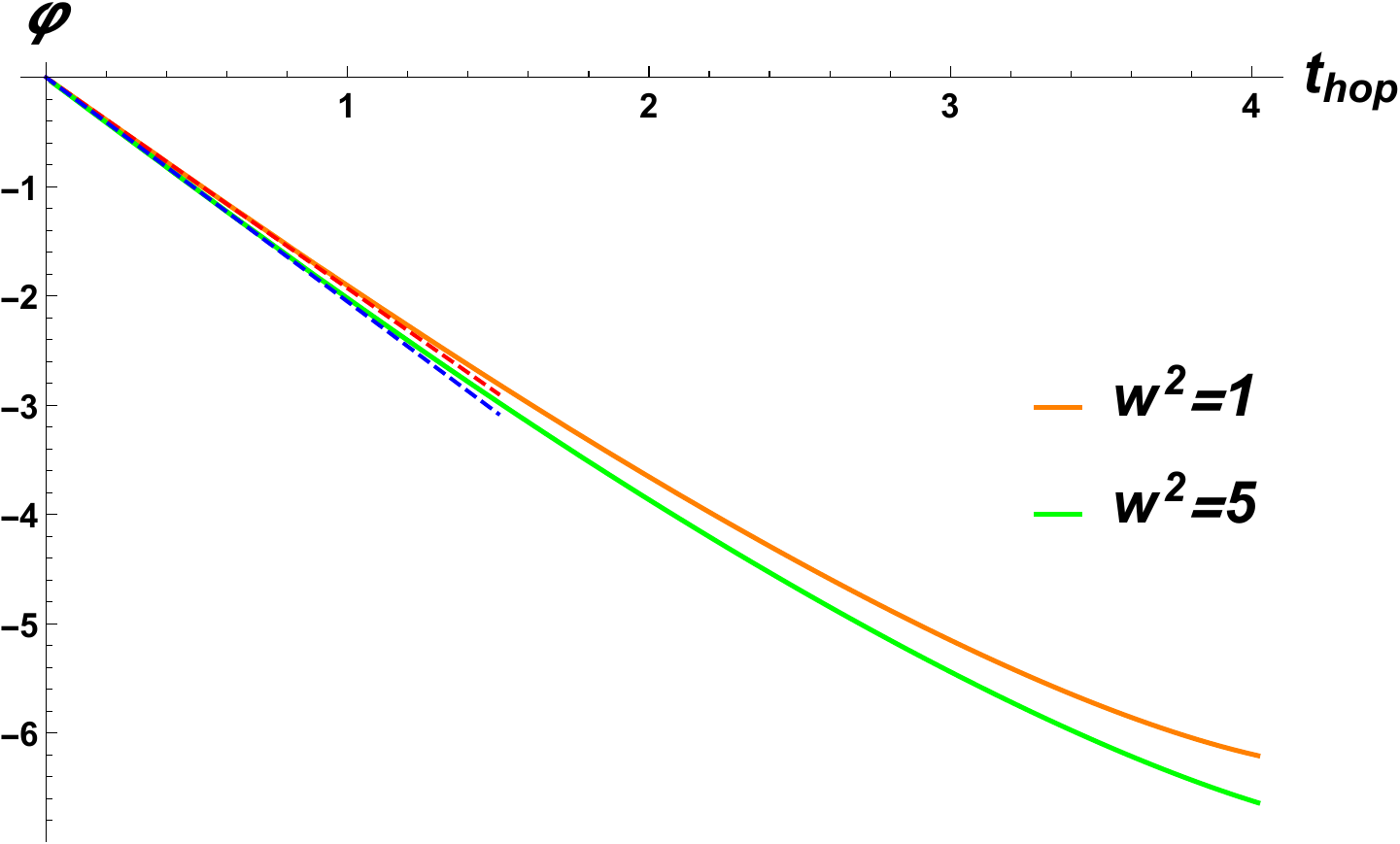} 
          \includegraphics[height=4.5cm,clip]{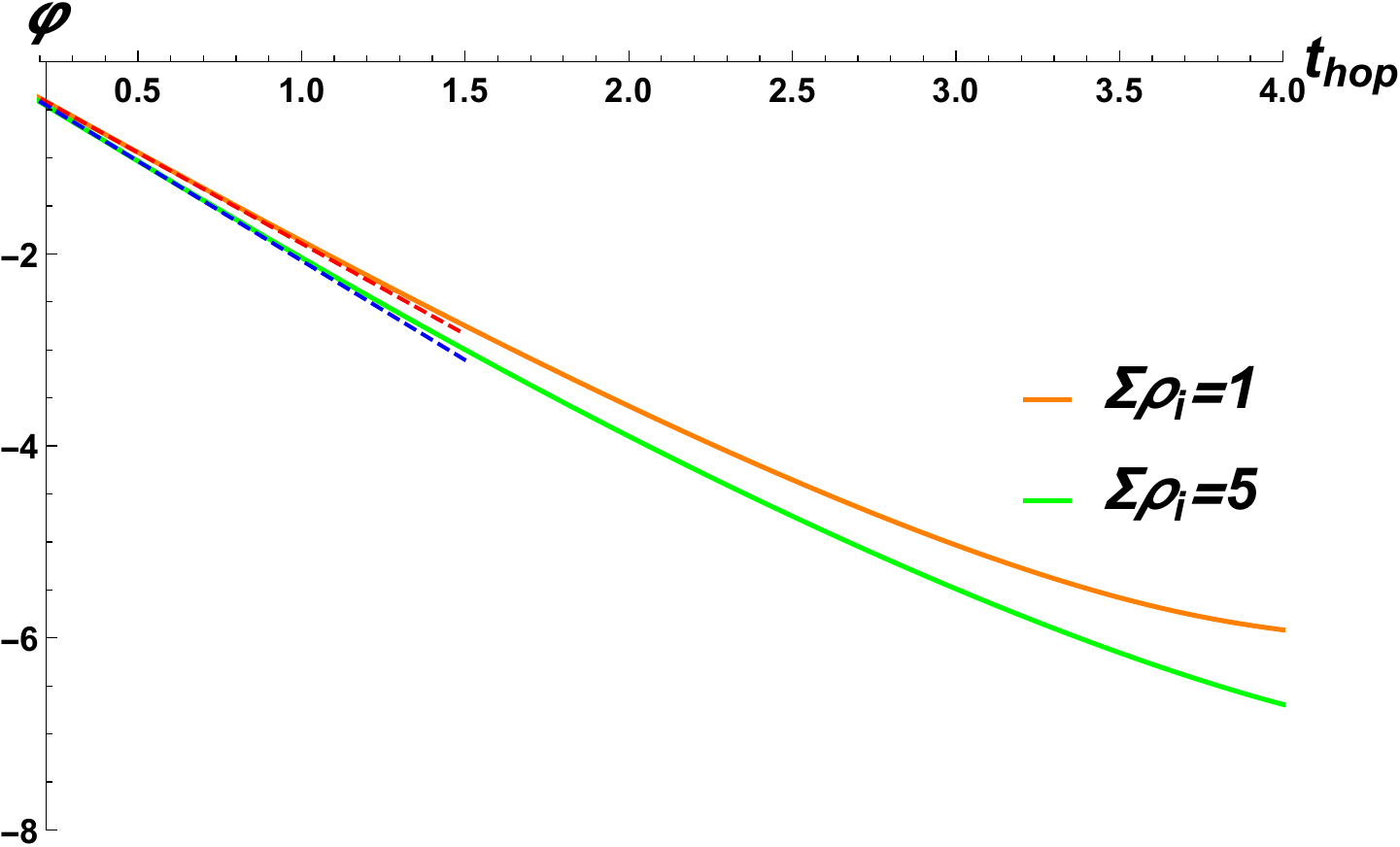} 
          \hspace{1.6cm}
           \caption{{The subleading term $\varphi$ as the function of $t_\mathrm{hop}$ in the non-homogeneous phase with the Neumann boundary condition ($\Lambda=1$, $u_\mathrm{h}=U=40$, and $q=\sqrt{6}/5$). {This term contributes to the total hopping kinetic energy in the superfluid phase.} Dashed lines mean analytic results \eqref{VEV57} obtained in Appendix \ref{FMNEQ}.   Left:  for $\lambda=1$, and all $\Lambda_{(p,r)}=0$. For small $t_\mathrm{hop}$, $\varphi$ is almost a linear function of $t_\mathrm{hop}$.  Right: for $\lambda=1$, $w=0$, $\Lambda_{(1,1)}=-3/2$ and other $\Lambda_{(p,r)}=0$. $\varphi$ is a linear function of $t_\mathrm{hop}$ for small $t_\mathrm{hop}$. There is a scaling relation, $(t_\mathrm{hop},\varphi )\to \sqrt{\Lambda}(t_\mathrm{hop},\varphi )$.}} 
    \label{fig:vevrho}
    \end{center}
\end{figure}

\subsection{Non-Homogeneous Superfluid Phase}

In the non-homogeneous phase the gauge fields differ between the two sites,  
\ba
A_{t}^{(1)}\neq A_{t}^{(2)}.
\ea
In this case, due to the coupling between the axial combination $A_{t}^{(1)}- A_{t}^{(2)}$ and the bifundamental $\phi$, an analytic solution cannot be obtained and numerical methods are needed to solve \eqref{EOL27}-\eqref{EOL28}. 
The solutions to \eqref{EOL27}-\eqref{EOL28} satisfy the near $AdS$ boundary expansion
\ba\label{ASY213}
&\phi\sim t_\mathrm{hop}u^{\beta_{\phi}}+{\cal O}(u^{3\beta_{\phi}})+\varphi u^{-1-\beta_{\phi}}(1+\dots),\nonumber \\
&A_{t}^{(k)}\sim \mu+\rho_{(k)}u+\dots,
\ea
where $\beta_{\phi}=(-1+\sqrt{1-4q^{2}\delta \rho^{2}})/2$. From this expansion it is found that {the dimension of the hopping kinetic energy is shifted by the charge difference $\delta \rho$} to be relevant,
\ba
\Delta = 1+\beta_\phi = \frac{1+\sqrt{1-4q^{2}\delta \rho^{2}}}{2}\,.
\ea
This shift arises from the coupling of the dual Bose-Hubbard model to the large $N$ CFT. In \cite{Fujita:2014mqa} a free boundary condition was imposed on $\phi$ at the IR wall $u=u_\mathrm{h}$, {such that the subleading piece $\varphi$ in the UV expansion \eqref{ASY213} vanished. In this way, the hopping kinetic VEV was completely generated by the contribution from the IR potential.} 

Alternatively, we can impose the following mixed Neumann boundary condition (a generalized version of~\cite{Nishioka:2009zj}) at $u=u_\mathrm{h}$ by requiring the boundary term to vanish at the hard wall boundary $u=u_\mathrm{h}$:
\ba\label{IRB216}
\dfrac{u_\mathrm{h}^2\phi'}{\Lambda}+\dfrac{\delta {I}^\mathrm{IR}_\mathrm{mixed}}{\delta \bar{\phi}}=0.
\ea
Since there may be many solutions for the hard wall boundary condition~\eqref{IRB216}, we also need to specify the behavior of $A_{t}^{(1)}-A_{t}^{(2)}|_{u=u_\mathrm{h}}\sim \delta \rho u_\mathrm{h}$ for small $t_\mathrm{hop}$ in order to pick the solution branch that can be continuously connected to the homogeneous phase with $A_{t}^{(1)}-A_{t}^{(2)}=0$.\footnote{{The other solution branches will have more nodes and hence higher free energy.}} The non-homogeneous solutions with these two conditions generate a very small {VEV piece} $\varphi$, i.e. are very close to the case of the free boundary condition employed in \cite{Fujita:2014mqa}. These conditions are also important to be consistent with the analysis of the level-crossing transition at $t_\mathrm{hop}=0$ where the Mott insulator phase is always favored. Solving the boundary condition \eqref{IRB216}, we have plotted the VEV piece $\varphi$ in the non-homogeneous phase in Fig. \ref{fig:vevrho}.  In both figures, $\varphi$ is a linear function of $t_\mathrm{hop}$ for small $t_\mathrm{hop}$, which confirms that we picked the correct solution branch.

The on-shell action is divergent at the $AdS$ boundary due to terms coming from the gauge fields that are cancelled by \eqref{homCT} as well as new divergencies coming from the bifundamental scalar. To cancel these new divergences, we add a counterterm 
\ba\label{CTphi1}
I_\mathrm{cut,2}=\dfrac{\beta_{\phi}}{\Lambda}\int_{u=u_\mathrm{max}}dt\sqrt{-h}\phi^{2}.
\ea
This is sufficient to render the on-shell action finite for $q>\sqrt{3}/4$. For $q\leq \sqrt{3}/4$  ($\beta_{\phi}\geq -1/4$), additional subleading divergencies appear that need to be cancelled separately. For vanishing mass $M=0$ we hence specify $q$ larger than $\sqrt{3}/4$, which is fulfilled by the value $q={\sqrt{6}\over 5}$ used both in \cite{Fujita:2014mqa} and in this work. The new counterterm \eqref{CTphi1} vanishes for $\delta \rho =0$ and hence is compatible  with the holographic renormalization of the Mott insulator phase. By adding two counterterms \eqref{homCT} and \eqref{CTphi1} to the action \eqref{ACT11}, we obtain the free energy
\ba
F=-(I+I_\mathrm{cut}+I_\mathrm{cut,2})/\beta.
\ea
 
 We analyze the phase structure by varying the {UV parameters}\footnote{{Usually, $\rho_i$ would be responses to $\mu$. Nevertheless, since the $\rho_i$ are quantized in our setup, we fix them to the corresponding values while varying $(\mu,t_\mathrm{hop})$.}} $(\mu, \rho_{i},t_\mathrm{hop})$ and by minimizing the free energy. {Going through the phase transition, in particular the internal energy $E\equiv F-\mu \sum_{k}\rho_{(k)}$ changes with $\mu$ due to the presence of the IR potential describing the interaction between the gauge field and $\phi$.} The inequality between $A_{t}^{(1,2)}$ in the non-homogeneous phase implies that the occupation number per site is not equal.  The non-homogeneous phase arises when the kinetic energy becomes large and the bosons become delocalized, which occurs in the large $t_\mathrm{hop}$ region of Fig. \ref{fig:mix}. In Fig. \ref{fig:mix}, the region surrounded by the lobe is the Mott insulator phase with equal occupation numbers $\rho_{(1)}=\rho_{(2)}$. Proper parameters in the IR potential with coupling among $\phi$ and the gauge fields realize the lobe-shaped phase structure of the Bose-Hubbard model,  as well as the $1/\rho$ behavior of the $t_\mathrm{hop}$ value at the tips of the lobe-shape.\footnote{The choice of parameters for each numerical result is mentioned in the caption of the corresponding figure. Generally, there exists a window of IR parameters in which the lobe-shaped phase structure is realized. A complete mapping of the possible phase structures for different parameter choices is left for future work.} In particular, the term with the coefficient $\Lambda_{(1,1)}$ in~\eqref{ACT11} can be approximated as a potential energy $\phi^{2}\rho^{2}_{i}$ depending on the kinetic energy $t_\mathrm{hop}$. With this same choice of IR potential as in \cite{Fujita:2014mqa}, almost the same phase structure as in the case of the free boundary condition \cite{Fujita:2014mqa} is obtained, since the generated VEV $\varphi$ under the IR boundary condition \eqref{IRB216} is small.~\footnote{{It can be shown that a cusp between two Mott insulating lobes is almost attached to the $\mu$-axis using the numerical computation of the integral.}} {In conclusion, the choice of boundary condition at the hard wall does not matter significantly as long as the generated VEV piece in the UV expansion of $\phi$ \eqref{ASY213} is small.}\footnote{It is noteworthy that the non-homogeneous phase (non-zero $A_{t}^{(1)}-A_{t}^{(2)}$ case) is similar to the analysis of the axial vector in hard/soft wall AdS/QCD models~\cite{Erlich:2005qh,Karch:2006pv}.}

{Another interesting fact is a scaling relation in the equations of motion which can be used to related different phase structures as in Fig. \ref{fig:mix},}
\be\label{rescaling}
(\phi,w^2,\lambda,\Lambda_{(p,r)})\to (\sqrt{\Lambda} \phi,w^2/\Lambda , \lambda/\Lambda^2, \Lambda_{(p,r)}/\Lambda^{p}).
\ee
Using this relation, one sees that the phase structure at $\Lambda =1$ has almost same phase structure as in \eqref{ACT11} after rescaling $t_\mathrm{hop}\to \sqrt{\Lambda}t_\mathrm{hop}$. This happens because the VEV $\varphi$ is a linear function of $t_\mathrm{hop}$ and hence obeys the same scaling relation as $\phi$. In the following section, the parameter $\Lambda$ is used to match the behavior of the effective hopping parameter (kinetic energy) with those in the field theory side. {We will in particular use this scaling relation to ensure a nearly unchanged phase structure at different values of $\Lambda$.}

\section{{Bulk scalar mass \& Hopping anomalous dimension}}\label{sec3}

In this section, we investigate the mass $M$ dependence in the lagrangian \eqref{ACT11}. {Since a bulk mass for the bifundamental scalar changes the scaling dimension of the hopping kinetic energy away from marginality, introducing this bulk mass effectively allows us to study the hopping kinetic term in the spirit of conformal perturbation theory as a UV perturbation away from the state with strong Coulomb repulsion. Tuning the anomalous dimension of the hopping energy operator to (in the RG sense) relevant or irrelevant values while keeping the dimension of the on-site conserved charge fixed, one would  expect an enhanced/decreased tendency to form the non-homogeneous superfluid phase. We will see in this section that this is not necessarily so, as this UV argument neglects the contribution to the free energy from the IR potential term \eqref{IRpotential}. Taking both effects into account, the picture becomes more involved.} 

For the ground state we consider  again the same ansatz as in sec.~\ref{sec2} of the background fields and bifundamental scalar depending on $u$ only, but now keep the mass term in \eqref{bottomupmodel}. Taking the gauge $A_u=0$ again, the EOM for the background now read 
\ba\label{MA317}
& (u^{2}\phi')'-M^{2}\phi+\dfrac{q^{2}}{u^{2}}(A_{t}^{(1)}-A_{t}^{(2)})^{2}\phi =0, \\
& A_{t}^{(m)\prime\prime}-\dfrac{2q^{2}\abs{\phi}^{2}}{u^{2}}(A_{t}^{(m)}-A_{t}^{(m+1)}) = 0,\quad m=1,2,3\,.
\ea

\subsection{Effective hopping in the homogeneous phase}\label{sec32}

If the gauge fields on both sites are equal, $A_{\mu}^{(1)}=A_{\mu}^{(2)}$, we find analytic solutions 
\ba\label{PHI16}
\phi=t_\mathrm{hop} u^{-\delta_M}+\varphi u^{-1+\delta_M}, \quad A_{t}^{(1)}=\mu +\rho u,
\ea
where $\delta_M=1/2-\sqrt{1+4M^2}/2$ and $\varphi =\Lambda \tilde{\varphi}/(1-2\delta_M)$ containing  the condensate $\tilde{\varphi}$ of the Bose-Hubbard model side (for a derivation c.f.~app.~\ref{HOL1}). The scaling dimension of the hopping kinetic energy dual to $t_\mathrm{hop}$ now becomes {relevant or irrelevant depending on the sign of $M^2$,}
\ba\label{anomdimM}
\Delta = 1-\delta_M = \frac{1+\sqrt{1+4M^2}}{2}\,.
\ea
The action \eqref{bottomupmodel} is invariant under a $\phi\mapsto - \phi$ symmetry, the action evaluated on the solution \eqref{PHI16} is invariant under the simultaneous sign change of $t_\mathrm{hop}$ and $\varphi$.  To agree with the Bose-Hubbard model \eqref{BHmodel}, $t_\mathrm{hop}$ is assumed to be negative $t_\mathrm{hop}<0$ in the remaining sections, and some plots are in terms of the positive parameter $U/{t_\mathrm{hop}}$.

 Following~\cite{Fujita:2014mqa}, we impose the same generalized Dirichlet boundary conditions \eqref{IRBChom} on the gauge potential as in the zero bulk mass case. Imposing the general Dirichlet boundary condition, $\mu$ and $\rho$ become a UV input~\cite{Erlich:2005qh}. Simultaneously, we can impose Dirichlet or Neumann IR boundary condition on the bi-fundamental scalar. In the Mott insulator phase, we impose the following general Dirichlet boundary condition as
\ba\label{BPH320}
\phi|_{u=u_\mathrm{h}}=t_\mathrm{hop} u_\mathrm{h}^{-\delta_M}.
\ea
{The subleading term $\sim \varphi$ in \eqref{PHI16} is switched off by this boundary condition, which is preferred in the homogeneous phase. }

\begin{figure}
     \begin{center}
          \includegraphics[height=4.5cm,clip]{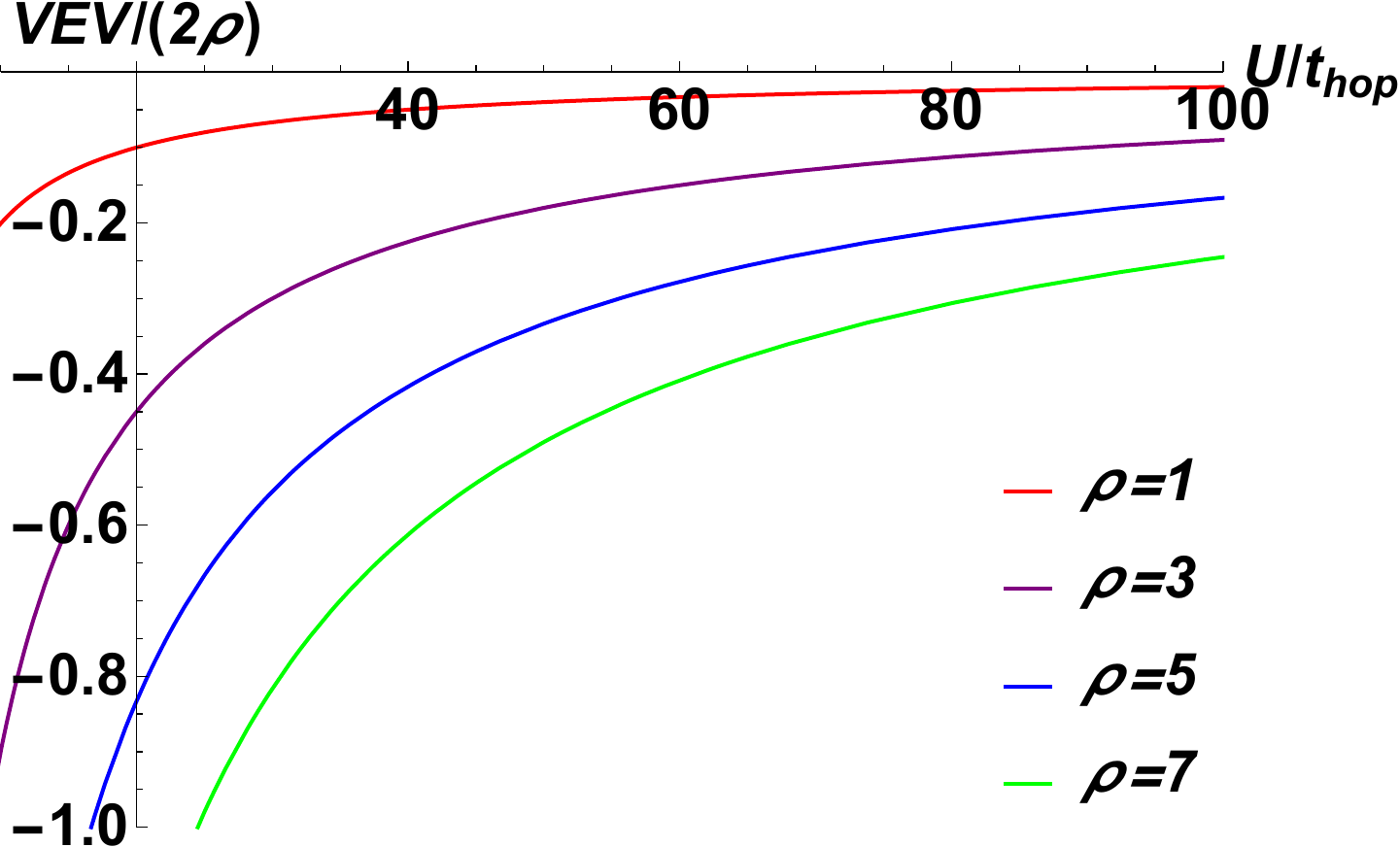}  
       \includegraphics[height=4.5cm,clip]{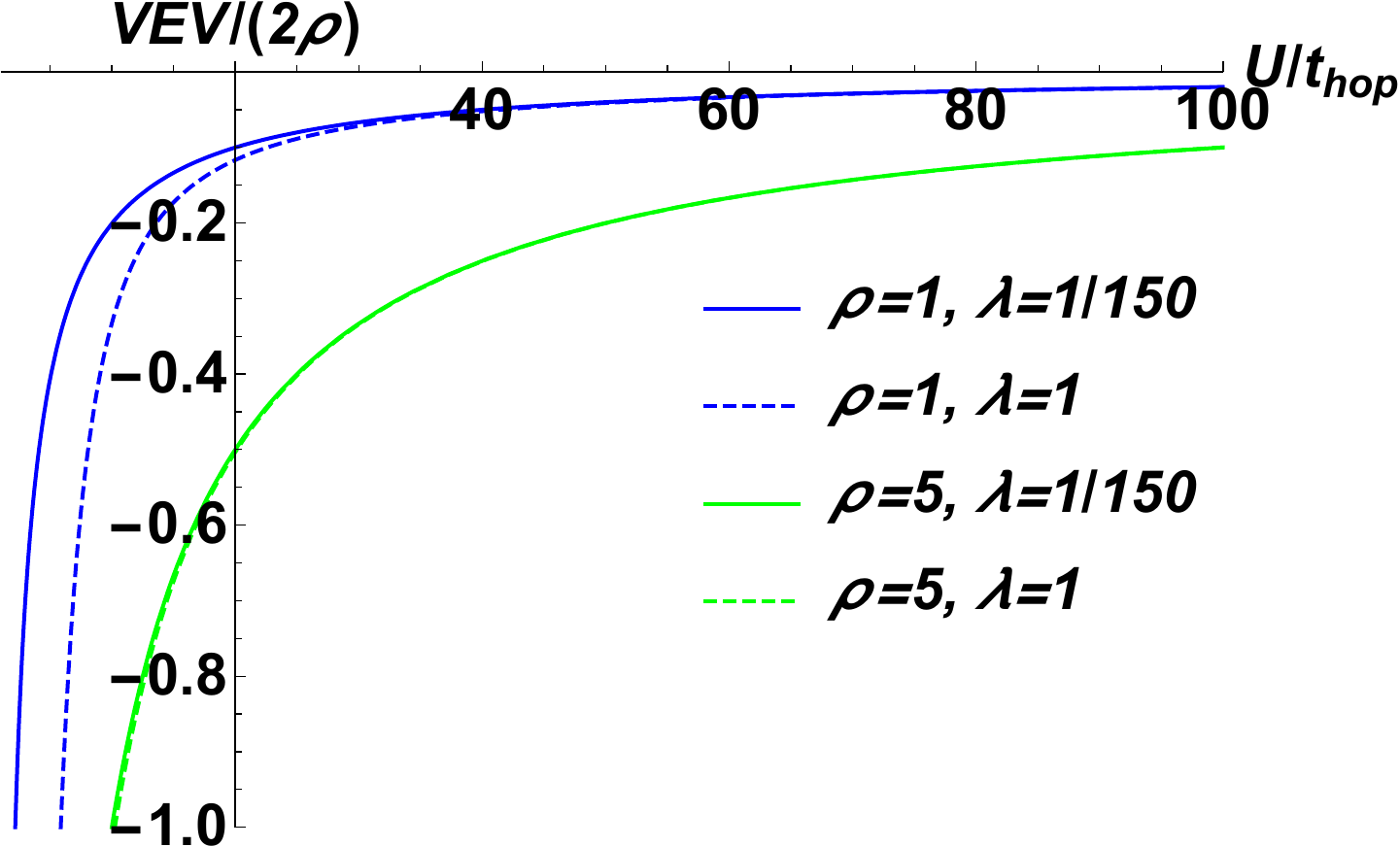} 
          \hspace{1.6cm}
        \end{center}
         \caption{The normalized VEV of the bi-fundamental is plotted as the function of $U/{t_\mathrm{hop}}$ for fixed $M^2=0$, $w=0$, $\lambda =1/\Lambda^2$, $\Lambda_{(1,1)}=-3/(2\Lambda)$, and other $\Lambda_{(p,r)}=0$, where the parameter $\Lambda$ is specified by eq. \ref{PAR328}.  {The normalized VEV shows the universal $t_\mathrm{hop}/U$ behavior at large $U$, consistent with second order perturbation theory in the single component two site Bose-Hubbard model. The coefficient coincides with the $\rho^2$ behavior in perturbation theory at large $\rho$, c.f. \eqref{VEV325}. } Left: the occupation number $\rho$ is changed. The absolute value of the VEV increases as $\rho$ increases. Right: the $\lambda$ dependence is plotted when $u_\mathrm{h}=10$. For small $U/t_\mathrm{hop}$, the absolute value of the VEV increases when $\lambda$ increases. When $\rho$ is larger than the other couplings, $\lambda$ becomes unimportant.}
    \label{fig:evenparticle}
\end{figure}

\begin{figure}
     \begin{center}
          \includegraphics[height=4.5cm,clip]{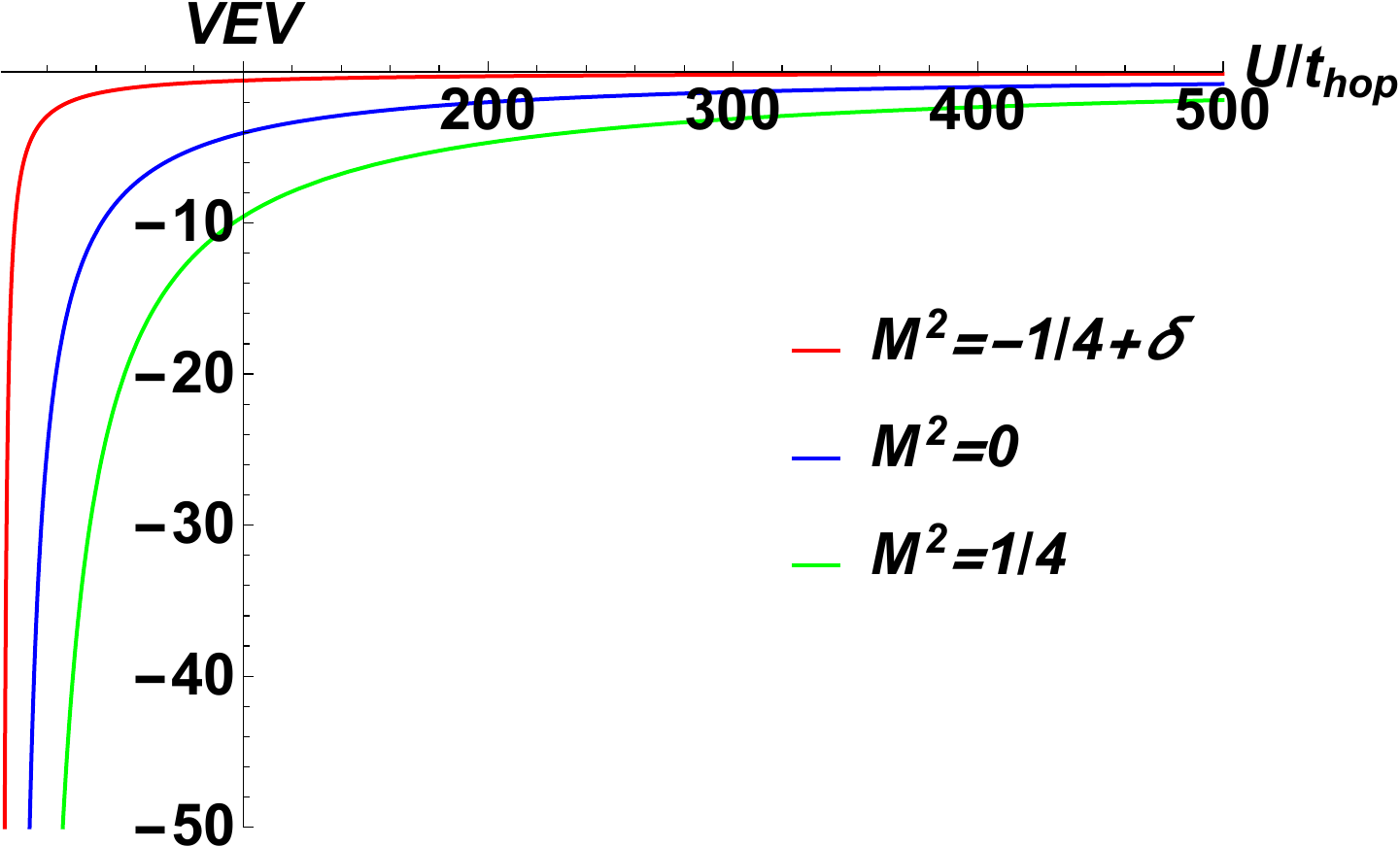} 
          \includegraphics[height=4.5cm,clip]{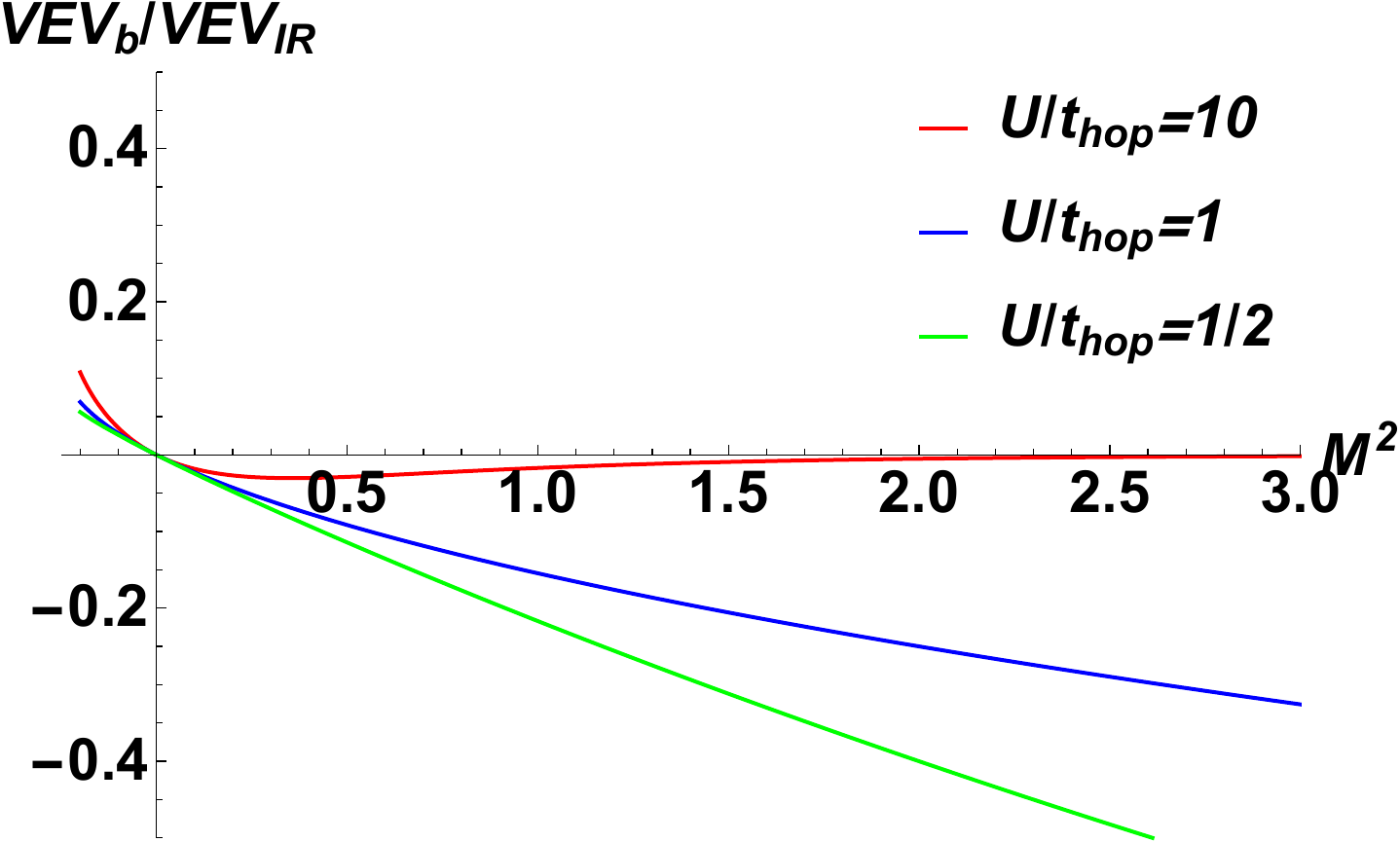}  
          \hspace{1.6cm}
        \end{center}
         \caption{With varying $M$ and for any $\rho$, the normalized VEV of the bi-fundamental is plotted as a  function of $U/{t_\mathrm{hop}}$ for fixed $u_\mathrm{h}=U=10$, $w=\lambda=1$, and $\Lambda_{(p,r)}=0$. Now, $\Lambda$ is specified to be 1, and the IR boundary condition is \eqref{BPH320} in the homogeneous phase. Left: $\delta$ is a small positive parameter. The absolute value of the VEV increases as $M^{2}$ increases. The VEV approaches $0$ in the large $U/t_\mathrm{hop}$ limit as expected in the Mott insulator phase of the Bose-Hubbard model. Right: The ratio of the bulk contribution to the IR contribution. The bulk contribution, which is the VEV as read off from the UV via the AdS/CFT dictionary, vanishes when $M=0$. The bulk contribution is suppressed at large $U$ and hence the major part of the VEV comes from the IR contribution. }
    \label{fig:mass}
\end{figure}

{We then holographically renormalize the action \eqref{ACT11} by adding the counterterms \eqref{homCT} and \eqref{CTphi1}. For nonvanishing bulk scalar mass we in particular need to include \eqref{CTphi1} even in the homogeneous phase due to the nontrivial RG running of $\phi$.} The free energy $F$ is computed from the holographic renormalized action in Euclidean signature, for details c.f. appendix~\ref{HOL1}. We define the VEV of the operator dual to the bi-fundamental scalar field to be equal to the derivative of the so-defined free energy w.r.t. to the hopping parameter $t_\mathrm{hop}$,
\ba\label{HoppingVEV}
&\langle b_{i}^{a\dagger}b_{ja}+c.c. \rangle \equiv \dfrac{dF}{dt_\mathrm{hop}}=-\dfrac{1}{\beta}\Big(\dfrac{\delta \bar{\phi}}{\delta t_\mathrm{hop}}\dfrac{\delta (I+I_{\mathrm{cut},m})}{\delta \bar{\phi}}+c.c.\Big)\,.
\ea
 This VEV then corresponds to the hopping kinetic energy on the dual large $N$ Bose-Hubbard model side. Even if the subleading piece $\varphi$ of in solution \eqref{PHI16} is zero, a non-trivial VEV is generated by the IR potential $\mathcal{I}_\mathrm{mix}^\mathrm{IR}$.~\footnote{Note that we also require $\delta \varphi =0$ at both the $AdS$ boundary and the hard wall cutoff.} 
Furthermore, the hopping kinetic energy is an order parameter for the Mott insulator to superfluid phase transition in the two-site Bose-Hubbard model. 

\bigskip

In the $\abs{t_\mathrm{hop}/U}\ll 1$ limit, one finds using \eqref{FMT324} that the VEV \eqref{HoppingVEV}  
behaves like
\ba\label{VEV325}
&\langle b_{i}^{a\dagger}b_{ja}+c.c. \rangle &={t_\mathrm{hop}Y_\mathrm{h}}\cdot \Big(\dfrac{1-\sqrt{1+4M^2}}{\Lambda}+4\sum_{r\ge 1}\Lambda_{(1,r)}(-\rho^2)^r+4w^2\Big) \nonumber \\
&+{\cal O}(t_\mathrm{hop}^2). 
\ea
where $t_\mathrm{hop}<0$ and $Y_h$ is defined below \eqref{FMT324}. {That the hopping VEV \eqref{VEV325} is proportional to $t_\mathrm{hop}$ is an expected behavior of the VEV in the $SU(N)$ Bose-Hubbard model at small hopping.} After matching the coefficient of the above leading term to data from the corresponding $SU(N)$ Bose-Hubbard model, {we can fix the parameters as a function of $N$ and $\rho$.}\footnote{Another parameter is the 't Hooft coupling of the hopping degrees of freedom to the gapless SU(N) gluon sector. The parameters of the model such as the charge or bulk mass will implicitly depend on it.} {This is  expected from a top-down string theory point of view (c.f. e.g. the top-down model of sec.~\ref{holohubbard}), where a natural scaling with $N$ exists for all quantities in the dual quantum theory. } 
For large occupation number $\rho$, the hopping kinetic energy \eqref{VEV325} can be further approximated as 
\be\label{VEVsmallthoplargerho}
\langle b_{i}^{a\dagger}b_{ja}+c.c. \rangle = dF/dt_\mathrm{hop} =  4t_\mathrm{hop}Y_\mathrm{h}\cdot \sum_{r\ge 1}  \Lambda_{(1,r)}(-\rho^2)^r\,.
\ee
We now proceed to match this result to the Bose-Hubbard model in second order perturbation theory.

We first compare our holographic result \eqref{VEVsmallthoplargerho} with perturbation theory in $t_\mathrm{hop}$ for the single component Bose-Hubbard model on two sites with an even number of particles ($N=1$ and $\rho_{(1)}=\rho_{(2)}=\rho$ in the Hamiltonian eq. \eqref{BHmodel} ). The total particle number is restricted to be even in order to have a Mott insulating ground state on two sites.\footnote{{Otherwise there would be a particle that could hop between the sites to first order in perturbation theory already.}} In second order perturbation theory, the VEV of the hopping term then behaves in the small $t_\mathrm{hop}/U$ limit as
\ba\label{pertVEVsinglecomponent}
\langle b_1^{\dagger}b_2 +c.c.\rangle \sim -\dfrac{4t_{(b)\mathrm{hop}}}{U}\rho (\rho +1)\sim -\dfrac{4t_{(b)\mathrm{hop}}}{U}\rho^2\quad (\rho\to \infty),
\ea
where $t_{(b)\mathrm{hop}}$ denotes the hopping integral in the Bose-Hubbard model \eqref{SUNBoseHubbard}. By comparing the coefficient of the $\rho^2$ term, one parameter of the IR potential is fixed to 
\ba\label{PAR328}
\Lambda_{(1,1)} =- u_\mathrm{h}^{2\delta_M-2}, \quad \mbox{for $\Lambda_{(1,1)}\neq 0$}.
\ea
 Accepting this $\rho^2 t_\mathrm{hop}/U$ behavior, moreover, the parameter $r$ in the summation in \eqref{IRpotential} is restricted to be 1, i.e. only terms quadratic in the field strength are allowed. \footnote{The term linear in $\rho$ is missing in the IR potential \eqref{IRpotential}. It could be generated by either a term $\phi^2 F_{ut}$ which breaks bulk diffeomorphisms, or $\phi^2 \sqrt{F^2}$, which does not. We could include such a term, but it does not affect the phase structure much.} 
 In this way we can match the result of a single species Bose-Hubbard model even if our hopping kinetic term has an anomalous dimension, while the first term in \eqref{SUNBoseHubbard} is of standard dimensionality. 

 We now consider the $SU(N)$ Bose-Hubbard model on two sites. In this case an $N^2t_\mathrm{hop}/U$ behavior is expected for the system with two  particles per species $N$, where $\rho=\rho_{(1)}=\rho_{(2)}$ is fixed to be $N$ such that there are exactly $N$ particles on each site and the total number of particles is even again. The hopping term of the $SU(N)$ Hubbard model itself will be discussed in more detail in sec.~\ref{sec4}. Here we consider the large N scaling: When $\rho =N$, the hopping VEV is proportional to $N^2$, which is different from the power $N$ usually obtained in the probe brane theory in the top-down approach~\cite{Mateos:2007vn}.~\footnote{The baryon vertex operator corresponds to such a theory in the gravity dual since $N$ fundamental strings end on it~\cite{Witten:1998xy,Sonnenschein:2016pim}.} { Moreover, in this case one can not ignore the backreaction of the gauge fields and bifundamental scalar onto the background geometry.} These issues are resolved by changing the field strength at on-shell($\equiv\rho$) to $\rho /N$ in the top-down approach of sec.~\ref{holohubbard}. In a nutshell, in the top-down approach the number of F1 strings ending on the defect brane is proportional to $N\rho$, which is quantized to be an integer and hence $\rho$ becomes an integer divided by $N$. Replacing $\rho\mapsto {\rho \over N}$, the large $N$ scaling expected from the $SU(N)$ Bose-Hubbard model is then consistent with the top-down string theory construction of sec.~\ref{holohubbard}. {Moreover, the fluctuation around the half filling state does not affect the leading large $N$ scaling of the probe brane when the fluctuation is ${\cal O}(1)$.} {Taking into account this lesson from the top-down construction, we can now trivially match the small $t_\mathrm{hop}/U$ large $\rho$ behavior \eqref{VEVsmallthoplargerho} of the holographic bottom up model to the second order perturbation theory of the $SU(N)$ Bose-Hubbard model \eqref{BHmodel} in the limit of small $t_\mathrm{hop}/U$. The result is again given by the conditions \eqref{PAR328}. The additional $N^2$ scaling of the hopping kinetic term is canceled by the replacement $\rho\mapsto {\rho \over N}$. }

{After matching the most relevant parameters of the IR potential to the SU(N) Bose-Hubbard model, we can study the dependence of the hopping kinetic energy on other parameters such as the charge density or the anomalous dimension, as well as the other still unfixed parameters of the IR potential.}  By using \eqref{rescaling} for parameters realizing the lobe-shaped phase structure to scale out $\Lambda$,  for $\Lambda_{(1,1)}\neq 0$, $\Lambda_{(1,1)}$ is restricted by \eqref{PAR328}. On the left-hand side of Fig. \ref{fig:evenparticle}, the VEV of the bi-fundamental is plotted for different choices of the charge density $\rho$, with the other parameters $M^2=0$, $\lambda =1/\Lambda^2$, $w=0$, $\Lambda_{(1,1)}=-3/(2 \Lambda)$ and other $\Lambda_{(p,r)}=0$ chosen to realize the lobe-shaped phase structure of the left hand side of Fig.~\ref{fig:mix}.\footnote{In Fig.~\ref{fig:mix}, $\Lambda_{(1,1)}$ was chosen to vanish. In Fig.~\ref{fig:evenparticle}, $u_h=10$ and hence $\Lambda_{(1,1)}=-\frac{1}{100}$ for $M^2=0$. In our experience, the phase structure of the left hand side of Fig.~\ref{fig:mix} does not change significantly under such a small change of $\Lambda_{(1,1)}$.} {In fig.~\ref{fig:evenparticle}, the VEV behaves as $t_\mathrm{hop}/U$ at large $U/t_\mathrm{hop}$. This universal behavior is consistent with second order perturbation theory in the single species two site Bose-Hubbard model. At large $\rho$, the coefficient of the VEV is $\rho^2$ as expected in the Bose-Hubbard model, c.f.~\eqref{pertVEVsinglecomponent}. 
The absolute value of the VEV increases when $\rho$ increases as expected.} On the right-hand side of Fig. \ref{fig:evenparticle}, the $\lambda$ dependence of the hopping VEV is plotted when $u_\mathrm{h}=10$. $\lambda$ is the quartic self-coupling of the hopping field $\phi$ in the IR potential \eqref{IRpotential}, which is absent in the top-down approach of sec.~\ref{holohubbard}. It is hence important to show that for realistic parameter choices in the bottom-up model, the resulting hopping VEV does not depend very sensitively on $\lambda$. The absolute value of the hopping VEV increases as $\lambda$ increases for small $U/t_\mathrm{hop}$. When $\rho$ becomes larger than the other couplings, the $\lambda$ dependence disappears altogether. 

{We hence conclude that we can simultaneously realize the lobe-shaped structure of Fig.~\ref{fig:mix} and qualitative behavior of the hopping kinetic energy of the Bose-Hubbard model \eqref{pertVEVsinglecomponent}.}   The influence of $\lambda$ is negligible in the bottom up model for a range of parameters around the ones chosen in this section, and hence the bottom up model has a chance to agree with the top-down model of sec.~\ref{holohubbard}. The other parameters on the right hand side of Fig.~\ref{fig:evenparticle} are unchanged compared to the left hand side plot.

Since the qualitative features of the phase structure of the left hand side of Fig.~\ref{fig:mix} is not sensitive to small changes of $\Lambda_{(1,1)}$ and $\Lambda$, we choose $\Lambda_{(1,1)}=0$ and $\Lambda =1$ in the following to compute the VEV of the bi-fundamental numerically.\footnote{Note that it is impossible to reinstate $\Lambda_{(1,1)}$ from $\Lambda_{(1,1)}=0$ by the scaling \eqref{rescaling}. Nevertheless, we find that even in this seemingly disconnected case, the phase structure and the behavior of the VEV are qualitatively unchanged.} This has the advantage that the VEV becomes independent of the charge density $\rho$ since the $\rho$-dependent terms in the IR potential vanish in the homogeneous Mott insulating phase. We study the dependence of the hopping VEV on the hopping field mass $M$, or equivalently on the anomalous dimension of the hopping kinetic energy operator. The bulk mass determines the anomalous dimension of the bi-fundamental operator via \eqref{anomdimM}. On the left side of Fig. \ref{fig:mass}, the normalized VEV of the bi-fundamental  is plotted as a function of $M^{2}$ above {the BF bound $M_\mathrm{BF}^{2}=-{1}/{4}$~\cite{Breitenlohner:1982bm,Breitenlohner:1982jf}} and for  $u_\mathrm{h}=10$, $\lambda =w=1$, and $\Lambda_{(p,r)}=0$. We find that the absolute value of the VEV increases as $M^{2}$ increases. The absolute value of the VEV approaches zero in the small $t_\mathrm{hop}$ limit, which is expected for the Mott insulator phase.  
At first, an increasing VEV for an operator whose dimension becomes more irrelevant as $M^2$ increases seems counterintuitive. However, the hopping VEV in our model receives two contributions, one UV contribution from the asymptotic behavior of the bulk field $\phi$, as well as contribution from the IR potential \eqref{IRpotential}. To understand the apparent conundrum, we compare the IR contribution to the VEV with the UV contribution,
\begin{eqnarray}\label{VEVtotal}
VEV_{total} &\equiv& dF/dt_\mathrm{hop} = VEV_\mathrm{b}+VEV_\mathrm{IR}\,, \\
VEV_\mathrm{IR} &=& -(\partial I^\mathrm{IR}_\mathrm{mixed})/(\beta\partial t_\mathrm{hop})\,, \\
VEV_\mathrm{b} &=& 2\delta_M Y_\mathrm{h}t_\mathrm{hop}\,.
\end{eqnarray} 
On the right hand side of Fig. \ref{fig:mass}, the ratio of the bulk contribution to the IR contribution to the VEV is shown. When $M=0$, the bulk contribution  ${VEV}_\mathrm{b}$ vanishes due to the vanishing of $\delta_M$ defined in \eqref{anomdimM} together with the boundary condition \eqref{BPH320}. The two contributions are the opposite sign of each other for $M^2>0$. The bulk contribution is much smaller than the IR contribution at large $U$, as it is suppressed by $t_\mathrm{hop}/U$. As will be discussed in sec.~\ref{sec:discussion} (c.f.~Fig.~\ref{fig:co1}), the parameter space given by $(M,w,u_h,\Lambda_{(1,1)})$ is sufficient to match the holographic model with the numerical results for the hopping kinetic energy of the $SU(N)$ Bose-Hubbard model for not too large values of $t_{\mathrm{hop}}/U$.

\subsection{Effective hopping in non-homogeneous phase}

We now turn to discuss the behavior of the hopping kinetic energy in the non-homogeneous phase, which is the holographic version of the superfluid phase of the Bose-Hubbard model. In the non-homogeneous phase where $A_{t}^{(1)}\neq A_{t}^{(2)}$, a numerical approach is required to derive the effective hopping parameter. The details of the derivation of  the effective hopping VEV in the non-homogeneous phase are given in Appendix~\ref{FMNEQ}. Solving the EOM \eqref{MA317}, the fields in the non-homogeneous phase are expanded near the $AdS$ boundary as 
\ba\label{PER28}
&\phi \sim t_\mathrm{hop}u^{\delta_{\phi}}(1+\dots) + \varphi_{v} u^{-1-\delta_{\phi}}(1 + \dots) , \nonumber \\
&A_t^{(l)}\sim\mu +\rho_{(l)}u+ \dots,
\ea
where $\delta_{\phi}=(-1+\sqrt{1+4M^2-4q^2\delta\rho^2})/2$, and the dots denote subleading corrections depending on the two integration constants $t_{\mathrm{hop}}$ and $\varphi_{v}$ which are computable {numerically.} The scaling dimension of the hopping kinetic energy dual to $t_\mathrm{hop}$ now depends on both $M$ and $\delta\rho = \rho_{(1)} - \rho_{(2)}$, and can be relevant, marginal or irrelevant in the RG sense,
\ba\label{anomdimMrho}
\Delta = 1+\delta_\phi = \frac{1+\sqrt{1+4M^2-4q^2\delta\rho^2}}{2}\,.
\ea
Note that the bi-fundamental scalar dual to the hopping kinetic term is charged under the axial combination $A_{t}^{(1)} - A_{t}^{(2)}$, and hence its anomalous dimension depends on the difference in charge density $\delta\rho$ between both sites. 

The fields at $u=u_\mathrm{h}$ have to satisfy a hard wall boundary condition, and the result for the hopping kinetic energy will slightly depend on it. Considering either the Neumann boundary condition \eqref{IRB216} (in Fig.~\ref{fig:nonh2}), or a Dirichlet condition $\varphi_v=\mbox{const}$ (and $\delta \varphi_v=0)$ and $A_t^{(1)}-A_t^{(2)}\sim \delta \rho u_\mathrm{h}$ as $t_\mathrm{h}\to 0$ (in Fig.~\ref{fig:nonh3}), we again compute the finite on-shell action by adding counter-terms at the $AdS$ boundary, c.f.~App.~\ref{FMNEQ} for details. The free energy $F$ is the holographically renormalized on-shell action in Euclidean signature. We compute the variation of the  free energy w.r.t. $t_{\mathrm{hop}}$ to compute the effective hopping kinetic energy. The details of the derivation can be found in App.~\ref{FMNEQ}.

\begin{figure}
       \begin{center}
       \includegraphics[height=4cm,clip]{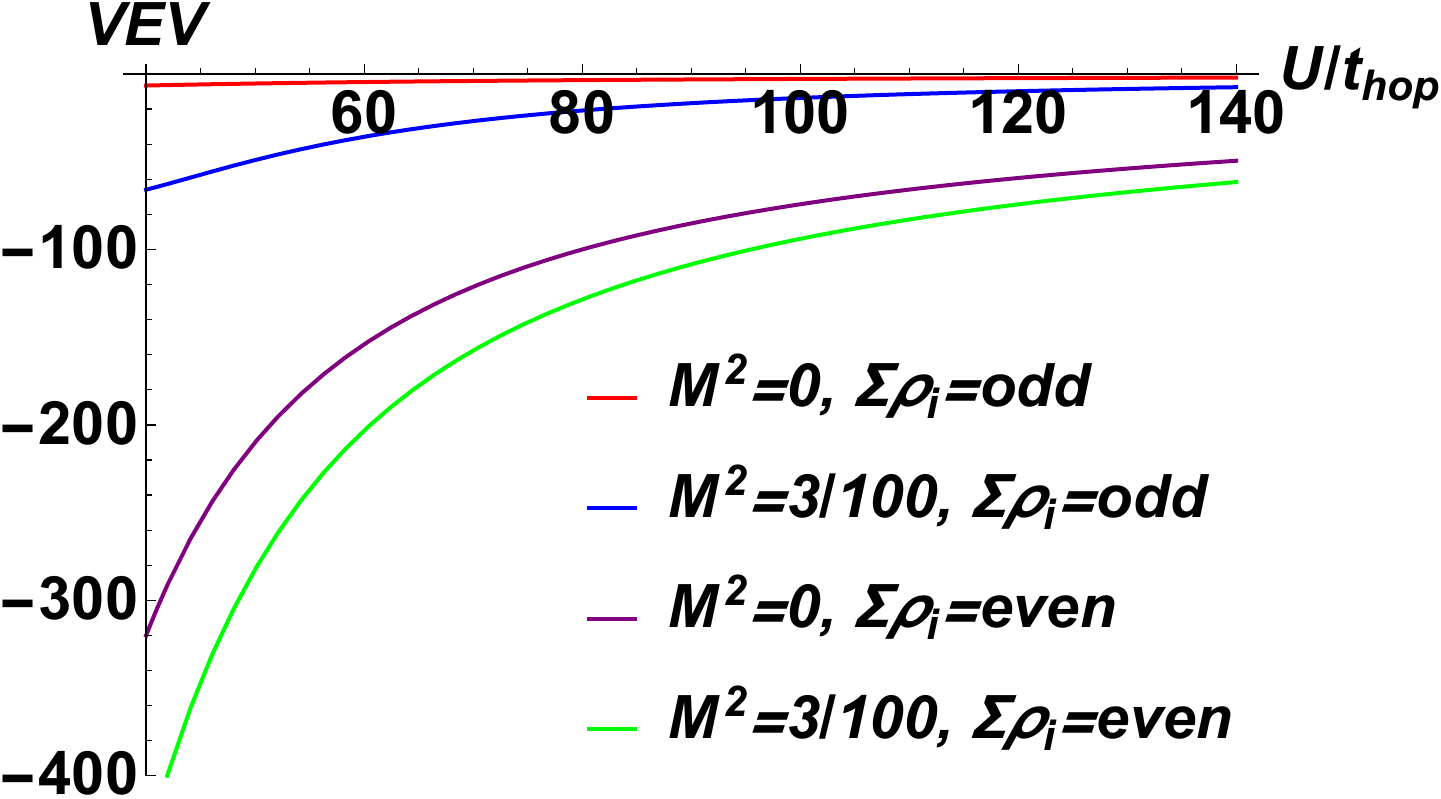} 
       \includegraphics[height=4cm,clip]{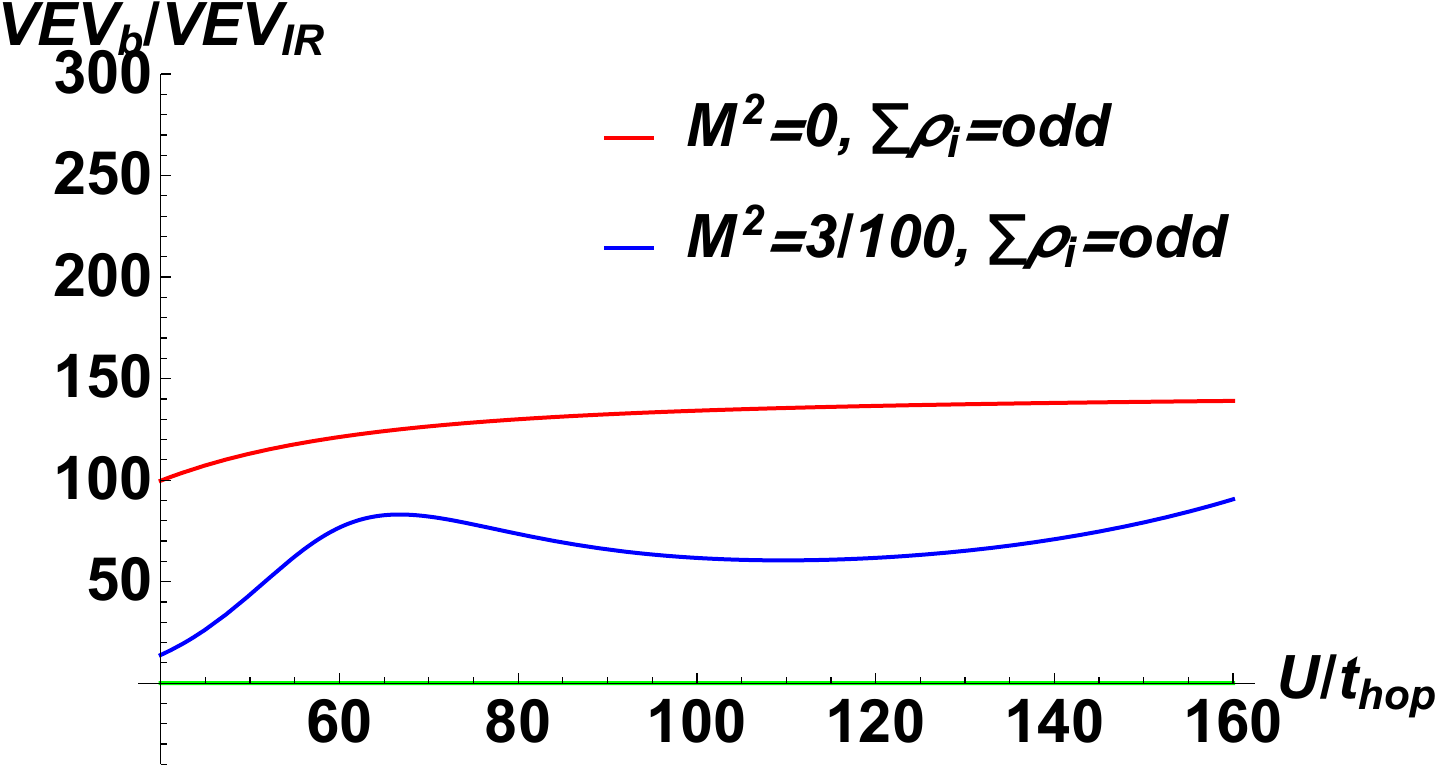}  
          \hspace{1.6cm}
        \end{center}
         \caption{The VEV in the non-homogeneous phase at the finite coupling $q=\sqrt{6}/5$ and $\abs{\delta \rho}\le 1$ and the Neumann IR boundary condition~\eqref{IRB216}. For any $\rho_{(i)}$ satisfying $\abs{\delta \rho}\le 1$ and varying $M$, the normalized VEV is plotted as the function of $U/{t_\mathrm{hop}}$ when $u_\mathrm{h}=40$, $\lambda =w=\Lambda=1$, and $\Lambda_{(p,q)}=0$. The absolute value of the VEV in the non-homogeneous phase (red and blue curve) is smaller than the one of the homogeneous phase (purple and green curve). $VEV_\mathrm{b}/VEV_\mathrm{IR}$ as a function of $U/t_\mathrm{hop}$ for curves with the same parameters (colors). $VEV_\mathrm{b}/VEV_\mathrm{IR}$ is zero when $M^2=0,\ \sum\rho_i=\mbox{even}$. The ratio becomes large and with the opposite sign compared to the homogeneous phase.
}
    \label{fig:nonh2}
\end{figure}

 The effective hopping (VEV) defined by $dF/dt_\mathrm{hop}$ is plotted as the function of $U/t_\mathrm{hop}$ for a charge of the bi-fundamental scalar~\footnote{The choice of this value is mostly for technical reasons explained in sec.~3 of \cite{Fujita:2014mqa}. Recall that, from \eqref{anomdimMrho}, when $q\ll 1$, many values of $\delta \rho$ can be allowed. The phase structure of these holographic models may change for $\abs{\delta \rho} \le 1$ as reviewed in section \ref{sec2}. Values of $\delta\rho$ can be restricted by choosing the bulk mass $M$ properly.} of $q=\sqrt{6}/5$ { and with the Neumann boundary condition \eqref{IRB216} in Fig. \ref{fig:nonh2}. The absolute value of the VEV becomes smaller than those of the homogeneous phase, because the contribution from the IR potential is suppressed by the Neumann boundary condition \eqref{IRB216}. The absolute value of the VEV is also smaller when $q$ is small. The small values imply that the free energy in the non-homogeneous phase changes more smoothly compared to the homogeneous phase. {Finally, from Fig.~\ref{fig:nonh2} we conclude that the VEV in the non-homogeneous phase (blue and red curves) is much more sensitive to changes of the anomalous dimension (changes of $M^2$) than in the homogeneous phase (purple and green curve). We think that this is due to the reduced contribution from the IR to the hopping VEV due to the Neumann boundary condition \eqref{IRB216}, which means increased sensitivity to changes in the UV contribution to the VEV: The Neumann boundary conditions \eqref{IRB216} allows the IR potential adjust itself dynamically towards a minimum of the IR potential, which is however bought by an additional contribution coming from the UV part of the VEV. On the other hand, the hard wall boundary condition \eqref{BPH320} is constructed to set the subleading piece of the UV expansion of $\phi$ to zero, and hence, since $\phi$ corresponds to an irrelevant operator, the UV contribution as defined in \eqref{VEVtotal} is strongly suppressed compared to the IR contribution.}
 
\begin{figure}
     \begin{center}
          \includegraphics[height=4cm,clip]{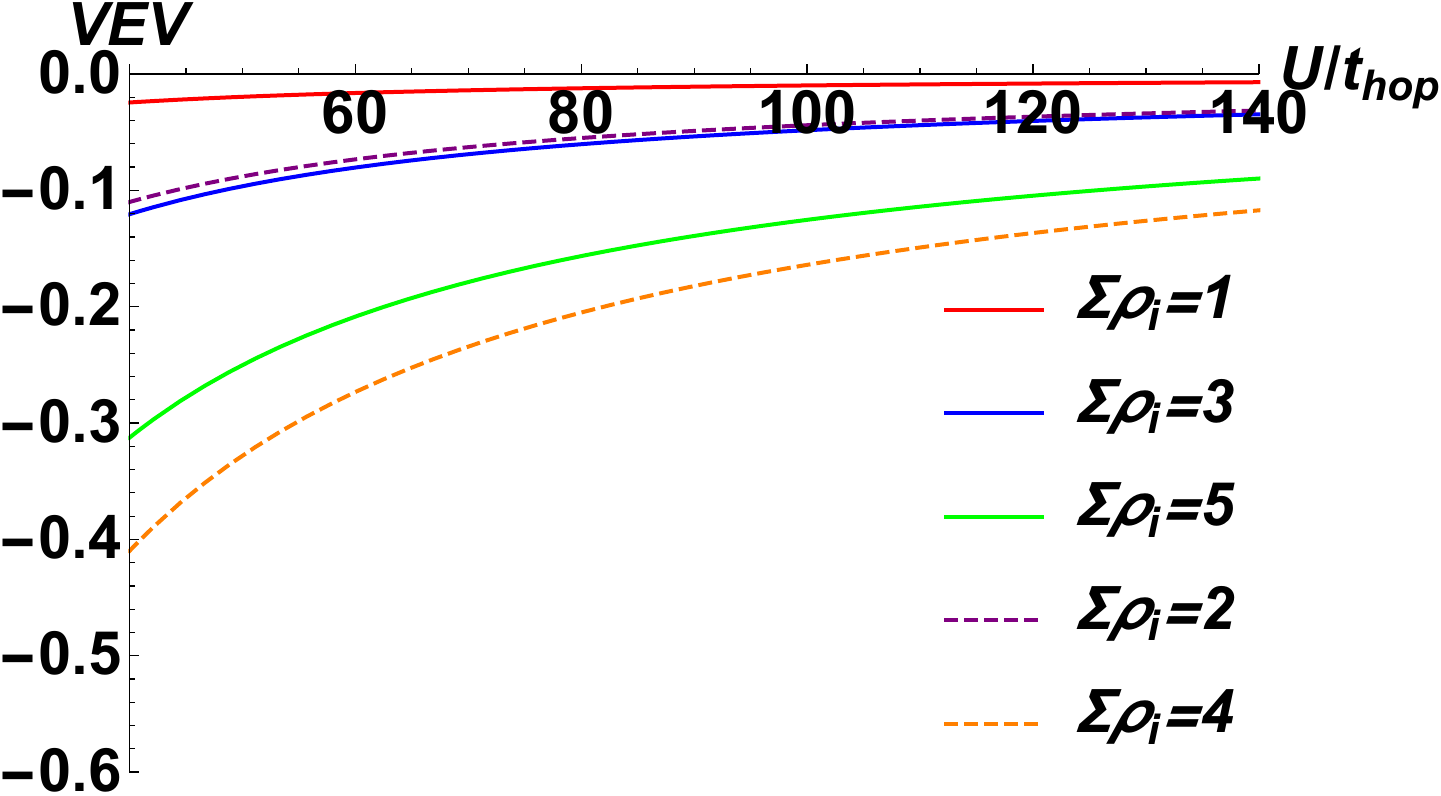} 
           \includegraphics[height=4cm,clip]{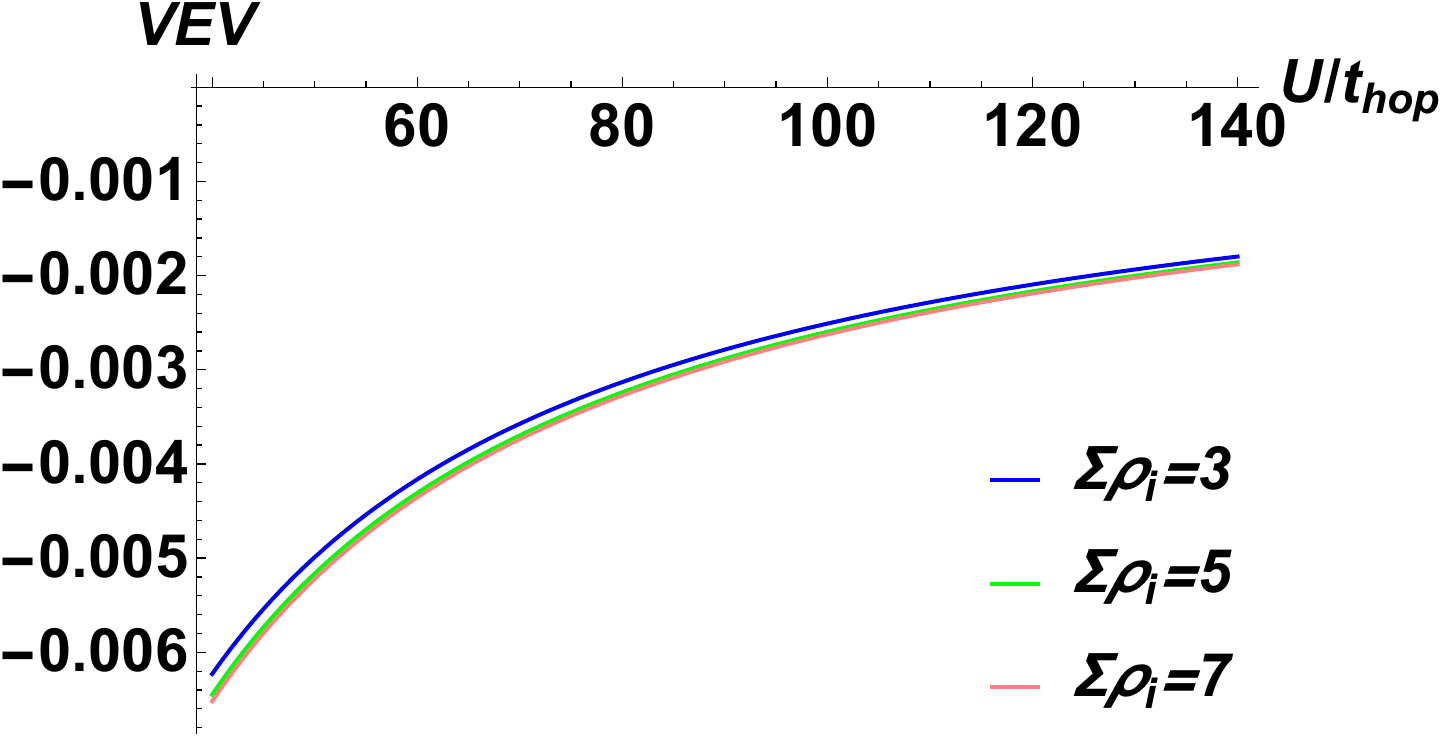} 
          \hspace{1.6cm}
        \end{center}
         \caption{The normalized VEV is plotted as the function of $U/{t_\mathrm{hop}}$ when $M^{2}=-21/100$ and $q^2=3/100$ (weak coupling). Other parameters are fixed to be $u_\mathrm{h}=40$, $\lambda =1/\Lambda^2$, $w=0$, $\Lambda_{(1,1)}=-3/(2\Lambda)$, other $\Lambda_{(p,q)}=0$. Now, $\Lambda =120\times 2^{1/5}\times 5^{2/5}$.  Left: The IR boundary condition $\varphi=0$ is imposed.  Dashed lines represent curves in the Mott insulator phase. Dashed curves are parametrized by $\sum \rho_{i}= 2,\ 4$. Right: The VEV for the same parameter choices, but now with Neumann boundary condition \eqref{IRB216}. We find that the absolute values of the VEV in the non-homogeneous phase are larger for Dirichlet than for Neumann boundary conditions. Larger VEVs are expected in the superfluid phase, which favors the Dirichlet condition in the superfluid phase as well.}
    \label{fig:nonh3}
\end{figure}

We also found that the effective hopping in the non-homogeneous phase is similar to that of the homogeneous Mott insulator phase discussed in sec.~\ref{sec32}, if we impose the Dirichlet boundary condition fixing $\varphi_v=0$  in~\eqref{PER28} instead of the Neumann boundary condition \eqref{IRB216}. With Dirichlet conditions, the absolute value of the VEV increases as the total occupation number increases, as can be seen from the left hand side of Fig.~\ref{fig:nonh3}, which is plotted with parameters $q^2=3/100$, $M^{2}=-21/100$,  $u_\mathrm{h}=40$, $\lambda =1/\Lambda^2$, $w=0$, $\Lambda_{(1,1)}=-3/(2\Lambda)$, other $\Lambda_{(p,q)}=0$. Now, $\Lambda \approx 262$. By contrast, the right hand side figure employs the Neumann condition \eqref{IRB216}, but otherwise same parameter choices as the left hand side figure. By comparing both figures, we find that the Neumann boundary condition leads to much smaller values for the VEV in the non-homogeneous phase compared to the Dirichlet condition. Since we expect the VEV in the non-homogeneous superfluid phase to be large, this observation favors the use of a Dirichlet condition in the non-homogeneous phase as well as in the homogeneous phase.\footnote{In the homogeneous phase the Dirichlet condition was chosen to ensure the vanishing of the subleading part $\varphi_v=0$ in the UV expansion \eqref{PER28}.}

\section{A holographic $SU(N)$ fermionic Hubbard model}\label{holohubbard}

{In this section, we construct a top-down holographic model dual to the $SU(N_c)_k$ {Fermi} Hubbard model at half filling, i.e. a Hubbard model in which fermions hop between lattice sites, but themselves transform in the fundamental representation of a $SU(N_c)$ gauge group whose dynamics is of Yang-Mills-Chern-Simons type with Chern-Simons level $k$.}  The field theory side is given by the {low energy limit of a} D3-D5-D7 system of the type IIB superstring theory. We start from the holographic dual to the level-rank duality built from the D3-D7 system \cite{Fujita:2009kw}, which we review in App.~\ref{LEV} as well as briefly below. We insert into this model a stack of $n_F$ coincident D5 branes carrying a $U(n_F)$ gauge theory. {The D5 branes end on the D7 brane in the IR. The rank $n_F$ can be interpreted as the number of the lattice sites after separating the D5 branes in the boundary spatial directions. Separating the D5 branes breaks the $U(n_F)$ symmetry down to $U(1)^{n_F}\subset U(n_F)$, and the low energy effective action in terms of the unbroken  gauge fields and corresponding modes coming from open strings connecting the separated D5 branes will be mapped both to the operator content and interactions of the bottom-up model \eqref{bottomupmodel} as well as  of the {Fermi} Hubbard model at half filling \eqref{eq:twosite_hamiltonian}. }

\subsection{Multiple D5-branes with non-Abelian symmetry}

Probe D5-branes with non-Abelian symmetry are considered on the $AdS_5$ soliton background with metric~\eqref{SOL14}. At energies below the gap, the solitonic geometry describes the confining vacuum of non-supersymmetric 3+1 dimensional $SU(N_c)$ Yang-Mills theory. The confinement scale is set by the radius of the circle on which the D3 branes giving rise to the solitonic background in the large $N_c$ limit are compactified on. The D5-branes wrapping $(t,u)$ and $S^4$ directions inside the transverse $S^5$ are introduced in the probe limit, i.e. without considering their backreaction. In order to attach the D5-branes on the tip of the $AdS$ soliton, by flux conservation~\cite{Strominger:1995ac} they need to end on another D brane. We engineer this by letting them end on D7-branes at the tip of the soliton. The setup is summarized in table~\ref{tab:setup}. 
    
\begin{table}[t]
\caption{D brane setup of the top-down construction of a $SU(N_c)_k$ Fermi-Hubbard model. $\tau$ is the compactified cigar direction, $u$ is the holographic direction, and $x^{5\dots 9}$ are the $S^5$ coordinates. We choose an embedding of $S^4$ into $S^5$ such that $x^{5\dots 8}$ parametrizes the $S^4$ wrapped by the D5 branes.}
\begin{equation*}
  \begin{array}{|c||c|c|c|c|c|c|c|c|c|c|}
    \hline
     & x^0 & x^1 & x^2 & x^3=\tau & x^4=u & x^5 & x^6 & x^7 & x^8 & x^9 \\
    \hline
    N_c \,\,\,{\rm D3} & \times & \times & \times & \times & \cdot & \cdot & \cdot & \cdot & \cdot & \cdot \\
    \hline
    k \,\,\,{\rm D7} & \times & \times & \times & \cdot & \cdot & \times & \times & \times & \times & \times \\
    \hline
    n_F \,\,\,{\rm D5} & \times & \cdot & \cdot & \cdot & \times & \times & \times & \times & \times & \cdot \\
    \hline
  \end{array}
  \end{equation*}
\label{tab:setup}
\end{table}

{
From the field theory point of view, the spectrum of the D3-D7 strings is non-supersymmetric and contains $k$  massive 2+1 dimensional Dirac fermions (with mass of order of the gap scale) transforming in the fundamental of the $SU(N_c)$ gauge group as well as a $U(k)$ global flavor symmetry. At low energies, the Dirac fermions can be integrated out, giving rise via the parity anomaly to a level $k$ Chern-Simons term for the $SU(N_c)$ gauge field. This is the holographic dual relevant for level rank duality. It was first described in \cite{Fujita:2009kw}, and its domain walls (which are supersymmetric) were recently analyzed in \cite{Fujita:2016gmu}. The effective theory at energies below the gap scale is hence a $SU(N_c)_k$ non-abelian Chern-Simons theory. \footnote{Note that the absence of supersymmetry is not a big drawback of this construction, since the gapped background breaks supersymmetry by itself, and the embedding of the D7 branes is stabilized by the occurrence of the gap - the D7 branes are simply staying at the bottom of the $x^3$-cigar.}

The D3-D5 strings on the other hand are 8 ND and hence supersymmetric \cite{Gomis:2006sb}, and contain $n_F$ 0+1-dimensional fermionic degrees of freedom transforming in the fundamental of the $SU(N_c)$ gauge group. This defect has recently been used to model the spin impurity in the holographic Kondo construction of \cite{Erdmenger:2013dpa}. In our context, these fermions are the hopping degrees of freedom residing on the 0+1-dimensional defects. We are hence describing a holographic dual to a Fermi-Hubbard model with gauge group and Chern-Simons level $SU(N_c)_k$.\footnote{For general $k$ and $N_c$, the presence of the Chern-Simons level influences the statistics of the D3-D5 strings. Following the discussion in \cite{Fujita:2009kw}, a single string stretching between the D5 brane and $N_c$ D3 branes in a definite color picks up a phase $e^{\frac{\pi i}{k}}$ when interchanged with another F1 string of the same color. When $k=1$, these strings have fermionic statistics. As far as the statistics of two D5 branes with charge density, as necessary for the construction of the Fermi-Hubbard model, is concerned, as explained around \eqref{quantizationcondition}, the charge density on the branes and hence the number of strings is quantized together with the embedding angle on the transverse $S^5$.}

Finally, the D5-D7 intersection is 4 ND and hence supersymmetric. We can hence trust the DBI-Wess-Zumino actions describing the coupling of these branes to the Ramond-Ramond gauge fields sourced by each other, and the corresponding flux conservation arguments~\cite{Strominger:1995ac}. These results ensure that the D5 brane can end on the D7 brane at the cigar tip.\footnote{The AdS soliton solution breaks sypersymmetry. Nevertheless, our system is locally supersymmetric since at vanishing temperature the tip of the cigar is locally flat, and the D5-D7 solution was supersymmetric in flat space-time. At finite temperature,  the D7 brane is too heavy to be pulled up by the D5 branes.}}

The remainder of this section is concerned with analyzing the dynamics of the multiple D5 branes as they are separated in the non-compact $x^1$ or $x^2$ directions in order to realize a multi-site version of the above-described bottom up Hubbard model construction. For $n_F=2$ we construct a top-down version of the two-site $SU(N_c)_k$ Fermi-Hubbard model. The embedding of the D5 branes into the D3-D7 model of \cite{Fujita:2009kw} is summarized in table \ref{tab:setup}. In order to proceed, we need to introduce some geometric structures on the D5 branes: In the Abelian case of a single D5 brane, the pull-back of the spacetime metric onto the brane is defined  as
\ba
P[G_{ab}]=\partial_a x^{\mu} \partial_b x^{\nu} g_{\mu\nu}=g_{ab}+(2\pi\alpha')^2\partial_a \Phi^{i} \partial_b \Phi^{j} g_{ij},
\ea
where $\mu$ is running through all 10 space-time dimensions, $a,b=t,u,S^4$ are the wrapped directions, $i,j$ denote the transverse directions and $g_{ai}=0$. The transverse scalars of the D5-branes are given by  $(2\pi\alpha')\Phi^i$ where ($i=x_1,x_2,\tau,\theta$) and have dimension of length. In the non-Abelian case, the partial derivative should be replaced by the $U(n_F)$ covariant derivative  $D_a\Phi^i=\partial_a\Phi^i+i[A_a,\Phi^i]$~\cite{Myers:1999ps}. The derivatives in the pull-backs are consequently replaced by $D_a\Phi^i$. The Neveau-Schwarz B-field $B^{NS}$ is switched off in our background. 
The non-Abelian D5 worldvolume action becomes 
\ba\label{D524}
&I_{D5}=-T_5\mbox{Str}\Big(\int d^6x\sqrt{-\mbox{det}(P[G_{ab}+G_{ai} (Q^{-1}-\delta)^{ij}G_{jb}]+(2\pi\alpha')F_{ab})\mbox{det}(Q^i{}_j)} \nonumber \\
&+\sum_{\alpha}\int P[e^{i(2\pi\alpha')i_{\Phi}i_{\Phi}}C_{\alpha}]e^{2\pi\alpha' F}\Big)+S_{fermion}, 
\ea
where $T_5=1/(2\pi)^5g_{s}\alpha^{\prime 3}$ and $Q^i{}_j\equiv \delta^i{}_j+i(2\pi\alpha')[\Phi^i,\Phi^k]g_{kj}$. Indices $i,j$ such as e.g. on $(\delta-Q^{-1})^i{}_j$ are lowered in terms of the transverse metric $g_{ij}$. $i_{\Phi}$ is the interior product in the direction of $\Phi^i$.  The meaning of the symmetric trace prescription $\mbox{Str}$ is to symmetrize over the gauge indices neglecting all commutators of the field-strength $F_{ab}$,  $[\Phi^i,\Phi^j]$, and single $\Phi^i$, i.e. $\mbox{Str}(\Phi^{i_1}\dots \Phi^{i_n})\equiv \mbox{Tr}(\Phi^{i_1}\dots \Phi^{i_n}+\mbox{all permutations})/n!$. In the non-Abelian case, $C_{\alpha}$ also has the $\Phi_i$ dependence. 
A consistent non-Abelian Taylor expansion for the RR-fields is \cite{Garousi:1998fg,Kabat:1997im}
\ba
C_{\alpha}(\Phi^i)=\sum_{n=0}^{\infty}\dfrac{(2\pi \alpha')^n}{n!}\Phi^{i_1}\dots \Phi^{i_n}\partial_{x^{i_1}}\dots \partial_{x^{i_n}}C_{\alpha}(\phi^i)|_{\phi^i=0}.
\ea
{In our case, the only RR background field will be $C_4$, sourced by the D3 branes.}

\subsection{Homogeneous Phase}\label{sec42}

\begin{figure}
     \begin{center}
          \includegraphics[height=3.5cm,clip]{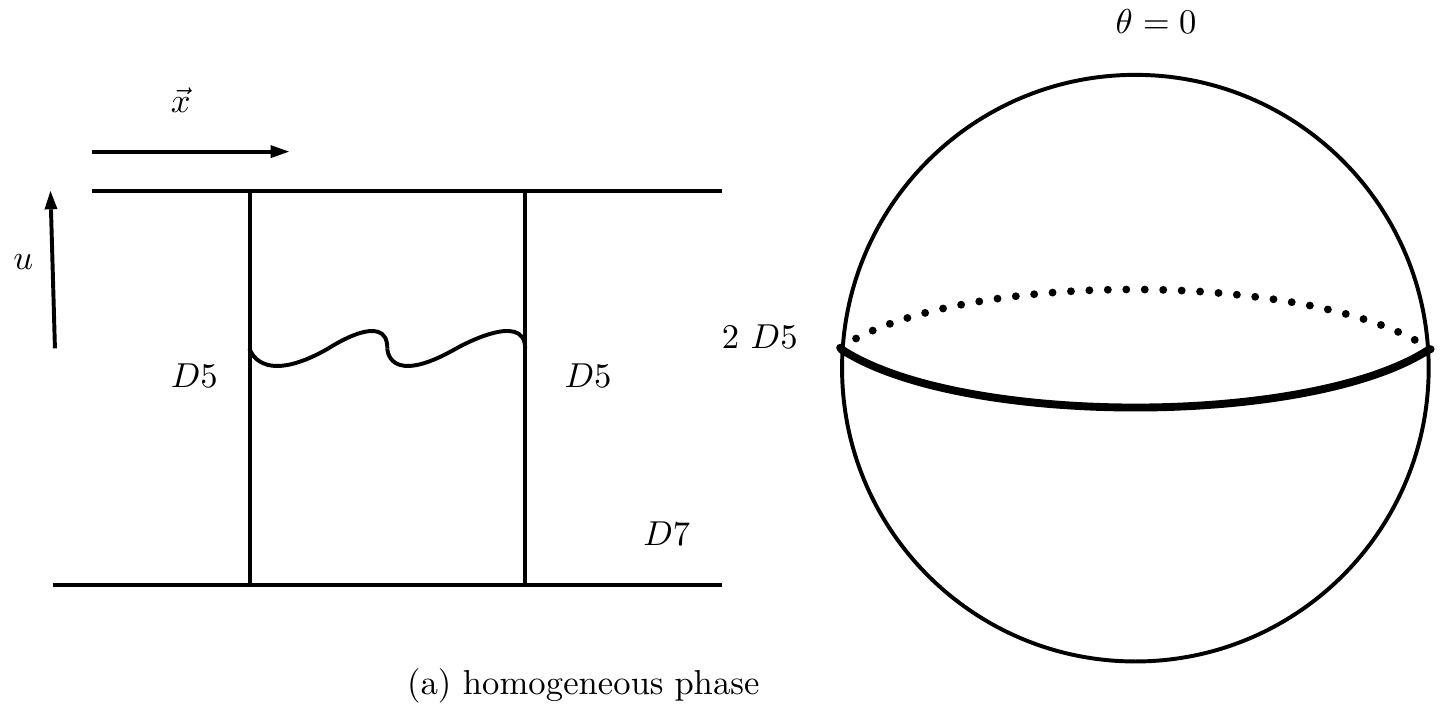} 
          \includegraphics[height=3.5cm,clip]{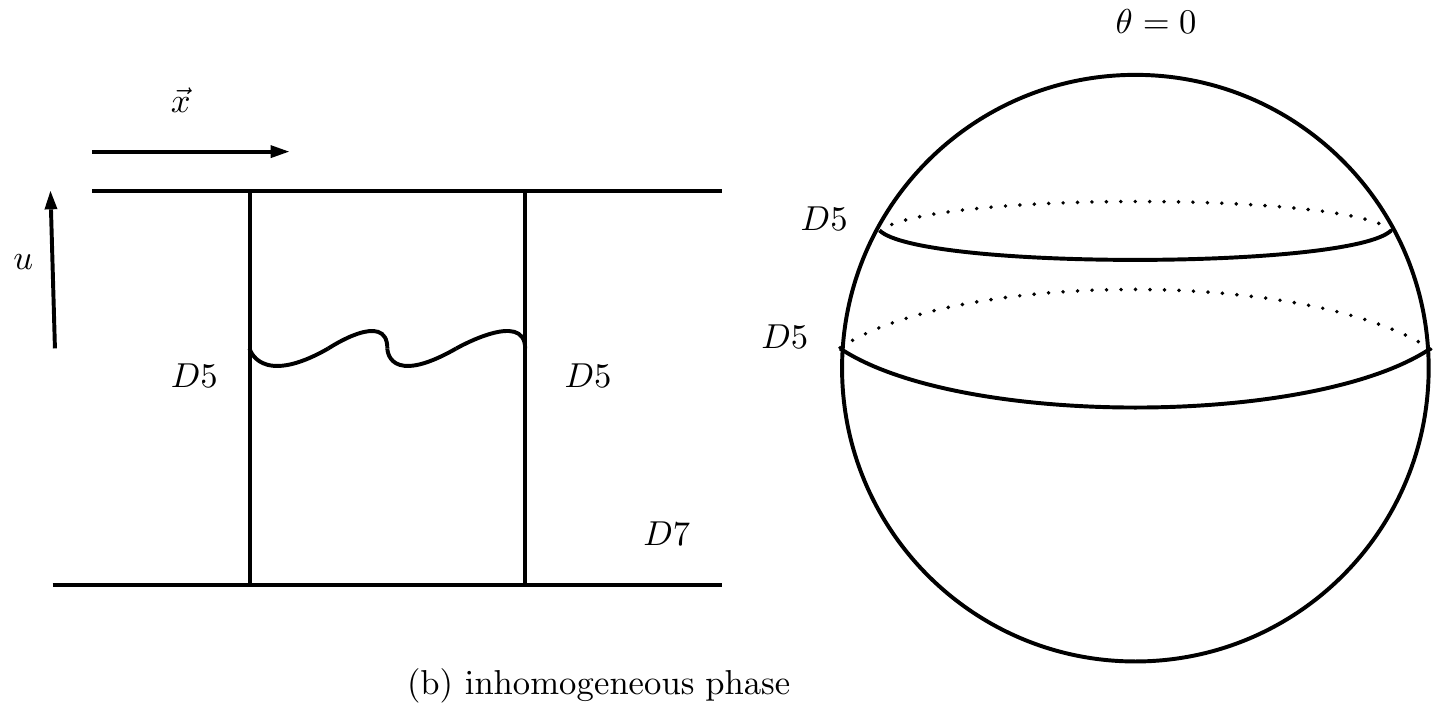}   
          \hspace{1.6cm}
        \end{center}
\caption{Phases of the top-down model: {(a) The homogeneous phase is described by the two D5 branes wrapping the $S^4\subset S^5$ at the same polar angle. The charge quantization condition \protect\eqref{quantizationcondition} then implies that the number density is the same on both defects. (b) In the inhomogeneous phase, the branes wrap different angles on the $S^5$, corresponding to a difference in charge density. The D5 branes wrapping angles close to the equation corresponds to half filling.}}
    \label{fig:topdown}
\end{figure}

In the top-down model with $n_F=2$, we make an Ansatz for the gauge fields and transverse scalars as 
\ba\label{VEVs}
A_b=\mbox{diag}(a_b^{(1)},a_b^{(2)}),\quad \theta=\mbox{diag}(\theta_{11},\theta_{22}),\quad  \Phi^{x_1}=\begin{pmatrix}0 & w^{x_1} \\ \bar{w}^{x_1} & 0 \\
\end{pmatrix},
\ea
with all other transverse scalars to be zero. 
One could also choose $w^{x_2}$ instead of $w^{x_1}$, due to the rotation symmetry in the $x_1x_2$-plane. As pictured in Fig.~\ref{fig:topdown}, the branes can be separated in the field theory direction $x_1$ as well as in the polar angle transverse to the $S^4\subset S^5$. As we will see, these fields are sufficient to describe a holographic Fermi-Hubbard model: Separation only in the field theory direction corresponds to the homogeneous Mott phase, while additional separation in the sphere direction corresponds to the inhomogeneous superfluid phase.

In the homogeneous phase, the D5 branes are separated in the field theory direction but not in the polar angle $\theta_{11}=\theta_{22}$,  and the gauge fields are constrained to be equal, $a_b^{(1)}=a_b^{(2)}$. The eigenvalues of the matrix $\Phi^{x_1}$ are $\pm \abs{w^{x_1}}$,\footnote{Diagonalization of $\Phi^{x_1}$ corresponds to a $SU(2)$ gauge transformation compatible with the Ansatz.} which yield the separation in the field theory direction $x_1$. 
 Note that in the homogeneous phase, $[\theta,\Phi^{x_1}]=0$ and $D_b\Phi^{x_1}=\partial_b \Phi^{x_1}$. 
 Since the off-diagonal components are only from $\Phi^{x_1}$, the action reduces to that of two separated D5 branes, each with Abelian gauge symmetry.  Considering the relation $[\theta,\Phi^{x_1}]=0$, the field $w^{x_1}$ does not receive a mass from the commutator squared potential in \eqref{Sb4532} (c.f. also \eqref{TOT436}). The field $w^{x_1}$ is the analogue of the W bosons in the standard model of particle physics. A constant solution $w^{x_1}=t_\mathrm{hop}$ is allowed in the homogeneous phase and does not contribute to the on-shell action, but breaks the $U(2)$ gauge symmetry down to the {$U(1)=\frac{1}{\sqrt{2}}\left(U(1)_{(1)}+U(1)_{(2)}\right)$} baryonic symmetry, corresponding to the vector symmetry of the bottom-up model. The mechanism of this symmetry breaking is similar to the one of chiral symmetry breaking in the AdS/QCD model of \cite{Erlich:2005qh,Karch:2006pv}, where the constant $t_\mathrm{hop}$ corresponds to the explicit breaking via a quark mass, and the hopping kinetic VEV corresponds to the spontaneously generated chiral condensate. 
From this construction we see that  $w^{x_1}$ corresponds to the hopping scalar of the bottom-up model. The Z boson ($a_b^{(1)}-a_b^{(2)}$) is also massless and corresponds to the axial gauge field $U(1)_{(1)}-U(1)_{(2)}$ of the bottom-up model, which does not contribute in the homogeneous phase.\footnote{A finite value of $t_\mathrm{hop}$ explicitly breaks the axial symmetry. As with the axial symmetry breaking in QCD, this will yield to a gap in the excitations of the Z boson field, while the background value vanishes.} 

Relations  $\theta_{11}=\theta_{22}$ and $a_b^{(1)}=a_b^{(2)}$ imply via the following angle quantization condition \eqref{quantizationcondition} in the $S^5$ direction perpendicular to $S^4$ that the charge density must be equal for both D5 branes, and hence the charge density on both boundary defects coincides~\cite{Camino:2001at}. By a well-known argument~\cite{Camino:2001at}, the charge density on the D5 brane must be quantized as an integer multiple of the fundamental string tension (i.e. the number of open strings ending on the brane).  The EOM for the transverse scalar $\theta$ has a constant solution, which is quantized in terms of the charge density such that the fundamental string tension cancels exactly the amount of charge induced on the brane by the $C_4$ RR field. The quantization condition has been worked out in \cite{Camino:2001at}, reading
\ba\label{quantizationcondition}
\dfrac{\pi n}{N_c}=\bar{\theta}-\sin\bar{\theta}\cos\bar{\theta}\,,
\ea where $\bar{\theta}=\pi-\theta$. The charge density $n$ on the brane equals the number of $(1+0)$-dimensional fermions at a site~\cite{Kachru:2009xf}.

This discussion shows that the effective tension {of the D5 brane}~\cite{Karch:2003nh} is suppressed in the large $N$ limit when the occupation number is of order 1, i.e. when the D5 brane sits close to the pole of the $S^5$, $\theta\sim 0$. From a string theory point of view, vanishing tension is not desirable, as fluctuations of the brane fields will not be suppressed compared to their background values any longer. The natural other locus for the D5 branes on the $S^5$ is the equator $\theta_{11}=\theta_{22}=\frac{\pi}{2}$. Since the D3-D5 intersection is supersymmetric, and as we will see explicitly in sec.~\ref{sec44} below, the slipping mode of the D5 on the equation will not violate the Breitenlohner-Freedman bound in $AdS_2$. Hence, expanding around the equator, the fluctuations of the brane will be suppressed compared to the background in the natural large $N$ scaling. The state at $\theta=\pi/2$ has an occupation number $N_c/2$. When comparing with the large $N$ Fermi-Hubbard model at half filling below, we will exactly recover this large $N$ scaling in the top-down model, and be able to match the operators and their sources to the corresponding terms in the Fermi-Hubbard lagrangian \eqref{eq:twosite_hamiltonian}.

\subsection{Inhomogeneous Phase}\label{sec43}

The inhomogeneous phase is characterized by a difference in the charge density on both defects. The D5 branes are consequently separated both in the field theory as well as in the $S^5$ direction due to the charge quantization \eqref{quantizationcondition}, $\theta_{11}\neq \theta_{22}$ and $a_b^{(1)}\neq a_b^{(2)}$. In the inhomogeneous phase, there are couplings between W and Z bosons from the nonabelian covariant derivative. 
As in the homogeneous phase, $w^{x_1}$ is the hopping scalar. Considering $[\theta,\Phi^{x_1}]\propto (\theta_{11}-\theta_{22})\neq 0$, the W boson now acquires the mass proportional to the difference in angles on the $S^4\subset S^5$. Due to the charge quantization condition \eqref{quantizationcondition}, the W boson mass and hence the hopping VEV now is related to the difference in charge density on both defects.\footnote{Due to the non-Abelian structure of fields \eqref{VEVs}, we expanded the non-Abelian D5 brane action up to ${\cal O}(\alpha'{}^4)$ in appendix \ref{NONA}. For the arguments given in this section, ${\cal O}(\alpha'{}^2)$ are sufficient. Note that one needs to match orders of $\xi$ to analyze thermodynamic stability between the two phases.}

\subsection{Comparison with the bottom-up model}\label{sec44}

\begin{table}
\caption{Comparison of top-down with bottom-up construction.}
  \label{TU2}
  \begin{center}
    \begin{tabular}{|l|l|} \hline
Top-Down Fermi-Hubbard   & Holographic Bose-Hubbard \protect\eqref{bottomupmodel}\\ \hline
$U(1)^{n_F}$ gauge symmetry &$U(1)^{n_F}$ gauge symmetry  \\ \hline
Bi-fundamentals $w_{i}$ & Bi-fundamentals $\phi_{ij}$ \\\hline
Bi-fundamental mass $M^2 L^2= 2$ & \textit{A priori} arbitrary $M^2 L^2$ \\\hline
Quantized number of F1 string &  Quantized occupation number \\ \hline
Near half filling state $(n_i=N/2)$ & $n_i\sim {\cal O}(1)$ or ${\cal O}(N)$ depending on normalization of \protect\eqref{bottomupmodel} 
\\ \hline
Brane tension ${\cal O}(N)$ close to equator & $S\sim {\cal O}(1)$ or ${\cal O}(N)$ depending on normalization of \protect\eqref{bottomupmodel} \\ \hline
    \end{tabular}
  \end{center}
\end{table}

In this section, we compare the bottom-up model \eqref{bottomupmodel} with the top-down holographic model of table~\ref{tab:setup}. We start with the {homogeneous phase,} the ansatz for which is \eqref{VEVs} with $\theta_{11}=\theta_{22}$ and $a_b^{(1)}=a_b^{(2)}$. The distance modes in $\theta$ are the eigenvalues, in matrix form (see also \eqref{AW78}), 
\ba\label{AW34}
\theta =\Big(\dfrac{\pi}{2}+\xi \varphi^{v}\Big)\mathbf{1}_{2\times 2},
\ea
where $\xi = 2\pi\alpha'$ and $v=\theta$ denotes the transverse scalar for the polar angle on the $S^5$. This Ansatz and gauge fields proportional to the unit matrix do not break the $U(2)$ symmetry. Since the adjoint scalar $\Phi^{x_1}$ commutes with the other matrices, the action reduces to the one for two separate D5-branes with Abelian symmetry. The mass of the W-boson (an open string mode) is zero. 

{In order to match to the bottom-up model \eqref{bottomupmodel},} we expand the action in terms of the bi-fundamental scalar $w^{x_1}$, the gauge field $a_b^{(1)}$, a transverse scalar $\varphi^v$ around the background $\theta=\pi /2 \mathbf{1}_{2\times 2}$ up to O($\xi^2$) (c.f. Appendix \ref{DBIap} for details). In the Hubbard model side, only one hopping term appears in the Hamiltonian. Hence, one needs only one bi-fundamental scalar in the lagrangian, which can be achieved by a rotation in the $x^1-x^2$ plane which aligns a general $w^i$, $i=x^1,x^2$, along $x^1$. The quartic potential in the EOM \eqref{EOMD32} vanishes in this case, and the EOM becomes
\ba\label{EOM32}
\dfrac{1}{\sqrt{-\bar{g}}}\partial_c(\sqrt{-\bar{g}}\partial^cw^{x_1}\bar{g}_{x_1x_1})=0.
\ea
Using the explicit form of $\bar g$, which is the metric induced from \eqref{SOL14} onto the D5 brane in $(t,u)$ direction at the embedding $\theta=\pi/2$, one finds that after a redefinition {$w^{x_1} = v^{x_1}/u$}, the EOM becomes that of a canonically normalized scalar $v^{x_1}$ in AdS$_2$ with $M^2L_{AdS}^2 = 2$. {The general homogeneous solution is now of the form $w^{x_1}=t_\mathrm{hop}+ \varphi u^{-3}$.} {The Neumann boundary condition $w^{x_1 \prime }|_{u=u_\mathrm{h}}=0$ can be employed at the tip of the soliton, which sets the VEV piece $\varphi=0$ and hence has the same effect as the generalized Dirichlet condition employed in sec.~\ref{sec32}.} The solution describing a vanishing VEV piece is then constant, $w^{x_1}=t_\mathrm{hop}$.\footnote{Note that the top-down model as it is does not seem to produce any IR potential terms. However, since the D5 brane ends on the D7 brane at the tip of the cigar, nontrivial field configurations in the D7 worldvolume theory could yield such terms. We will investigate this in future work.}

In the non-homogeneous phase, adjoint fields in the non-Abelian action do not commute anymore. The non-Abelian action should be expanded in terms of $\alpha'$ and the symmetric trace evaluated. The details of the non-Abelian Taylor expansion can be found in Appendix \ref{DBIap}.
To compare with the bottom-up model it is convenient to perform the rescalings
\ba
\bar{g}_{ab}\to L_1^2 \tilde{g}_{ab},\ \bar{g}_{ij}\to L_1^2 \tilde{g}_{ij},\  \Phi^i=\dfrac{\Phi_x^i}{\xi},
\ea
where $L_1$ is the AdS radius, and $\xi = 2\pi \alpha'$. The transverse scalars have dimension of the length ($i\neq \theta$) or dimensionless ($i=\theta$), coinciding with the bottom up model \eqref{bottomupmodel}.

As will become clear from the following arguments, the action of the rescaled theory is proportional to $N_c$ like in typical probe brane models. At the equator of the $S^5$, $\theta =\pi/2$, the coupling constant $q_t$ on the $AdS_2$ induced hard wall geometry is ${1}/{q_t^2}=4N_c/(3\sqrt{\lambda_t})$, which is much smaller than the effective $5d$ gravitational coupling $1/2\kappa^2=N_c^2/(8\pi^2L_1^3)$ by dimensionally reducing on $S^5$. The other coefficients of the probe brane are also of order $N_c$, and hence the probe limit is valid in our regime of interest. {We hence find that we can truncate the top-down model for D5 branes at $\theta={\pi\over 2}$ to the bottom-up approach of sec.~\ref{sec2} with small anomalous dimensions\footnote{Smallness of $q^2 \delta\rho^2$ seems important in order to prevent hitting the unitarity bound in \eqref{anomdimMrho}. One could also try to scale $\delta \rho \sim N_c^{\frac{1}{2}}$ in order to achieve finite anomalous dimensions.} if $\delta \rho \sim {\cal O}(1)$ in the large $N_c$ limit. This is consistent with small fluctuations of the brane embedding at the equator of the $S^5$, as $\rho_{total} = \rho_1+\rho_2 \sim N_c$.} 

As discussed around \eqref{EOM32}, the quartic interaction of $w_{x}^I$ vanishes in the EOM \eqref{EOMD32} when only a single component is switched on. Such a quartic interaction resembles a $\phi^4$ interaction at the hard wall. The top-down model implies that no quartic interactions correspond to the bottom-up model with $\lambda =0$ in \eqref{ACT11}. Actually, vanishing quartic interaction terms are consistent with the small tension limit (large $\Lambda$) of \eqref{ACT11}, under the rescaling $(\phi, \lambda, w^2, \Lambda_{p,q})\to (\sqrt{\Lambda}\phi, \lambda/\Lambda^2, w^2/\Lambda, \Lambda_{p,q}/\Lambda^p) $ {for large $\Lambda$.}

Also the cubic interaction $[\Phi^i,\Phi^j]F_{ab}F^{ab}$, which resembles the quartic irrelevant term $\phi^2 F_{ab}F^{ab}$ in  the bottom-up model \eqref{ACT11}, is not present in \eqref{bigeq} due to the symmetric trace prescription. The higher order action   \eqref{S2b} includes the following interaction
\ba\label{Sb4532}
&S_2^b\supset - \mbox{Str}\int d^2x \sqrt{-\det (\tilde{g}_{ab})}\Big(\dfrac{1}{8q_t^2}D_c\Phi^i_xD^c\Phi_{x,i}F_{ab}F^{ab}-\dfrac{1}{\Theta_{t,1}}\delta\theta^2 D_a\Phi^i_xD^a\Phi_{x,i} \nonumber \\
&+\dfrac{1}{16\Theta_{t,1}}[\Phi_x^j,\Phi_{k,x}][\Phi_x^k,\Phi_{j,x}]F_{ab}F^{ab} \Big),
\ea
where coefficients are given by in terms of the 't Hooft coupling {$\lambda_t$}
\ba\label{eq:coeffs}
&\dfrac{1}{q_t^2}=\dfrac{T'\xi^2}{L_1^2}=\dfrac{4N_c}{3\sqrt{\lambda_t}},
\quad\dfrac{1}{ \Theta_{t,1}}=T'L_1^2=\dfrac{N_c\sqrt{\lambda_t}}{3\pi^2},
\ea
These terms are similar to the quartic irrelevant interaction \eqref{ACT11} at the hard wall since when valuated on the homogeneous solution they include $t_\mathrm{hop}^2\rho^2_{(i)}$. In components, the third term of \eqref{Sb4532} becomes similar to $(\phi_v-L_v)^2\abs{w}^2F_{ab}F^{ab}$. Since the third term vanishes for $\phi_v=L_v$, such a term does not play the role of the quartic irrelevant term in  the hard wall \eqref{ACT11}. 

In conclusion, comparing the top-down model \eqref{bigeq} with the bottom-up model \eqref{ACT11}, we find that the most relevant IR potential parameters in \eqref{ACT11} vanish, $\lambda=0$ and $\Lambda_{(1,1)}=0$, but less relevant terms are generated by the non-abelian structure of the DBI action \eqref{D524}.

\subsection{Matching the Fermi-Hubbard model at half filling}\label{sec45}

With the holographically renormalized free energy of the top-down model in units of $\alpha'=1$,\footnote{The free energy \eqref{FRE446} is an extension of the free energy of a single D5 brane eq. (2.18) in \cite{Kachru:2009xf} to an $AdS_2$ hard wall-like geometry induced by the AdS$_5$ soliton background. The first term of \eqref{FRE446}, which  is not present in the black hole embedding of the D5-brane, is generated from the hard wall boundary condition on the gauge field $a_t^{(i)}|_{u=u_h}=const$. }
\ba\label{FRE446}
&F=\sum_{i=1}^2\Big(\mu\Big(\dfrac{N_c}{2}+\rho_{(i)}\Big)+c_\mathrm{os}\dfrac{u_h}{N_c}\rho_{(i)}^2-\dfrac{8c_\mathrm{os}}{\pi^2}N_cu_h+\dots\Big),
\ea at hand, we can now match its parameters to the two-site $SU(N)$ Fermi-Hubbard model \eqref{eq:twosite_hamiltonian} by comparing the corresponding large $N$ scaling properties ($N=N_c$). The large-N scaling of \eqref{eq:twosite_hamiltonian} is the same as for the two-site $SU(N)$ Bose-Hubbard model \eqref{SUNBoseHubbard} described in footnote~\ref{BHNscaling}: Requiring the Hamiltonian \eqref{eq:twosite_hamiltonian} to be ${\cal O}(N)$ and $t_\mathrm{hop}$ as well as the chemical potential $\mu$ to be color independent, one finds that the charge density scales as $n_j \sim {\cal O}(N)$, as does the hopping operator VEV $\langle c^\dagger_{1,\alpha} c_{2,\alpha}
+ c^\dagger_{2,\alpha} c_{1,\alpha}\rangle \sim {\cal O}(N)$. 

Comparing to the form of the free energy \eqref{FRE446}, {the chemical potential $\mu$ is ${\cal O}(1)$. The charge density $N_c/2+\rho_{(i)}$ near the half filling state is ${\cal O}(N_c)$. The on-site interaction $c_\mathrm{os}u_h/N_c$ is ${\cal O}(N_c^{-1})$, where $c_\mathrm{os}=\sqrt{\pi}\Gamma (3/4)/(24\Gamma(1/4))=0.024961$. Higher power interactions of $\rho_{(i)}$ are suppressed in the large $N_c$ limit, when $\rho_{(i)}/N_c\ll 1$, i.e. $\rho_{(i)}$ scales slower than $N_c$. The hopping term vanishes in the free energy in the background of the homogeneous phase $w^{x_1}=t_\mathrm{hop}$. However, the expansion of the bi-fundamental scalar around the background gives the kinetic term corresponding to the hopping term, c.f. Appendix \ref{DBIap}. The solution of the bi-fundamental scalar has the asymptotic behavior $\xi w^{x_1}\sim c_1u^0$ or $c_2u^{-3}$ near the $AdS$ boundary.  The coefficient of the kinetic term is ${\cal O}(N_c)$ and the bi-fundamental scalar is ${\cal O}(1)$ like $t_\mathrm{hop}$. } In summary, we  find the following identification between Fermi-Hubbard and top-down model parameters. 
\begin{eqnarray}\label{TopDownFermiHubbard}
\mu_{FH} &=& - \mu\,,\quad t_\mathrm{hop,FH} = t_\mathrm{hop}\,,\\
n_j &=& \frac{N}{2} + \rho_{(j)}\,,\\
\langle c^\dagger_{1,\alpha} c_{2,\alpha}
+ c^\dagger_{2,\alpha} c_{1,\alpha}\rangle &=& \dfrac{2Nc_2}{\pi^2\lambda} \,,\\
U &=& \frac{c_\mathrm{os}u_h}{N}\,
\end{eqnarray}
where $\lambda$ is 't Hooft coupling.
Hence, at the two-derivative level, the top-down construction presented in this section completely reproduces the two-site Fermi-Hubbard hamiltonian \eqref{eq:twosite_hamiltonian} at half filling. The higher order terms discussed at the end of sec.~\ref{sec44} will of course yield higher order contributions to the dual Hamiltonian, but are suppressed if the corresponding field values are not too large. The only thing left to determine is the large $N$ scaling of the deviations from half filling $\rho_{(i)}$. Following the arguments given in sec.~\ref{sec42} to restrict $\rho_{(i)} = {\cal O}(1)$ is natural from the point of view of string theory, in order to ensure supressed fluctuations around a classical saddle point in the large-N limit.

An analogous match has been achieved in \cite{Fujita:2014mqa} between the $SU(N)$ Bose-Hubbard model \eqref{SUNBoseHubbard} and the bottom-up construction \eqref{bottomupmodel}, where however the information about the large-N scaling was not used explicitly. Here we see that due to the natural expansion of the top-down model around the half-filling state, matching the large-N scaling properties is basically forced upon us. A similar large-N scaling would have been achieved in \cite{Fujita:2014mqa} if the action of the bottom-up model \eqref{bottomupmodel} would have been normalized to be ${\cal O}(N)$, which is the normalization natural for probe brane models in AdS/CFT.

\section{Effective hopping in the $SU(N)$ Bose-Hubbard model}\label{sec4}

\subsection{Numerical Simulations}

\begin{figure}
     \begin{center}
          \includegraphics[width=0.45\linewidth]{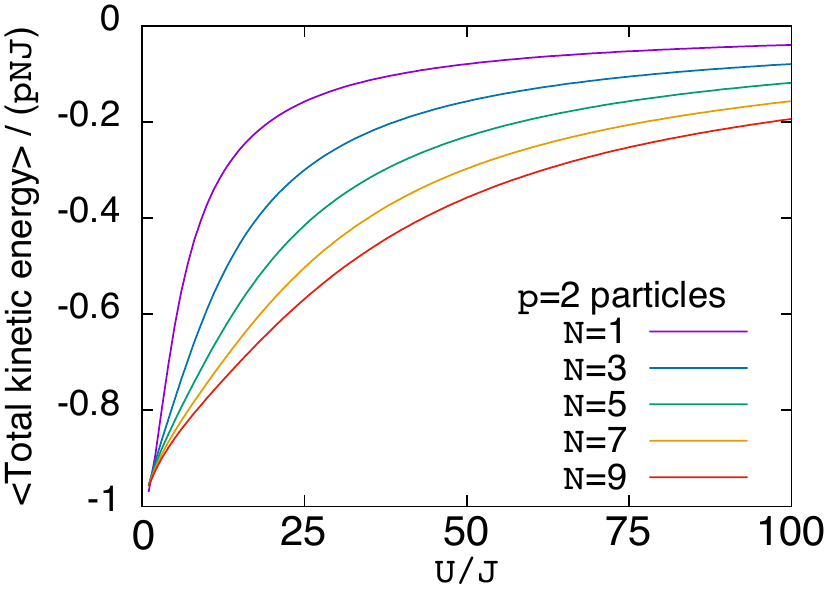} 
          \includegraphics[width=0.45\linewidth]{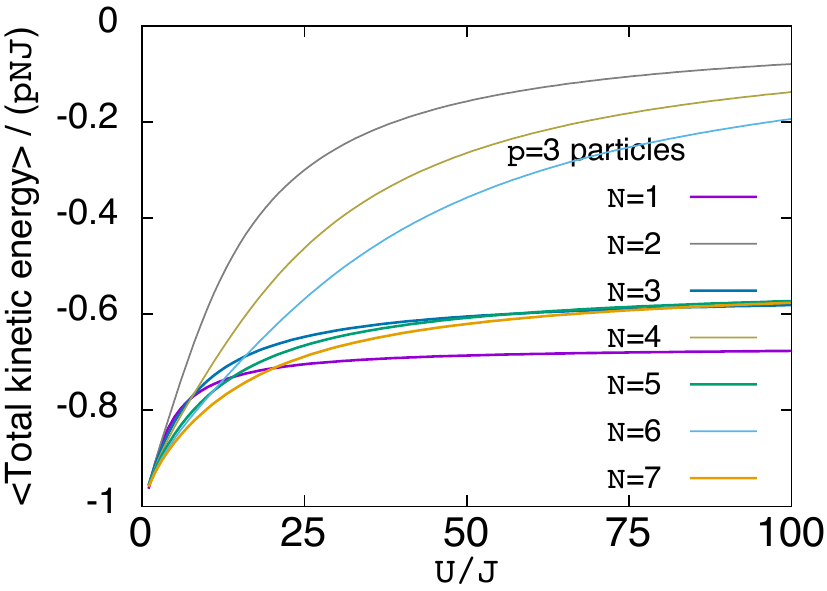}   
          \hspace{1.6cm}
        \end{center}
         \caption{Left: Numerical plot of the effective hopping with 2 particles per species. Only the cases with odd values of $N$ is plotted, but for an even $N$, the value appears between the plots for neighboring odd numbers. Mott insulator phase appears in this system of even number of particles at the large $U$ limit. Right: Numerical plot of the effective hopping with 3 fixed particles. The VEV is non-zero at large $U$ in the system of odd number of particles and odd number of components $N$. Otherwise, the VEV decreases to zero at large $U$.}
    \label{fig:EFF1}
\end{figure}

In this section, we compute the VEV of the hopping term by using numerical simulations and fixing the total occupation number.
We obtain the ground state wavefunction $\vert \Psi \rangle$ for $p$ particles for each of the $N$ components
by exact diagonalization using the Lanczos method for \eqref{BHmodel}, and obtain the total kinetic energy
\be
E_\mathrm{K} = - t_\mathrm{hop}\sum_{\langle i j \rangle}\left \langle \Psi \left \vert b_i^{a\dag} b_{ja} + \mathrm{c.c.} \right \vert \Psi \right \rangle.
\ee
In the limit of $\vert U / t_\mathrm{hop} \vert \to 0$, $E_\mathrm{K}$ approaches $-pN t_\mathrm{hop}$, because each particle occupies the bonding orbital formed from the two lattice sites.
As the repulsive interaction $\vert U\vert$ is increased, the absolute value of the VEV of the hopping term decreases.
In the case of two lattice sites and $p=2$ particles per component, the total kinetic energy divided by the number of particles continues to increase and approaches zero as $\vert U/t_\mathrm{hop}\vert \to \infty$. In this case, a single particle of each component at each site could be localized at each lattice site, but there are exponentially many (as a function of $N$) localized configurations of particles having the same number, $pN/2$, of particles at each site, therefore the ground state wavefunction can be approximated by a linear combination of such localized states.
On the other hand, for $p=3$ and an odd number of species, the total number of particles is an odd number, so that it is not possible to place the same number of particles at the two lattice sites. In this case, the total kinetic energy divided by $pN$ approaches a constant value.

We have placed results of the plot in Fig. \ref{fig:EFF1}. For even number of particles, we find the Mott insulator phase where the effective hopping behaves like $1/U$ at large $U$. For odd number of particles, the effective hopping approaches a non-zero constant at large $U$. This behavior is different from the gravity dual in which it is difficult to have the non-zero VEV at large $U/t_\mathrm{hop}$ in the non-homogeneous phase.

Dependence of the behavior of the effective hopping similar to the two-site case is also observed for larger numbers of sites:
for 3 lattice sites, the Mott-insulator-like behavior is only observed when $pN$ is a multiple of three, and the superfluid-like behavior is observed otherwise, \textit{i.e.} when $p$ nor $N$ is a multiple of the prime three.
Then interesting is what happens for $4(=2\times2)$ sites:
is it required that either $p$ or $N$ is a multiple of $4$, or is it enough if $pN$ is divisible by $4$, for the $pN$ bosons to behave like a Mott insulator at large $U/t_\mathrm{hop}$?
The latter is the answer: we observe that the total kinetic energy for $p=N=2$, for example, approaches zero as $U/t_\mathrm{hop}$ is increased.

\subsection{Quantum Mechanics for $p=1$} \label{sec5}

In the following sections, we calculate the quantum mechanical
hopping VEV in the perturbative
limit for the case of $p=1$ Bose-Hubbard model for all $N$. 
This gives us an exact point of reference to compare the results from holography. For the
$p=1$ case, it turns out that the Hilbert space for the fermionic
Hubbard model is same as the Bose-Hubbard model. We also show that
the hopping VEV for the two cases are equal to each other by use of a
two-site version of the Jordan-Wigner transformation. 
The hopping VEV for odd $N=2m+1$ is $-(m+1)t$ proportional to $N$,
while for even $N=2m$ is $-m(m+1)t^2/U$ proportional to $N^2$.
This $N$ dependence should be seen in a holographic model
in the large $N$ limit.

For the fermionic
case, we go beyond $p=1$ condition and show that the ground state still
lies in the $p=1$ sector of the Hilbert space. Thus for the fermionic
case, the perturbative answer for the hopping VEV is fully general,
while for bosons, our conclusions are restricted to fixed $p$ as in
earlier sections.

The two site Hamiltonian may also be written as
\begin{align}
\mathcal{H} & = \mathcal{H}_t + \mathcal{H}_U \nonumber \\
\mathcal{H}_t & = - t \sum_\alpha (c^\dagger_{1,\alpha} c_{2,\alpha}
+ c^\dagger_{2,\alpha} c_{1,\alpha}) \nonumber \\
\mathcal{H}_U & = U \sum_{i=\{1,2\}} \left( n_i - \frac{N}{2} \right)^2 \nonumber \\
n_i & = \sum_{\alpha} n_{i,\alpha} =\sum_{\alpha} c^\dagger_{i,\alpha} c_{i,\alpha} 
\label{eq:twosite_hamiltonian}
\end{align}
where $\alpha$ indexes the $N$ flavours. The $N/2$ present in the expression
of $\mathcal{H}_U$ fixes the half-filling condition as $U/t \rightarrow \infty$.
The operators in Eq. \ref{eq:twosite_hamiltonian} have bosonic commutation
relations for Bose-Hubbard model, and fermionic commutation relations for the
Fermi case. We start by discussing the fermionic case, and point out the
direct connection it has to the $p=1$ Bose-Hubbard model discussed in the previous
sections.

When $U = \infty$, we should stay exactly in the subspace of half-filled Hilbert space
which has lowest energy cost for $\mathcal{H}_U$. 
We will call this subspace as lowest Hubbard energy subspace or LHES.
All states in the LHES are degenerate to each other and have zero energy (up to
a constant shift in the Hamiltonian)
as no hops are permitted. 
We are eventually interested in the limit
$t/U \rightarrow 0^+$. In this limit, it still suffices to stay in the LHES and 
work out the matrix elements of $\mathcal{H}$ in this subspace in a
perturbative sense. By staying in the LHES throughout, we also usefully have
$\langle \text{g.s.} | \mathcal{H} | \text{g.s.} \rangle = 
\langle \text{g.s.} | \mathcal{H}_t | \text{g.s.} \rangle$
since all states in LHES are degenerate with respect to $\mathcal{H}_U$
and give $\langle \mathcal{H}_U \rangle = 0$ up to a trivial constant.

For even $N=2m$, the LHES corresponds to all combinations of $m$ fermions 
on both sites. 
For odd $N=2m+1$, the LHES corresponds to 
all combinations of $m$ fermions on one site and $m+1$ fermions
on the other. Now it is clear that in the odd case, the system can gain hopping energy
at linear order
while staying within the LHES due to hops that result from the imbalance in fermion
occupation on the two sites. These hops at linear order
directly give rise to a hopping expectation value
in the ground state.
For the even case, the system can not gain hopping
energy at linear order since any hop necessarily takes the system out of the
LHES, but it can gain hopping energy at quadratic order. We will restrict our
attention to these lowest orders for odd and even $N$ cases since we are interested
in understanding the $t/U \rightarrow 0^+$ limit. Now it remains to work
out the details of these lowest order hopping processes to evaluate the hopping
expectation value in this limit. The number of states in the 
LHES is given by $\binom{N}{m}^2$ for even $N$
and $2 \binom{N}{m} \binom{N}{m+1}$ for odd $N$ respectively, but this counting
is not going to be important in the following arguments. 

\subsection{Odd $N$} 
Let us start with
odd $N$ case.
We can further classify
the states by how many of the fermion flavours are commonly occupied
on the two sites. For example for $SU(3)$ with 1 fermion on one site 
and 2 fermions on other site, we have only two possibilities :
1) 0 common flavours, e.g. $|\text{first site} : 
1,0,0 \; ; \;  \text{second site} : 0,1,1\rangle$, 2) 1 common flavour, e.g.
$|\text{first site} : 
1,0,0 \; ; \;  	\text{second site} : 1,0,1\rangle$. Similarly there are 
$m+1$ possibilities for $SU(N=2m+1)$.

First we observe that the states categorized in this way form disjoint blocks in the 
lowest order Hamiltonian as $t/U \rightarrow 0^+$ in the 
LHES. This is because any flavour that is commonly occupied can not hop,
and thus there is no way to change this commonly occupied flavour number by performing
hops in the rest of the flavours. In fact the disjointedness of these blocks remains
true for bosons as well, the only difference being the commonly occupied 
bosonic flavours can hop unlike fermions.

Now we ask which of these blocks can gain most in hopping energy, or in other words,
which block will contain the ground state ? Intuitively it should be the block
that allows for most number of hops (within that block). This corresponds to the block
with 0 common flavours, and that is the indeed the answer as we will see. 
At this point, we note that this block is \emph{exactly} same
as the Hilbert
space of the $p=1$ Bose-Hubbard model in terms of the Fock 
occupations (See. Eq. \ref{eq:statesSU3} as an example for
$N=3$). The difference
lies in the statistics, and in the following we will show that this does not
make a difference to the hopping VEV. This allows us to compute the hopping VEV for both
bosonic and fermionic cases simultaneously.

As an aside, blocks with $> 0$ commonly occupied flavours further sub-divide into smaller
blocks indexed by which flavours in particular are commonly occupied. This will give
rise to degeneracies in the excited states in the $t/U \rightarrow 0^+$ problem
that can be counted if desired.

Next we observe that in the 0 commonly occupied flavour block, we can
have exactly $(m+1)$ hops while staying in the LHES since we necessarily
hop from the site with $(m+1)$ fermions to the site with $m$ fermions to
stay in the LHES. All of these are off-diagonal processes.
Furthermore we can convince ourselves that we can reach any
state in the block from any other state through a finite number of
the allowed hops. Thus all states are symmetrical to each other with respect
to $\mathcal{H}_t$, i.e. each ket is connected to exactly $(m+1)$ bras.
Thus $\mathcal{H}_t$ matrix will have the general structure such that
there are only $(m+1)$ non-zero entries in all rows/columns.
 Since we are
restricted to $t/U \rightarrow 0^+$, the non-zero entries have equal
magnitudes $|t|$ at lowest order. This matrix structure is verified easily for
the low values of odd $N$, but we keep in mind that this matrix structure
is valid for all odd $N$. We show in Eq. \ref{eq:matSU3}
this matrix structure in the 0 commonly
occupied block for $SU(3)$ with $m=1$ \emph{without
paying attention to} the fermion anti-commutation signs.
This has $m+1=2$ non-zero entries
in all rows/columns.
Below in
the subsection on
fermion signs, we elaborate
why the following arguments still apply.
Briefly, we may apply
a two-site version of the Jordan-Wigner transformation to ``bosonize" the Hamiltonian.
Indeed for the $p=1$ Bose-Hubbard case
where there are no fermion signs to begin with, this is precisely the
$\mathcal{H}_t$ in the LHES.
\begin{align}
\mathcal{H}_t = -t \left( 
\begin{array}{cccccc}
0 & 0 & 0 & 1 & 1 & 0 \\
0 & 0 & 0 & 1 & 0 & 1 \\
0 & 0 & 0 & 0 & 1 & 1 \\
1 & 1 & 0 & 0 & 0 & 0 \\
1 & 0 & 1 & 0 & 0 & 0 \\
0 & 1 & 1 & 0 & 0 & 0 
 \end{array} \right)
\label{eq:matSU3}
\end{align}
The states in this 0 commonly occupied block are 
\begin{align}
|1\rangle = |\text{first site} : 1,0,0 \; ; \;  \text{second site} : 0,1,1\rangle \nonumber \\
|2\rangle = |\text{first site} : 0,1,0 \; ; \;  \text{second site} : 1,0,1\rangle \nonumber \\
|3\rangle = |\text{first site} : 0,0,1 \; ; \;  \text{second site} : 1,1,0\rangle \nonumber \\
|4\rangle = |\text{first site} : 1,1,0 \; ; \;  \text{second site} : 0,0,1\rangle \nonumber \\
|5\rangle = |\text{first site} : 1,0,1 \; ; \;  \text{second site} : 0,1,0\rangle \nonumber \\
|6\rangle = |\text{first site} : 0,1,1 \; ; \;  \text{second site} : 1,0,0\rangle 
\label{eq:statesSU3}
\end{align}
with $p=1$ condition seen explicitly.

Due to this symmetry of the states in the 0 commonly occupied block of
LHES, the lowest energy 
state in this block is an uniform superposition
of these states \emph{independent} of the number of the states in this
block. This state can be written as $\frac{1}{\sqrt{\text{Norm}}}
\left(1 \; 1\; 1 \; \ldots \; 1 \right)^T$, and the energy of this state
is $-(m+1)t$ as can be checked by application of $\mathcal{H}_t$.
Any other superposition will increase the energy.

For the bosonic case,
blocks with commonly occupied flavours are not part of the Hilbert space
due to the $p=1$ condition. Thus the hopping VEV is $-(m+1)t$,
which is the main result for odd $N$. When we approach the perturbative
limit by holding $t$ fixed and letting $U\rightarrow \infty$,
a hopping VEV of $-(m+1)t$ implies different plateaus for different $N$
as $U\rightarrow \infty$  as discussed in the earlier sections.

\begin{figure}[h]
\includegraphics[width=0.7\columnwidth]{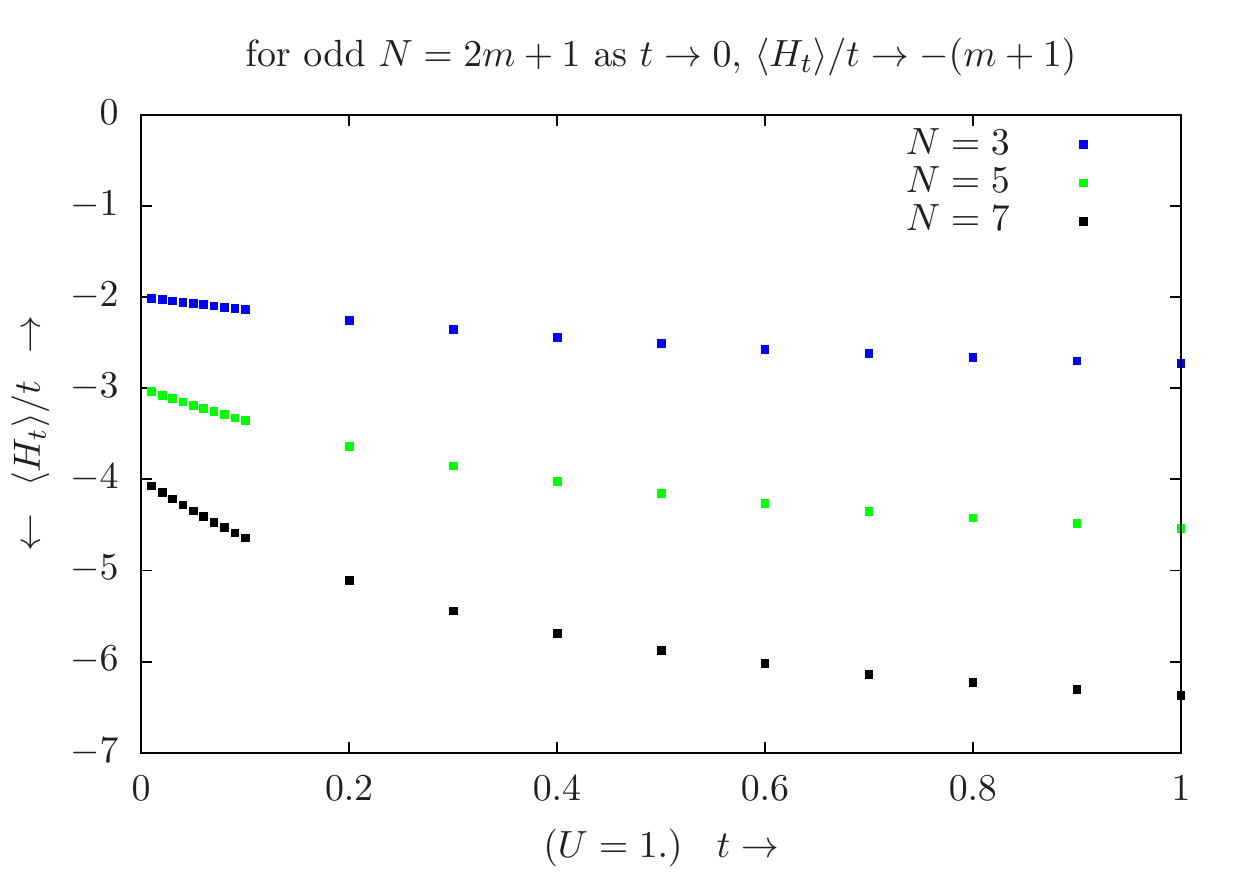}
\caption{ \label{fig:oddNfermion}
Hopping VEV for odd $N=2m+1$ case}
\end{figure}

For fermions we will now show that blocks with non-zero commonly
occupied flavours have greater hopping energy and do not
contribute to the hopping VEV. It is clear that for the blocks with
non-zero commonly occupied flavours, in each sub-block of given common
occupation (say $\alpha=1$ is commonly occupied on both sites
for $SU(3)$, etc.), the hopping matrix will have the same structure as above except
the number of non-zero entries will be $(m+1)-c$ where $c$ is the number of commonly 
occupied flavours. This is because those $c$ commonly occupied flavours
are forbidden to hop due to Pauli's exclusion.  Again states in each such
sub-block will be symmetrical to each other with respect to $\mathcal{H}_t$.
Thus for a block with a non-zero value of commonly occupied flavours $c$,
the lowest energy would be $-((m+1)-c)t$ up to some calculable degeneracies
coming from the number of sub-blocks indexed by given common occupations.
Thus we have shown that among all the blocks,
the ground state lies in the 0 commonly occupied block 
in the $t/U \rightarrow 0^+$ limit, and the hopping VEV
(remember $\langle \text{g.s.} | \mathcal{H} | \text{g.s.} \rangle 
= \langle \text{g.s.} | \mathcal{H}_t | \text{g.s.} \rangle$ in LHES) 
is exactly $-(m+1)t$ up to lowest order in $t$ for all $SU(N=2m+1)$. This 
is confirmed from Exact Diagonalization
in Fig. \ref{fig:oddNfermion} below.

\subsection{Even $N$} 
Now we tackle the even $N=2m$ case
where we have $m$ fermions on each site in the LHES. All single 
hops will take
us out of the LHES, and we have no contribution at linear order
in $t/U$. At quadratic order we can have contributions, since
there can be two successive hops which can take us out of and afterward return
us back to the LHES. Any such process will generate a matrix element 
$\propto -t^2/U$ in degenerate perturbation theory where the degenerate
states are the states of the LHES.
We will restrict our attention to these lowest order
processes as $t/U \rightarrow 0^+$. This is a little subtler than odd $N$ case
in terms of hopping expectation value,
and here it becomes very useful to use 
$\langle \text{g.s.} | \mathcal{H} | \text{g.s.} \rangle 
= \langle \text{g.s.} | \mathcal{H}_t | \text{g.s.} \rangle$
when
we restrict our attention to the LHES. The idea is thus to get a knowledge
of the hopping expectation value in the ground state in the perturbative limit 
by using this equality.

Again as before, we have disjoint
blocks depending on the number of commonly occupied flavours since
hops conserve flavours. 
To find out which block contains the ground state, we may run through
similar arguments as for the odd $N$ case. We seek that block which allows
for the maximal number of these quadratic order two-hop process. This again
is the block with 0 commonly occupied flavours, and thus again as before
we can simultaneously compute for the hopping VEV for both Bose-Hubbard
and Fermi-Hubbard cases.

For even $N$ we can have both
diagonal and off-diagonal two-hop processes.
The number of \emph{off-diagonal} two-hop processes starting 
from a given state and ending in a \emph{different} state in this block
corresponds to $m^2$. This is because we have $m$ choices for the first hop
from one of the sites, and $m$ choices for a different flavour
from the second site such that we end in a different state. 
The matrix element corresponding to a single two-hop process
is $-t^2/2U$ (the factor of 2 in the denominator
is due to the chosen form of Hubbard interaction
which leads to a difference in energy $=2U$ between
a LHES state and intermediate excited state).
Thus off-diagonal matrix elements $= -2 \times t^2/2U = -t^2/U$.
This is because given an initial state and a different final state
in the LHES, there are two possibilities for the two-hop processes
that connect the two states owing to the choice of the two sites
for the annihilation site of the first hop (the annihilation site
for the second hop is subsequently fixed). 
Furthermore this implies
there are $m^2$ off-diagonal non-zero entries in all rows/columns in this block of LHES
after doing perturbation theory up to lowest order
$\propto 
\mathcal{H}_t (E_{\text{LHES}} - \mathcal{H}_U)^{-1} \mathcal{H}_t$.
Now, we again have
the irrelevance of fermion signs due to the same arguments as mentioned
in the 
previous subsection and elaborated in the subsection below, 
i.e. $\mathcal{H}_t$ and $\mathcal{H}_U$ may be thought
of as the Jordan-Wigner transformed Hamiltonian.

The counting of diagonal two-hop processes is $2m$ for a given state
in LHES, i.e. we can choose any one of the $2m$ flavours on both sites
for the first hop and afterward we hop the same flavour back. The matrix element
of a diagonal two-hop process is thus $=-2m t^2/2U = - m t^2/U$
for all states in this block of the LHES. All states are again symmetrical
to each other and the lowest energy state in this block
can be written as $\frac{1}{\sqrt{\text{Norm}}}
\left(1 \; 1\; 1 \; \ldots \; 1 \right)^T$, with the energy
given by $-m^2 t^2/U - m t^2/U = -m(m+1) t^2/U$. We see that this
is in agreement with the familiar form for the case of $SU(2)$
\begin{align}
\mathcal{H}_t = -\frac{t^2}{U} \left( 
\begin{array}{cc}
1 & 1  \\
1 & 1 
 \end{array} \right)
\label{eq:matSU2}
\end{align}
which leads to a singlet ground state. The states in this 0 commonly occupied block are 
\begin{align}
|\text{first site} : 1,0 \; ; \;  \text{second site} : 0,1 \rangle \nonumber \\
|\text{first site} : 0,1 \; ; \;  \text{second site} : 1,0\rangle 
\label{eq:statesSU2}
\end{align}

\begin{figure}[ht]
\includegraphics[width=0.7\columnwidth]{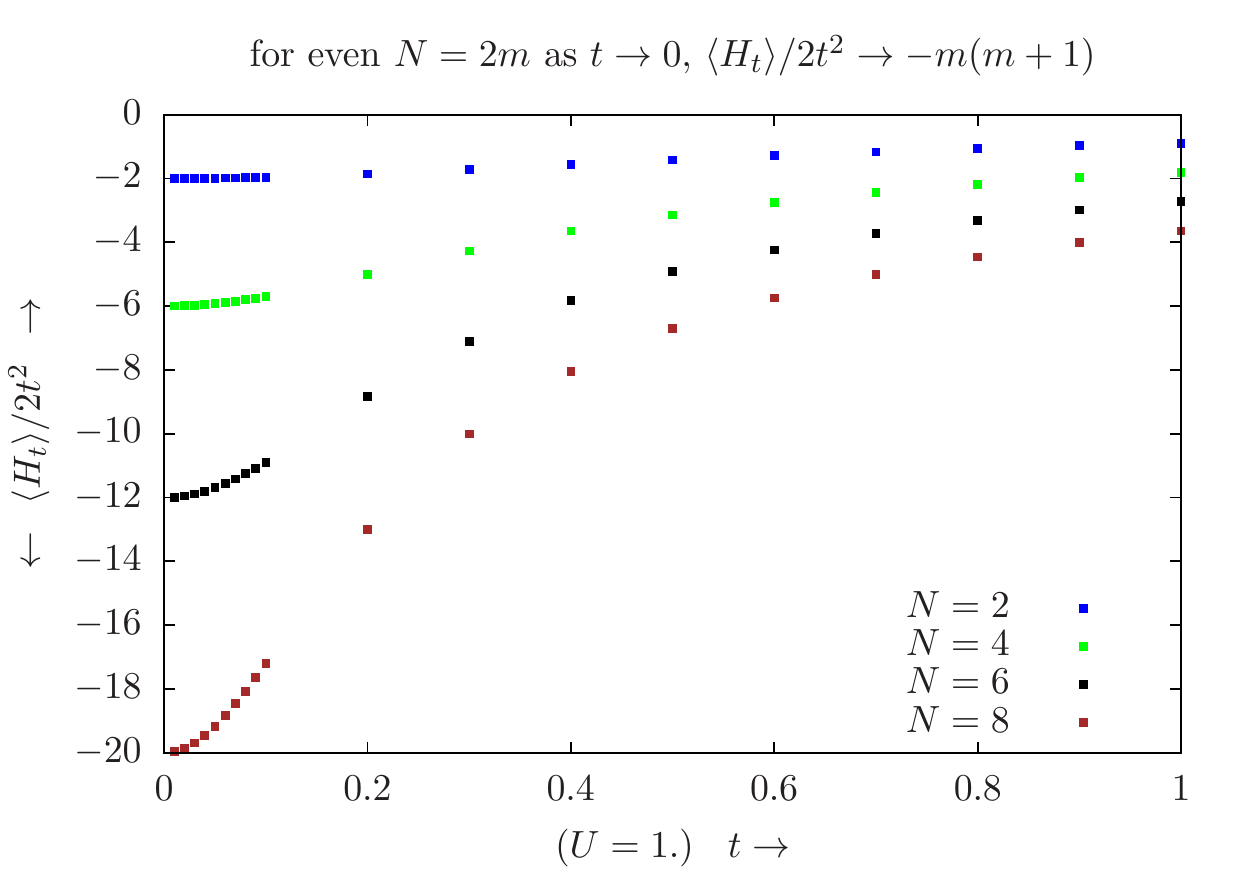}
\caption{ \label{fig:evenNfermion}
Hopping VEV for even $N=2m$ case. 
}
\end{figure}

For the bosonic case,
blocks with commonly occupied flavours are again not part of the Hilbert space 
due to the $p=1$ condition. Thus we have the hopping VEV as $-m(m+1)t^2/U$ which
is the main result for even $N$.
For fermions, again as before, the blocks with non-zero common occupations
will have higher hopping energy and do not contribute to the hopping VEV.
For these blocks with non-zero common occupations, the number of non-zero off-diagonal
entries will
similarly be $2(m-c)^2$ where $c$ is the number of common occupations.
The diagonal entries will equal $-(m-c) t^2/U$ in these blocks. 
The lowest energy will equal $-(m-c)(m-c+1) t^2/U$ and we can count the
degeneracies as well. This shows that the 
ground state lies in 0 common occupation block.
By using 
$\langle \text{g.s.} | \mathcal{H} | \text{g.s.} \rangle 
= \langle \text{g.s.} | \mathcal{H}_t | \text{g.s.} \rangle$
in the perturbative limit $t/U \rightarrow 0^+$,
we arrive at the hopping expectation value in the ground state being equal to
$- m(m+1) t^2/U$ for all even $N=2m$ up to lowest order
in $t$. The $m(m+1)$ proportionality 
of the hopping expectation value is confirmed from Exact Diagonalization
in Fig. \ref{fig:evenNfermion} below.

\subsection{Fermion Signs and Jordan-Wigner Transformation} 
The particular
convention that gives rise to
 the matrix in Eq. \ref{eq:matSU3}  
is : A state is obtained by first creating fermions of $N=3^\text{rd}$ flavour,
secondly of the $N-1=2^\text{nd}$ flavour, etc. all the way up till
$1^\text{st}$ flavour. For each flavour, the fermion on the $2^\text{nd}$
site
is created first followed by the $1^\text{st}$ site.
This may be summarized as
\begin{equation}
c^\dagger_{1,\alpha=1} c^\dagger_{2,\alpha=1} c^\dagger_{1,\alpha=2}
c^\dagger_{2,\alpha=2} c^\dagger_{1,\alpha=3} c^\dagger_{2,\alpha=3} 
|\text{vacuum}\rangle = 
+ |\text{first site} : 1,1,1 \; ; \; \text{second site} : 1,1,1 \rangle
\label{eq:JW_transformation}
\end{equation}

Now we show that this kind of fermion sign convention gives rise to
the matrix structure of the type in Eq. \ref{eq:matSU3} for
all $N$, by using a two-site version of the Jordan-Wigner transformation.
First, we need a combined site-flavour index $\mu$ which goes
from 1 to $2N$ in our fermion sign convention order, i.e.
$c^\dagger_{\mu=1} c^\dagger_{\mu=2} \ldots
c^\dagger_{\mu=2N-1} c^\dagger_{\mu=2N} 
|\text{vacuum}\rangle \equiv
c^\dagger_{1,\alpha=1} c^\dagger_{2,\alpha=1} \ldots
c^\dagger_{1,\alpha=N} c^\dagger_{2,\alpha=N} 
|\text{vacuum}\rangle = 
+ |\text{first site} : 1,\ldots,1 \; ; \; \text{second site} : 1,\ldots,1 \rangle$.
Let us define new creation/annihilation operators as following
in terms of this new site-flavour index
\begin{align}
b_\mu & = \prod_{\nu < \mu}(-1)^{n_\nu} c_\mu \nonumber \\
b^\dagger_\mu & = \prod_{\nu < \mu}(-1)^{n_\nu} c^\dagger_\mu 
\label{eq:Jordan_Wigner}
\end{align}
These new $b$ operators now commute when coming with different
site-flavour indices, e.g. with $\mu < \nu$ and keeping in 
mind that we can commute the $(-1)^{n_\mu}$ phase factors with 
$c_\nu$ operators when $\nu \neq \mu$
\begin{align}
 [b_\mu,b_\nu] &
=  b_\mu b_\nu - b_\nu b_\mu  \nonumber \\
& =  (c_\mu (-1)^{n_\mu} c_\nu - (-1)^{n_\mu} c_\nu c_\mu) 
\prod_{\mu < \rho < \nu} (-1)^{n_\rho} \nonumber \\
& =  (c_\mu (-1)^{n_\mu} + (-1)^{n_\mu} c_\mu) \; c_\nu
\prod_{\mu < \rho < \nu} (-1)^{n_\rho} \nonumber \\
& = 0   
\label{eq:abc}
\end{align}
where we have applied fermionic anti-commutation from second to third line,
and the last equality follows by observing that
(trivially) 
$(c_\mu (-1)^{n_\mu} + (-1)^{n_\mu} c_{\mu}
) |0\rangle = 0$ 
and 
$(c_{\mu} (-1)^{n_{\mu}} + (-1)^{n_{\mu}} c_{\mu}
) |1\rangle = - c_{\mu} |1\rangle + (-1)^{n_{\mu}} |0\rangle
= - |0\rangle + |0\rangle = 0$. 
The case of $\mu > \nu$ now follows since, when $[A,B]=0$, $[B,A]=0$ as well.
Similar arguments can be used to show
that $[b^\dagger_\mu,b^\dagger_\nu]=0$ and $[b_\mu,b^\dagger_\nu]=0$ with
$\mu \neq \nu$.
But these $b$ operators are not fully bosonic operators since for
the same site-flavour index $\mu$, 
their commutation relations are not bosonic; rather they are
still fermionic to respect Pauli's exclusion. 
 
This two-site version of Jordan-Wigner transformation keeps
$\mathcal{H}_U$ unchanged trivially
since $c^\dagger_\mu c_\mu = b^\dagger_\mu b_\mu$ for any $\mu$.
The fact that $\mathcal{H}_t$ remains unchanged can be seen as
\begin{align}
& b^\dagger_{1,\alpha} b_{2,\alpha} + b^\dagger_{2,\alpha} b_{1,\alpha} \nonumber \\
= \; \; & b^\dagger_{\mu} b_{\mu+1} + b^\dagger_{\mu+1} b_{\mu} \nonumber \\
= \; \; & c^\dagger_{\mu} (-1)^{n_\mu} c_{\mu+1} + c^\dagger_{\mu+1} (-1)^{n_\mu} c_\mu
\nonumber \\
= \; \; & c^\dagger_{\mu}  c_{\mu+1} + c^\dagger_{\mu+1}  c_\mu \nonumber \\
= \; \; & c^\dagger_{1,\alpha} c_{2,\alpha} + c^\dagger_{2,\alpha} c_{1,\alpha}
\label{eq:abc1}
\end{align}
where we have used
$c^\dagger_{\mu} (-1)^{n_\mu}  = c^\dagger_{\mu}$
and $(-1)^{n_\mu} c_\mu = c_\mu$ as easily checked
by their action on $|0\rangle$ and $|1\rangle$.
This exposes the reason for absence of fermion signs
in Eq. \ref{eq:matSU3} for our chosen sign convention.
Since we group flavours first and sites  secondly in the sign
convention, we do not get extra sign factors in the algebra
above, which allows us to see the equality of $\mathcal{H}_t$
in both $c$ and $b$ basis. Since $b$ operators commute for
different $\mu$, furthermore, the lack of fermion signs become manifest for all $N$.
One may note that 
the ordering of flavours while choosing our fermion sign convention
was not important, and we could have chosen any permutation of flavour
ordering for the Jordan-Wigner transformation to work. Furthermore
the spectrum is invariant with respect to fermion sign convention,
and the only difference that comes due to a different sign convention
are phase factors of $e^{i \pi} = -1$ in the coefficients
when the ground state is expressed
as a linear combination of the states with
zero commonly occupied flavours.

{
This equality of the $p=1$ Bose and Fermi hopping VEVs is in fact
a non-perturbative result following from the above discussions.
We note for completeness that the above can also be
generalized to a linear chain of sites with
open boundary conditions. We may summarize the generalized Jordan-Wigner
transformation as
\begin{align}
& \left(\ldots c^\dagger_{i-1,1} c^\dagger_{i,1} c^\dagger_{i+1,1} \ldots \right)
\left(\ldots c^\dagger_{i-1,2} c^\dagger_{i,2} c^\dagger_{i+1,2} \ldots \right)
\ldots
\left(\ldots c^\dagger_{i-1,N} c^\dagger_{i,N} c^\dagger_{i+1,N} \ldots \right)
|\text{vaccum} \rangle  \nonumber \\
&\equiv \mathbf{+} |\text{fully filled state} \rangle  \nonumber \\
& \Longrightarrow \prod^N_{\alpha=1} \left(\prod_i c^\dagger_{i,\alpha}  \ldots \right) 
|\text{vaccum} \rangle \equiv \mathbf{+} |\text{fully filled state} \rangle 
\label{eq:generalized_JW_transformation}
\end{align}
The above may be used to show the non-perturbative
equality of the hopping VEVs of the $p=1$ Bose and 
Fermi-Hubbard models 
for a linear chain of sites again
with effectively the same many-body Hilbert space, i.e.
\begin{align}
\mathcal{H} = -t \sum^N_{\alpha=1} \sum_{i} \left( c^\dagger_{i,\alpha} c_{i-1,\alpha} + 
c^\dagger_{i,\alpha} c_{i+1,\alpha} \right) + U \sum_i \left( \sum^N_{\alpha=1} 
c^\dagger_{i,\alpha} c_{i,\alpha} - \frac{N}{2} \right)^2
\end{align}
where $i$ is a site index for the linear chain
by an appropriate generalization of Eq. \ref{eq:Jordan_Wigner}.
}

\begin{figure}[t]
     \begin{center}
\includegraphics[height=4.5cm,clip]{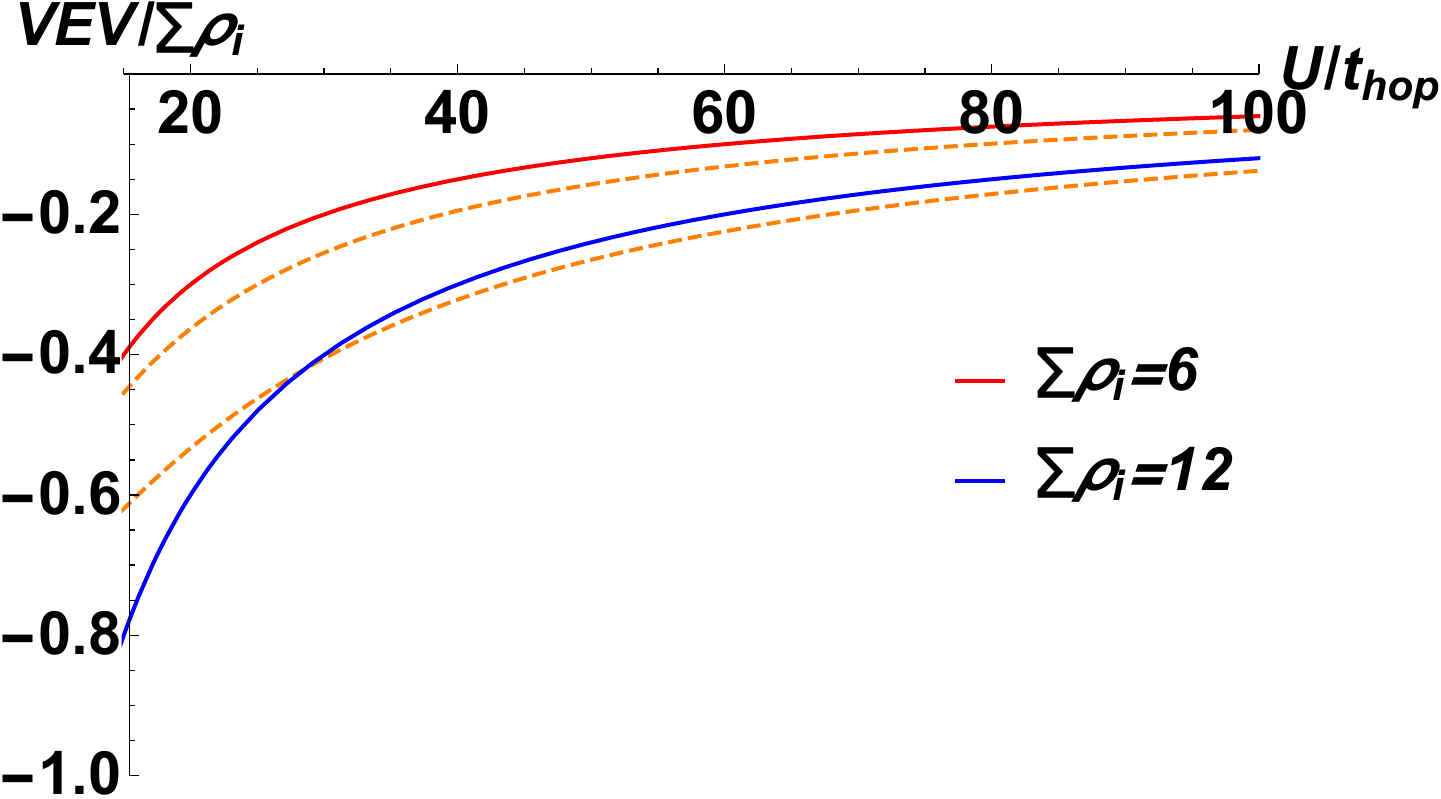} 
 \end{center}
 \caption{VEV$/\sum\rho_i$ in the Mott insulator phase as a function of $U/t_\mathrm{hop}$ with fixed parameters $M^2=0$, $w=0$, $\lambda =1/\Lambda^2$, $\Lambda_{(1,1)}=-3/(2\Lambda)$, and other $\Lambda_{(p,r)}=0$, where the parameter $\Lambda$ is specified by eq. \ref{PAR328}. The orange curves describe the effective hopping of the $SU(3)$ Bose-Hubbard model with fixed numbers of particles $p=2,\ 4$ per species. Total numbers of bosons are $6,\ 12$, respectively. The parameters were fixed by matching to second order perturbation theory at large $U/t_\mathrm{hop}$. We find near agreement when the hopping parameter is smaller than the other couplings. We believe that the small disagreement at large $U/t_\mathrm{hop}$ is due to the term linear in $\rho$ in \protect\eqref{pertVEVsinglecomponent} which is not present in the holographic model.}
\label{fig:co1}
\end{figure}

\section{Discussion \& Conclusions}\label{sec:discussion}

In the first part of this paper, we extended the study of the holographic duality proposed in~\cite{Fujita:2014mqa} between the large $N$ $SU(N)$ Bose-Hubbard model on multiple sites and the two-dimensional gravity on the $AdS_{2}$ hard wall. In the gravity dual, we computed and analyzed in sections~\ref{sec2}~and~\ref{sec3} the hopping VEV of the two-site model both in the homogeneous Mott insulator phase and the non-homogeneous superfluid phase. {In section~\ref{sec4}, we computed the hopping VEV fixing the particle number per component, $p$, in the large $N$ $SU(N)$ Bose-Hubbard model on two sites. When the particle number is $2N$ ($p=2$), VEV goes to zero like $1/U$. It is the behavior in the Mott insulator phase. When the particle number is $3N$ ($p=3$), VEV is non-zero at large $U$. We confirmed an analytic expression of the hopping VEV in the perturbative limit $t/U\ll 1$ for $p=1$. In the large $N$ limit, the hopping VEV for even $N$ and odd $N$ is proportional to $N^2$ and $N$, respectively. Moreover, as discussed later, the effective hoppings in Fermi and Bose cases are equivalent due to the Jordan-Wigner transformation. Paying attention to these large $U$ behavior, we compared results betweten the gravity dual and the large $N$ Bose-Hubbard model. }

In the homogeneous Mott insulating phase we find that at small $t_\mathrm{hop}$ the holographically computed hopping VEV in the presence of the bulk mass $M$ qualitatively agreed with that of the $SU(N)$ Bose-Hubbard model. The behavior at small $t_\mathrm{hop}/U$ (large $U/t_\mathrm{hop}$) matches the behavior of the VEV in the two-site Bose-Hubbard model at large $N$ (see Fig. \ref{fig:co1}). In order for the Mott insulating phase to have small hopping VEV, the total particle number had to be fixed to be even in the two-site Bose-Hubbard model.\footnote{{If it were odd, the unpaired particle could hop back and forth at first order in $t_\mathrm{hop}$, which neither matches expectations for the behavior of a Mott insulating phase, nor our holographic model. In the limit of small $t_\mathrm{hop}$, a  plateau was found for the hopping kinetic energy instead if the total occupation number is odd, c.f.~sec.~\ref{sec4}. It will be interesting to analyze the plateau in holographic models.}} We hence establish a relation between the large N two site Bose-Hubbard model at even total particle number, and the holographic model of \cite{Fujita:2014mqa}. At large $t_\mathrm{hop}/U$, i.e. close to the transition to superfluidity, the VEVs do not coincide. This is not surprising, since e.g. already the details of the phase boundary in Fig.~\ref{fig:mix} and the order of the phase transition differ from e.g. the mean field analysis of \cite{Fisher:1989zza}. We also analyze the dependence on the bulk mass $M$, which is related to the anomalous dimension of the hopping kinetic energy. The absolute value of the hopping VEV increases gradually as $M$ increases and as the occupation number increases. 

In the non-homogeneous superfluid phase, we compare the holographic hopping VEV for different hard wall boundary conditions with the one from the $SU(N)$ Bose-Hubbard model, as well as with the homogeneous phase. We find that Neumann boundary condition~\eqref{IRB216} (c.f.~Fig.~\ref{fig:nonh2}) for the bi-fundamental scalar generate a VEV smaller than the one in the homogeneous phase. This seems unphysical, since the VEV is expected to increase as $t_\mathrm{hop}$ increases into the superfluid phase. A Dirichlet condition instead produces  larger VEVs (c.f.~Fig.~\ref{fig:nonh3}), and hence seems to be closer to the situation in the actual Bose-Hubbard model. 

In our two-site model an interesting interplay between spontaneous and explicit symmetry breaking similar to the one in QCD, and in particular in AdS/QCD models, is at work. For two sites, there exist two global charge symmetries from the boundary perspective, the axial symmetry $U(1)_A = \frac{1}{\sqrt{2}}(U(1)_1-U(1)_2)$ and the vector symmetry $U(1)_V = \frac{1}{\sqrt{2}}(U(1)_1+U(1)_2)$. While the vector symmetry leads to total charge conservation, the axial symmetry implies conservation of the charge difference between the two sites. The hopping VEV is charged under the axial symmetry and hence breaks it spontaneously, while the hopping parameter $t_\mathrm{hop}$ is the source for the hopping VEV, and explicitly breaks the axial symmetry. In QCD, explicit chiral symmetry breaking is effected by a small bare quark mass, and spontaneous breaking by the chiral condensate. The interplay of explicit and spontaneous breaking leads to the  Gell-Mann-Oakes-Renner relation~\cite{Erlich:2005qh,Abidin:2009aj,Zuo:2009hz,Ballon-Bayona:2014oma}, which relates the mass of the Goldstone boson of the broken chiral symmetry (the pion) to the non-zero quark mass and the value of the chiral condensate in a universal way. Such a relation has also been derived in lower dimensional field theories such as graphene~\cite{Araki:2010gj}. By analogy, a similar relation should hold for small $t_\mathrm{hop}$ in the holographic Bose-Hubbard model \eqref{bottomupmodel}. We leave a careful investigation of the excitation spectrum of \eqref{bottomupmodel} in view of this interplay for future work.

In section \ref{holohubbard}, we present on a top-down construction of a $SU(N)_k$ Fermi-Hubbard model at half filling in terms of a D3/D5/D7 brane intersection, where the multiple D3-branes were replaced with the $AdS_5$ soliton $\times S^5$ background \cite{Horowitz:1998ha}. The massless spectrum of open strings stretching between the D3 and D5 branes constitutes the hopping fermionic degrees of freedom if there is a single D7 brane involved in the construction.\footnote{For more than one D7 brane one gets abelian anyons with relative statistics depending on $N_c$ and the number of D7 branes $k$ instead.} The D5 branes correspond to the lattice sites and are separated in the boundary spatial directions. Open strings stretching between the separated D5 branes then give rise to the modes which we identify with the hopping bifundamental field of the bottom-up model \eqref{bottomupmodel}. The Wess-Zumino term is present in the D5 action and, due to the quantization of induced worldvolume flux~\cite{Camino:2001at},  stabilizes the D5-brane embedding~\footnote{We would like to thank M. Shigemori for discussion about the stability of non-Abelian D5-branes. On the other hand, non-supersymmetric D3-branes wrapping on the asymptotic $AdS_2\times S^2$ will also be a candidate of the holographic dual to a Bose-Hubbard model since these D3-branes include a tachyonic mode, which should be bosonic, in the NS sector of the field theory side. However, it is shown that D3-branes do not couple with the pullback of the background RR 4-form and are unstable.} and leads to a relation between the quantized polar angle on the $S^5$ transverse to the D5 brane, and the charge density on the D5 brane, c.f.~eq.~\eqref{quantizationcondition}. We find that the homogeneous phase of equal charge density on all sites corresponds to the D5 branes wrapping $S^4\subset S^5$ at the same polar angle, while the inhomogeneous phase with unequal charge densities corresponds to separating the D5 branes in the polar angle direction as well as the boundary directions (c.f.~Fig.~\ref{fig:topdown}). 

We then compared the top-down construction with the bottom-up model, c.f.~table~\ref{TU2}. Expanding the DBI-WZ action for the D5 branes to second order in $2\pi\alpha'$, the matter content of the bottom-up model such as gauge fields and bi-fundamental fields linking the different lattice sites are reproduced in the top-down model. The symmetry breaking that occurs during the process of separating the $n_F$ D5 branes from a stack with nonabelian $U(n_F)$ symmetry in the boundary direction is crucial in obtaining the correct field content: Separating the D5 branes breaks $U(n_F)\rightarrow U(1)^{n_F}$ by giving a  VEV to the transverse scalars. The remaining abelian $U(1)^{n_F}$ corresponds to the conserved charge on each defect D5 brane. The bi-fundamental scalar fields dual to the hopping term are obtained from the off-diagonal component of the transverse scalars, i.e. from a mode of the string stretching between different D5 branes. Switching on the same hopping parameter $t_\mathrm{hop}$ between all the sites then breaks $U(1)^{n_F}\rightarrow \frac{1}{\sqrt{n_F}} \sum\limits_{i=1}^{n_F} U(1)_i$, i.e. down to the baryonic vector $U(1)$ symmetry that corresponds to total charge conservation. The hopping VEV of course also contributes to this breaking. This symmetry breaking pattern again is completely analogous to the one of the bottom-up model, as well as of chiral symmetry breaking in AdS/QCD~\cite{Erlich:2005qh,Karch:2006pv}).}

Finally, we were able to match the holographic bottom-up model \eqref{bottomupmodel} to the Bose-Hubbard model \eqref{SUNBoseHubbard}  (c.f.~sec.~\ref{sec32}), and the top-down model of sec.~\ref{holohubbard} to the Fermi-Hubbard model at half filling \eqref{eq:twosite_hamiltonian} (c.f.~sec.~\ref{sec45}) by large N scaling arguments. In particular, adjusting the scaling of \eqref{bottomupmodel} by multiplying it with $N_c$, one can see that all four models are mapped into each other: The top-down model to lowest order in $\alpha'$ is equivalent to the bottom up model \eqref{bottomupmodel} at half filling, and hence at large $N_c$ the two-site $SU(N_c)$ Bose-Hubbard model \eqref{SUNBoseHubbard} must be equivalent to the two-site Fermi-Hubbard model.  
This reinforces the intuition that fundamental bosons and fundamental fermions 
show rather similar behavior in the large $N_c$ limit: Due to the large number of possible additional quantum numbers (species or color quantum numbers, for example), the Pauli exclusion principle is not relevant any longer and fermions can occupy states in the same energy interval. The only obvious difference between the two holographic models is the absence of the Wess-Zumino term in \eqref{bottomupmodel},  in which charge quantization has to be imposed by hand (as done in \cite{Fujita:2014mqa}). Indeed,  
in section~\ref{sec5}, the Hilbert space for the Fermi-Hubbard model is shown to be equivalent to that of the Bose-Hubbard model in a sector fixing the number of particles with the help of the Jordan-Wigner transformation on two sites. It is also in sec.~\ref{sec4} where we provide the details of the numerical simulations of the $SU(N_c)$ two-site models that we then compare with our holographic constructions.

As an outlook for future work, it will be interesting to see whether one can take a continuum limit of a lattice of holographic defects similar to the ones described in this work, in this way obtaining a higher dimensional holographic theory. Within this description many issues such as the presence of charge density wave instabilities such as the Peierls instability or the emergence of quantum critical phases from e.g. Kondo lattices, generalizing the one- and two-defect models of \cite{Erdmenger:2013dpa,OBannon:2015cqy}, could be addressed. Another interesting question is what becomes of the anomalous dimension of the hopping kinetic energy that we introduced via the bulk mass parameter in such a continuum limit. Since deformations by non-marginal operators typically trigger RG flows, one may speculate whether in a continuum description nontrivial scaling exponents along the lines of \cite{Charmousis:2010zz,Huijse:2011ef,Hartnoll:2015sea} might appear. This would in particular be interesting from the point of view of quantum critical phases and Kondo physics. We plan to investigate this and related questions in future work.

\paragraph*{Acknowledgement:}
We would like to thank S. Das, A. Dymarsky, S. He, S. Lin, S. Moriyama, N. Ohta, M. Shigemori, S. Sugimoto, and T. Takayanagi for helpful discussions and comments. Especially, M.F. would like to thank A. Shapere for careful reading of the manuscript and valuable discussion.  This work was in part supported by JSPS KAKENHI Grants No. 26870284 and JP17K17822 (M.T.).  The work of R.M. was supported by World Premier International Research Center Initiative (WPI Initiative), MEXT, Japan, by the U.S. Department of Energy under Contract No. DE-FG-88ER40388, by the Alexander-von-Humboldt Foundation through a Feodor Lynen postdoctoral fellowship, and by the DFG via SFB 1170 "Tocotronics". 
S.P. was supported by NSF grant DMR-1056536, and thanks the hospitality of the TIFR for the final stage of this work.

\appendix
\section{The holographic renormalization in the homogeneous phase when $M\neq 0$}\label{HOL1}
In this appendix, we explain the holographic renormalization in the homogeneous phase when $M\neq 0$. 
The action \eqref{ACT11} evaluated at the above solution \eqref{PHI16} is still divergent at the $AdS$ boundary. To compute the finite on-shell action, we need to add counterterms on the constant $u(=u_\mathrm{max})$ slice of the $AdS$ boundary as
\ba
I_{\mathrm{cut},m}=\sum_{n}\dfrac{1}{2}\int_{u=u_\mathrm{max}}dt\sqrt{-h}A_{(n)t}A^{t}_{(n)}-\dfrac{\delta_M}{\Lambda }\int_{u=u_\mathrm{max}} dt\sqrt{-\gamma}\phi^{2}.
\ea
Note that the regularized action has the variation with respect to $t_\mathrm{hop}$: $\delta_{\phi}(I+I_{cut,m})=(1-2\delta_M)\int_{u=u_\mathrm{max}} dt (\delta t_\mathrm{hop}\varphi/\Lambda +c.c.)+\mbox{surface terms at $u=u_\mathrm{h}$}$.

The free energy is computed from the renormalized finite on-shell action  by adding the four pieces (see also \eqref{ACT11}) and taking the analytic continuation to  Euclidean signature as follows:
\ba
F_\mathrm{Mott}=-\dfrac{1}{\beta}(I+I_{\mathrm{cut},m}).
\ea
Substituting the solution \eqref{PHI16}, the analytic expression for the free energy in the homogeneous phase depends on the IR potential and is given by
\ba\label{FMO326}
&F_\mathrm{Mott}=2\mu\rho +u_\mathrm{h}\rho^{2}+ \nonumber \\
&\dfrac{1}{2 \Lambda Y_\mathrm{h}\,\sqrt {4\,{M}^{2}+1}}  (-4\,{M}^{2}t_\mathrm{hop}^{2}Y_\mathrm{h}^{2}+t_\mathrm{hop}^{
2}Y_\mathrm{h}^{2}\sqrt {4{M}^{2}+1}-8\,{ Y_\mathrm{h}}{M}^{2} t_\mathrm{hop}
\,{\varphi} \nonumber \\
&+2\,\sqrt {4\,{M}^{2}+1}{ t_\mathrm{hop}}\,{\varphi} Y_\mathrm{h} -t_\mathrm{hop}^{2} Y_\mathrm{h}^{2}+4\,{M}^{2}\varphi^{2} 
+\sqrt{4\,{M}^{2}+1}\varphi^{2} \nonumber \\
&-2\,t_\mathrm{hop}\,\varphi \,Y_\mathrm{h}+\varphi ^{2})+\mathcal{I}^\mathrm{IR}_\mathrm{mixed} \nonumber \\
&=2\mu\rho +u_\mathrm{h}\rho^{2}+\dfrac{1}{2\Lambda} { t_\mathrm{hop}^{2}Y_\mathrm{h}\, \Big( 1-\sqrt{1+4M^{2}} \Big) }+ \nonumber \\
&+\dfrac{\lambda t_\mathrm{hop}^{4}Y_\mathrm{h}^{2}}{u_\mathrm{h}}+2w^{2}t_\mathrm{hop}^{2}Y_\mathrm{h}+\sum_{p,r\ge 1}\Lambda_{(p,r)}u_h^{1-p}t_\mathrm{hop}^{2p}Y_h^p\sum_n(-\rho_{(n)}^2)^r, \label{FMT324}
\ea
where $Y_\mathrm{h}=u_\mathrm{h}^{1-2\delta_M }$ and $\varphi=0$ was used in the last line. Note that the variation of the above action includes surface terms at $u=u_\mathrm{h}$. 

We find that when parameters are chosen as 
\ba
\lambda=\dfrac{\lambda_1}{(u_\mathrm{h}Y_\mathrm{h})^2},\quad \Lambda_{(p,r)}=\dfrac{\Lambda_{(p,r),1}}{(u_\mathrm{h}Y_\mathrm{h})^p},\quad \Lambda^{-1}=\dfrac{\Lambda_1}{u_\mathrm{h}Y_\mathrm{h}},
\ea  
 the effective hopping parameter $dF_\mathrm{Mott}/dt_\mathrm{hop}$ evaluated at $\varphi =0$ becomes a function of $t_\mathrm{hop}/U$. Now, $\lambda_1,\ \Lambda_1,\ \Lambda_{(p,r),1}$ are constants and the power of the denominator becomes the power of $\phi$ interaction divided by 2. The above choice of parameters is consistent with the matching condition \eqref{PAR328}.

\section{The derivation of free energy and the effective hopping when $M\neq 0$}\label{FMNEQ}
In this section, we derive the formula of the free energy $F$ and the effective hopping $dF/dt_\mathrm{hop}$ in the presence of the bulk mass. These formulas are a straightforward generalization of $M=0$ case. Similar to main section, we  choose the following metric in the unit $AdS$ radius as
\ba\label{APP1}
ds^{2}=\dfrac{du^{2}}{f(u)u^{2}}-g(u)u^{2}dt^{2},
\ea
where we take $f(u)$ and $g(u)$ to be functions which approach to $1$ at the $AdS$ boundary quickly enough. One can also perform the coordinate transformation of $u$ to satisfy $g(u)=1$. For the $AdS_2$ hard wall, $f(u)=g(u)=1$. {In the $AdS_2$ hard wall geometry in the $AdS_5$ soliton described in section of discussion, $f(u)=1-(u_\mathrm{h}/u)^4$ and $g(u)=1$.}

Using the above metric and an ansatz for background fields depending on $u$ only, the action \eqref{ACT11}  in the radial gauge $A_u^{(n)}=0$ becomes
\ba\label{ACT26A}
&I=\int^{U_\mathrm{max}}_{u_\mathrm{h}} du \Big[\dfrac{1}{2}\sqrt{\dfrac{f(u)}{g(u)}}  ((A_{t(1)}^{\prime })^{2}+(A_{t(2)}^{\prime})^{2})+\dfrac{1}{\Lambda}\Big(-M^{2}\abs{\phi}^{2}\sqrt{\dfrac{{g(u)}} 
{{f(u)}}} \nonumber \\
&-\sqrt{f(u)g(u)}u^{2}{\abs{\phi'}^{2}}+q^{2}(A_{t(1)}-A_{t(2)})^{2} \dfrac{\abs{\phi}^{2}}{\sqrt{f(u)g(u)}u^{2}}\Big)  
\Big]+I_\mathrm{mixed}^\mathrm{iR}.
\ea

The EOM are derived from \eqref{ACT26A}. Solving the EOM in the presence of the bulk mass, the asymptotic expansion of fields at the $AdS$ boundary is changed in terms of exponents as
{\ba\label{ASY28}
&\phi \sim t_\mathrm{hop}u^{\delta_{\phi}}-\dfrac{4  \delta\rho^2 q^4
     t_\mathrm{hop}^3u^{
    {3} \delta_{\phi}}}{\Lambda (2\delta_{\phi}+1) \delta_{\phi} (-m^2+q^2\delta\rho^2 +
      {3} \delta_{\phi} +
      {9} \delta_{\phi}^2)} + {\cal O}(r^{5\delta_{\phi}}) + \varphi_{v} u^{-1-\delta_{\phi}}(1 + \dots) , \nonumber \\
&A_t^{(l)}\sim\mu +\rho_{(l)}u-(-1)^l\dfrac{\delta\rho q^2 u^{2\delta_{\phi}+1} t_\mathrm{hop}^2}{
 \Lambda (2\delta_{\phi}+1) \delta_{\phi}} +(-1)^l\dfrac{4q^2}{\Lambda} t_\mathrm{hop}\varphi_v\log (u)+ {\cal O}(u^{4 \delta_{\phi}+ 1}),
\ea
where $\delta_{\phi}=(-1+\sqrt{1+4M^{2}-4q^2\delta\rho^{2}})/2$}. The case $f(u)=g(u)=1$ is considered in both the EOM and the action \eqref{ACT26A}  from now on.  The EOM derived from \eqref{ACT26A} are solved numerically in the non-homogeneous phase with $A_{\mu}^{(1)}\neq A_{\mu}^{(2)}$.  Solutions satisfy the IR boundary condition \eqref{IRB216}. Solutions also satisfy $A_{t}^{(1)}-A_{t}^{(2)}|_{u=u_\mathrm{h}}\sim \delta \rho u_\mathrm{h}$ as $t_\mathrm{hop}\to 0$.  When $t_\mathrm{hop}\to 0$, the boundary condition \eqref{IRB216} is solved in terms of the asymptotic expansion \eqref{ASY28}. In this limit, the VEV $\varphi_v$ approaches to
\ba\label{VEV57}
\varphi = -\frac{t_\mathrm{hop} u_\mathrm{h}^{2 \delta_{\phi}+1} \left(-\Lambda\Lambda_{(1,1)} \left(\sum_i\rho_{(i)}^2 \right)-\delta_{\phi}+2\Lambda w^2\right)}{-\Lambda \Lambda_{(1,1)} \left((\sum_i\rho_{(i)}^2\right)+\delta_{\phi}+2 \Lambda w^2+1}+{\cal O}(t_\mathrm{hop}^2).
\ea
When the IR boundary condition is satisfied in the variation of $\phi$ and $\bar{\phi}$, the surface term at $u=u_\mathrm{h}$ is then canceled out  the IR potential.

On the other hand, the on-shell action is still divergent near the $AdS$ boundary. 
To compute the free energy, these divergences should be canceled. The following counterterms are used
\ba\label{APP29}
&I_\mathrm{cut,ap}^{\prime}=\sum_{n}\dfrac{1}{2}\int_{u=u_\mathrm{max}}dt\sqrt{-h}A_{(n)t}A^{t}_{(n)}+\dfrac{\delta_{\phi}}{\Lambda}\int dt\sqrt{-\gamma}\phi^{2} \nonumber \\
&=-\sum_{n}\dfrac{1}{2u_\mathrm{max}\sqrt{g(u_\mathrm{max})}} \int_{u=u_\mathrm{max}}dtA_{(n)t}^{2}+\dfrac{\delta_{\phi}}{\Lambda}u_\mathrm{max}\sqrt{g(u_\mathrm{max})} \int dt \phi^{2}. \ea
We need more counterterms to cancel the divergences of the on-shell action if the combination $q^2\delta \rho^2-m^2$ is small. When $m=0$, we need to take $q>{\sqrt{3}}/{4}$ to cancel the divergences.

The free energy is given by the summation with the above counterterms as
\ba\label{FA59}
F=-\dfrac{1}{\beta}(I+I_\mathrm{cut,ap}^{\prime}).
\ea

Hereby, we consider the variation principle since we obtain a finite renormalized action and the free energy. 
When we vary the action in terms of the gauge field, it is convenient to define the momentum of the gauge field as 
\ba
\pi_F^{(i)}=
\sqrt{\dfrac{f(u)}{g(u)}}F_{ut(i)}=\dfrac{\partial \mathcal{I}}{\partial \partial_u A_t^{(i)}},
\ea
where $\mathcal{I}$ is the lagrangian density of $I$.
The gauge variation at on-shell becomes
\ba\label{VARA60}
&\delta_A (I+I_\mathrm{cut,ap}')=\int_{u\to \infty} dt \sum_i\Big(\pi_F^{(i)}+\dfrac{\delta I_\mathrm{cut,ap}'}{\delta A_t^{(i)}}\Big)\delta A_t^{(i)} \nonumber \\
&-\int_{u=u_h} dt \sum_i\Big( \pi_F^{(i)}\delta A_t^{(i)}+\dfrac{\partial \mathcal{I}_\mathrm{IR}^\mathrm{mixed}}{\partial \partial_uA_t^{(i)}}\partial_u \delta A_t^{(i)}\Big),
\ea
where $\mathcal{I}_\mathrm{IR}^\mathrm{mixed}(=-\abs{\phi}^2\sum_nF_{ut}^{(n)2}\Lambda_{(1,1)}+\dots)$ is the lagrangian density of the IR potential. We focus on boundary terms of \eqref{VARA60} at $u\to \infty$. 
The gauge variation is not changed in the presence of the boundary term, while the momentum is changed into
\ba
\pi_F^{(i)}+\dfrac{\delta I_\mathrm{cut,ap}'}{\delta A_t^{(i)}}=-\dfrac{\mu}{u}+\dots,
\ea
which vanishes at $u\to \infty$. Thus, the variation is unlike the one of higher dimensional $AdS_{d+1}/CFT_d$ ($d\ge 3)$, where both $A_t^{(i)}$ and the density are finite at the $AdS$ boundary.~\footnote{Authors of \cite{Cvetic:2016eiv} introduced a conterterm making the variation principle similar to the one of higher dimensional $AdS_{d+1}/CFT_d$. We can consider the resembling  counterterm  \cite{Cvetic:2016eiv} as follows:  
\ba\label{APPis}
&I_\mathrm{cut,2}^{\prime}=\sum_{n}\dfrac{1}{2}\int_{u=u_\mathrm{max}}dt\Big(-\pi_F^{(n)}A_t^{(n)}+\dfrac{1}{2}\sqrt{-h}\pi_{F}^{(n)2}\Big)+\dfrac{\delta_{\phi}}{\Lambda}\int dt\sqrt{-\gamma}\phi^{2}.
 \ea
The above counterterm changes the gauge variation at on-shell into
\ba
\delta (I+I_\mathrm{cut,2})=\sum_i\int_{u\to \infty} dt \delta \pi_F^{(i)}(-A_t^{(i)}+\sqrt{-\gamma}\pi_F^{(i)})+\dots ,
\ea
where dots describe the variation at the IR. The above counterterm can give the variation of a finite density $\pi_F^{(i)}$, while the renormalized gauge field $A_t^{(i)}-\sqrt{-\gamma}\pi_F^{(i)}$ is still divergent like $t_\mathrm{hop}^2u^{2\delta_\mathrm{\phi}+1}$  at the boundary. To cancel the remaining divergences, we need more conterterms. 
}

For the bi-fundamental matter $\phi$, the on-shell variation becomes 
\ba\label{VARA64}
&\delta (I+I_\mathrm{cut,ap}')=\int_{u=u_\mathrm{max}}\dfrac{\delta \bar{\phi}}{\Lambda}  (-\sqrt{g(u)f(u)}u^2\phi' +\delta_{\phi}\sqrt{g(u)}u\phi) \nonumber \\
&+\int_{u=u_\mathrm{max}}\delta \bar{\phi} \Big(\sqrt{g(u)f(u)}u^2\dfrac{\phi'}{\Lambda}+\dfrac{\delta I_\mathrm{IR}^\mathrm{mixed}}{ \delta \bar{\phi}}\Big)+ c.c.  \nonumber \\
&=\int_{u=u_\mathrm{max}}dt(1+2\delta_{\phi})\dfrac{\delta t_\mathrm{hop}}{\Lambda} \varphi +c.c.,
\ea
where we have used the IR boundary condition \eqref{IRB216} in the first line of \eqref{VARA64}.

The effective hopping is then computed from the free energy \eqref{FA59} by differentiating the free energy in terms of $t_\mathrm{hop}$ as
\ba\label{VA60}
&\langle b_{i}^{a\dagger}b_{ja} \rangle +c.c. \equiv \dfrac{dF}{dt_\mathrm{hop}}=-\dfrac{1}{\beta}\Big[\Big(\dfrac{\delta \bar{\phi}}{\delta t_\mathrm{hop}}\dfrac{\delta (I+I'_\mathrm{cut,ap})}{\delta \bar{\phi}}+c.c.\Big)+\sum_i\dfrac{\delta A_t^{(i)}}{\delta t_\mathrm{hop}}\dfrac{\delta (I+I'_\mathrm{cut,ap})}{\delta A_t^{(i)}}\Big] \nonumber \\
&=-\Big(\dfrac{1+2\delta_{\phi}}{\Lambda}\varphi +c.c.\Big)+\sum_i \Big(\pi_F^{(i)}\dfrac{\partial A_t^{(i)}}{\partial t_\mathrm{hop}}+\dfrac{\partial \mathcal{L}^\mathrm{mixed}_\mathrm{IR}}{\partial \partial_uA_t^{(i)}}\dfrac{\partial \partial_u A_t^{(i)}}{\partial t_\mathrm{hop}}\Big)\Big|_{u=u_\mathrm{h}},
\ea
where the variation term of gauge fields vanishes at the $AdS$ boundary because $\partial A_t^{(i)}/\partial t_\mathrm{hop}$ starts with the order $u^{2\delta_{\phi}+1}$.  

Using \eqref{FA59} and \eqref{VA60}, we perform an analytic computation of the free energy and the effective hopping near the $\mu_b$ axis by substituting the boundary expansion \eqref{ASY28} in the approximation of small $t_\mathrm{hop}$ and large $u_\mathrm{h}$. Note that $f(u)$ and $g(u)$ depending on $u_\mathrm{h}$ (e.g. $f(u)=1-(u_\mathrm{h}/u)^4$)  are not considered in this limit since it requires more careful analysis at $u=u_\mathrm{h}$.  Choosing  parameters $q={\sqrt{6}}/{5}$ and $m=0$, the free energy of the non-homogeneous phase in the same limit is expanded as
\ba\label{VE73}
&F_{\rho,\rho+ 1}=\mu \sum_i\rho_{(i)}+\dfrac{u_\mathrm{h}\sum_i\rho_{(i)}^2}{2}+\dfrac{A_1t_\mathrm{hop}^2}{2}+{\cal O}(t_\mathrm{hop}^3), \nonumber \\
&\dfrac{dF_{\rho,\rho+ 1}}{dt_\mathrm{hop}}=A_1t_\mathrm{hop}+{\cal O}(t_\mathrm{hop}^2), \nonumber \\
&A_1=\dfrac{2u_\mathrm{h}^{\frac{1}{5}}}{\Lambda} \Big(\dfrac{24\log (u_\mathrm{h}) \left(-\Lambda\Lambda_{(1,1)}\sum_i\rho_{(i)}^2+2 \Lambda w^2 +\frac{2}{5}\right)}{25(-\Lambda\Lambda_{(1,1)}\sum_i\rho_{(i)}^2+2 \Lambda w^2 +\frac{3}{5})} \nonumber \\
&-\dfrac{14(-\Lambda\Lambda_{(1,1)}\sum_i\rho_{(i)}^2+2\Lambda w^2 +\frac{43}{70})}{5(-\Lambda\Lambda_{(1,1)}\sum_i\rho_{(i)}^2+2\Lambda w^2 +\frac{3}{5})}\Big).
\ea
{Evaluating the difference $\Delta F= F_{\rho,\rho +1}-F^\mathrm{Hom}$ in terms of $F^\mathrm{Hom}$ with $\rho_{(i)}=\rho$ (or $\rho_{(i)}=\rho +1$), one can show that the coefficient of $t_\mathrm{hop}^2$ in $\Delta F$ is negative for small $t_\mathrm{hop}$ and for parameters used in Fig. \ref{fig:mix}. In particular, for $\Lambda=1,\ u_\mathrm{h}=40,\ q=\sqrt{6}/5,\ \lambda=w^2=1,$ and $\Lambda_{(p,r)}=0$, $\Delta F\sim -79t_\mathrm{hop}^2$ around the $\mu - $ axis. This implies the occurrence of a cusp attached to the $\mu -$axis.}
,
Similarly, for $q=1$ and $m={\sqrt{7}}/{3}$, the free energy in the non-homogeneous phase is expanded as 
\ba
&F^{\rho,\rho +1}_1=\mu\sum_i\rho_{(i)}+\sum_i \rho^2_{(i)} \dfrac{u_\mathrm{h}}{2}+\dfrac{A_2t_\mathrm{hop}^2}{2}+{\cal O}(t_\mathrm{hop}^3), \nonumber \\
&\dfrac{dF^{\rho,\rho +1}_1}{dt_\mathrm{hop}}=A_2t_\mathrm{hop}+{\cal O}(t_\mathrm{hop}^2), \nonumber \\
&A_2=2\dfrac{u_\mathrm{h}^{\frac{1}{3}} }{\Lambda} \Big(\dfrac{\log (u_\mathrm{h}) \Big(-12\Lambda\Lambda_{(1,1)}\sum_i\rho_{(i)}^2+24 \Lambda w^2+4\Big)}{-3\Lambda\Lambda_{1,1}\sum_i\rho_{(i)}^2+6\Lambda w^2+2} \nonumber \\
&+\dfrac{26\Lambda\Lambda_{(1,1)}\sum_i\rho_{(i)}^2-52\Lambda w^2-\frac{53}{3}}{-3\Lambda \Lambda_{(1,1)}\sum_i\rho_{(i)}^2+6\Lambda w^2+2}\Big),
\ea
and, for $q={\sqrt{26}}/{5}$ and $m={2}/{\sqrt{5}}$, the free energy becomes
\ba\label{VE75}
&F^{\rho,\rho +1}_2=\mu \sum_i \rho_{(i)}+\sum \rho_{(i)}^2 \dfrac{u_\mathrm{h}}{2}+\dfrac{A_3t_\mathrm{hop}^2}{2}+{\cal O}(t_\mathrm{hop}^3), \nonumber \\
&\dfrac{dF^{\rho,\rho +1}_2}{dt_\mathrm{hop}}=A_3t_\mathrm{hop}+{\cal O}(t_\mathrm{hop}^2), \nonumber \\
&A_3=2\dfrac{u_\mathrm{h}^{\frac{1}{5}} }{\Lambda} \Big(\dfrac{\log (u_\mathrm{h}) \Big(-\frac{104}{5}\Lambda\Lambda_{(1,1)}\sum_i\rho_{(i)}^2+\frac{208}{5}\Lambda w^2+\frac{208}{25}\Big)}{-5\Lambda\Lambda_{1,1}\sum_i\rho_{(i)}^2+10\Lambda w^2+3} \nonumber \\
&+\dfrac{64\Lambda\Lambda_{(1,1)}\sum_i\rho_{(i)}^2-128\Lambda w^2-\frac{193}{5}}{-5\Lambda\Lambda_{(1,1)}\sum_i\rho_{(i)}^2+10 \Lambda w^2+3}\Big).
\ea
The difference $\Delta F_i= F^{\rho,\rho +1}_i-F^\mathrm{Mott}$ in terms of $F^\mathrm{Mott}$ with $\rho_{(i)}=\rho$ (or $\rho_{(i)}=\rho +1$) can be evaluated. Again, the coefficient of $t_\mathrm{hop}^2$ in $\Delta F$ can be shown to be  negative for small $t_\mathrm{hop}$ and for the above parameters ($\rho\ge 1$). It shows  the occurrence of a cusp around the $\mu -$axis.  In summary, the analytic expression was useful to find a cusp near the $\mu-$ axis. However, we can not use these expression of the free energy to obtain the phase structure of the holographic model since the asymptotic expansion is not good approximation near the hard wall $u=u_\mathrm{h}$.

While the phase structure is interesting at zero temperature, it will also be interesting to consider the theory at the finite temperature from $AdS$ black hole in the context of the $10d$ type IIB  supergravity.  The black hole phase corresponds to the deconfinement phase. It is known that the $AdS_5$ black hole geometry is thermodynamically stabler than the $AdS_5$ soliton when temperature increases as $T>M_{KK}$. At the critical temperature $T=M_{KK}$, the Hawking-Page phase transition takes place~\cite{Witten:1998zw} by comparing the free energy of both backgrounds. The Hawking-Page phase transition corresponds to the confinement/deconfinement phase transition of the dual theories at the strong coupling. After   introducing probe D5-branes in the $AdS$ black hole(the background of the $10d$ type IIB supergravity), more rich phase structure may be obtained like the chiral symmetry breaking in the deconfinement phase~\cite{Aharony:2006da}. Unlike the hard wall geometry, the dissipative effect is expected in the probe brane in the deconfinement phase because of the dissipation at the horizon. 

\section{Holographic dual to the level-rank duality}\label{LEV}

 The field theory side becomes the D3-brane theory compactified on the circle $\tau$ with the radius $R_1$. The anti-periodic boundary condition is imposed on adjoint fermions of the D3-brane theory. The adjoint fermions receive the tree level mass of order $1/R_1$ because of this supersymmetry breaking boundary condition. Furthermore, scalars acquire the mass after taking into account the quantum corrections of scalar's mass. In the low energy limit, $3d$ pure Yang-Mills theory is then obtained on $R^{1,2}$ after ignoring these massive modes. The pure Yang-Mills theory is a confining gauge theory, having a mass gap. The gauge coupling in $3d$  is identified with $g_{3}^2=g_s/(2\pi R_1)$. In addition, the background axion $\chi\sim k_1\tau$ is switched on. The Wess-Zumino term of the D3-brane is then replaced by the $3d$ Chern-Simons term as $
S^{WZ}_{D3}=\mbox{Tr}\int_{D3}\chi (F\wedge F)/(4\pi)=k_1\int_{R^{1,2}}\omega_3(a)/(4\pi).$
So, the field theory side is changed into the $U(N_c)_{k_1}$ Yang-Mills-Chern-Simons theory~\cite{Fujita:2009kw}.  The $U(N_c)_{k_1}$ Yang-Mills-Chern-Simons theory becomes $U(N_c)_{k_1}$ pure Chern-Simons theory in the IR limit.

In the type IIB gravity dual, on the other hand, the $AdS_5$ soliton times $S^5$ is considered~\cite{Witten:1998zw}. The $AdS_5$ soliton times $S^5$ background is obtained by considering the backreaction of the $N_c$ D3-branes and taking the near horizon limit. The $10d$ metric becomes 
\ba\label{SOL14}
&ds^2=\Big(\dfrac{L_1^2}{f( u )u^2}du^2+\dfrac{u^2}{L_1^2}(-dt^2+f( u )d\tau^2+dx_i^2)+L_1^2(d\theta^2 +\sin^2\theta d\Omega_4^2)\Big), \\
& C_4=\dfrac{u^4}{L_1^4}dt d\tau dx_1dx_2 -L_1^4 C(\theta) d\Omega_4 ,  \\
& C(\theta)=\dfrac{3}{2}(\pi -\theta +\sin\theta \cos\theta ) +\sin^3\theta\cos \theta,
\ea
where $f( u )=1-(u_h/u)^4$, $i=1,2$, and $L_1^4=4\pi g_s N_c\alpha^{\prime 2}$. The tip of the $AdS$ soliton is located at $u=u_h$, where the geometry becomes smooth. The spacetime $\tau$ direction is compactified as $\tau\sim \tau +2\pi R_1$, where $R_1=L_1^2/2u_h$. 

We then introduce $k_1$ D7-branes along $(tx_1x_2,S^5)$  at the tip of the soliton in the probe limit without considering the backreaction $N_c\gg k_1$. These D7-branes correspond to the configuration of the infinite separated D7-brane defects in~\cite{Fujita:2016gmu}. The $U(k_1)_{N_c}$ pure Chern-Simons theory arises from the Wess-Zumino term of  $k_1$ D7-branes at $u=u_h$ in the presence of the $RR$ 5-form flux in the IR limit. The level-rank duality of the Chern-Simons theory is then realized by using the holography of the $AdS_5$ soliton and $k_1$ D7-branes. That is, $U(N_c)_{k_1}$ Chern-Simons theory (D3-branes) $\leftrightarrow$ $U(k_1)_{N_c}$ Chern-Simons theory (D7-branes).

\section{The DBI expansion}\label{DBIap}
In this appendix, we give a useful formula to consider the expansion of the DBI action.  
 Ignoring the gauge indices in the tensor product, we decompose the determinant of $n\times n$ matrices  as
\ba
\det (g_{\alpha\beta}+\xi F_{\alpha\beta})= \det(E^i_{\alpha})^2\det (\delta_{ij}+\xi F_{ij})=\det(g_{\alpha\beta})\det(\delta^{\alpha}{}_{\beta}+\xi g^{\alpha\gamma}F_{\gamma\beta}),
\ea
where $E^i_{\alpha}$ is the cotangent frame field and $F_{ab}=E^{i}_aF_{ij}E^j_b$. In the right-hand side of the first equality, we make the anti-symmetric representation of $O(n)$ manifest in $F_{ij}$.

A useful formula of the Determinant  of $n\times n$ matrices becomes for $n>4$~\cite{Polchinski,Polchinski2}
\ba\label{DBA1}
\det (1+\xi X)=\exp \Big(\mbox{tr}\Big(\xi X-\dfrac{(\xi X)^2}{2}+\dfrac{(\xi X)^3}{3}-\dfrac{(\xi X)^4}{4}-\dots\Big)\Big),
\ea
where $X_{\mu}{}^{\nu}=F_{\mu\alpha}g^{\alpha\nu}$ and traces are over the $n\times n$ matrices. Coefficients of the expansion are defined by $\sqrt{\det (1+\xi X)}\equiv \sum \xi^ix_i$. It is known that these coefficients $x_i$ can be written as symmetric polynomials of the eigenvalue of $X$ if $X$ can be diagonalized.
Coefficients $x_i$ are obtained up to ${\cal O}(\xi^4)$ as follows:
\ba
&x_0= 1,\quad  x_1= \dfrac{1}{2}\mbox{tr}X,\quad  x_2=\dfrac{1}{4}\Big(\dfrac{1}{2}(\mbox{tr}X)^2-\mbox{tr}X^2\Big), \nonumber \\
& x_3= \dfrac{1}{6}\mbox{tr}X^3-\dfrac{1}{8}\mbox{tr}X^2\mbox{tr}X+\dfrac{1}{48}(\mbox{tr}X)^3, \nonumber \\
&x_4=-\dfrac{1}{8}{\mbox{tr}X^4}+\dfrac{1}{12}{\mbox{tr}X}{\mbox{tr}X^3}-\dfrac{1}{32}{\mbox{tr}X^2}{({\mbox{tr}X})}^{2}+\dfrac{1}{32}{{(\mbox{tr}X^2)}}^{2}+\dfrac{1}{384}{{(\mbox{tr}X)}}^{4}. \nonumber \\
\ea

\subsection{The non-Abelian Taylor expansion}\label{NONA}
It is known that the $\alpha'$ expansion of the non-Abelian DBI action reproduces the effective action of the non-Abelian superstring theory up to the quartic order of the field strength~\cite{Tseytlin:1997csa,Hashimoto:1997gm}. Even for the bosonic string theory, coefficients of the quartic terms are different only with the excess of the commutator terms (or derivative dependent terms).
To analyze the full non-Abelian action \eqref{D524}, in this section, we expand the lagrangian based on a the non-Abelian Taylor expansion and take the symmetrized trace. We restrict to $U(2)$ non-Abelian symmetry and first consider the DBI part. The problem is that , when we choose $\theta =0, \pi$, the effective tension of the DBI action goes to zero.
One can show that $\theta =\pi/2\mathbf{1}_{2\times 2}$ becomes the solution of the EOM and the vacuum,   where the tension of the brane remains finite and does not go to zero. Choosing $v=\theta$ and using $\theta= \pi/2\mathbf{1}_{2\times 2}+\delta \theta$, we then expand the non-Abelian action around the vacuum.
 Using the DBI expansion in appendix \ref{DBIap}, we expand the non-Abelian DBI action to include higher derivative terms. From \eqref{D524}, we can choose $X$ in \eqref{DBA1}  as
\ba
&X^{(1)c}{}_{a}= g^{ca}\Big(\xi(D_a\Phi^iD_b \Phi_i +D_a\Phi_i(Q^{-1}-1)^{ij}D_b\Phi_j) + F_{ab}\Big), \nonumber \\
&X^{(2)i}{}_j=i [\Phi^i,\Phi^k]g_{kj},
\ea
where $\xi= 2\pi \alpha'$. Note that the second term of $X^{(1)}$ is of order $\xi^2$. Besides, the determinant of $g$ and $RR$ 4-form $C_4$ depend on $\theta$. These are expanded around $\theta =\pi/2$ as 
\ba
\dfrac{\sqrt{-\mbox{det}(g_{ab})}\sqrt{f(r)}}{L_1^4}= \sin^4 (\theta)\sim 1-2\,{{\delta \theta}}^{2}+\dfrac{5}{3}\,{{\delta \theta}}^{4},\quad 
C_4\sim \dfrac{3}{4}\pi-4 \delta\theta+\dfrac{8}{3}\delta\theta^3.
\ea
We perform the dimensional reduction of the expanded action since we are  interested in the zero modes along the angular direction $S^4$. By remaining terms up to $\xi^4$ in the dimensional reduction, the $U(2)$ YM theory in addition to higher derivative terms is obtained on the asymptotic $AdS_2$ as
\ba\label{S2b}
&S^{b}=-T'\mbox{Str}\int d^2x\sqrt{-\mbox{det}(\bar{g}_{ab})}\Big(\xi^2\Big(\dfrac{1}{4}F_{ab}F^{ab}+\dfrac{1}{2}D_a \Phi^iD^a \Phi^j\bar{g}_{ij} \nonumber \\
&-\dfrac{1}{4}[\Phi^i,\Phi^j][\Phi^k,\Phi^l]\bar{g}_{ik}\bar{g}_{jl}\Big)-2\,{{\delta \theta}}^{2}\Big) \nonumber \\
&+S^b_{WZ}+S_{2}^b , \label{TOT436} \\
&S^b_{WZ}=T'\xi\int \mbox{Str}\Big(\Big( \dfrac{3}{4}\pi-4 \delta\theta+\dfrac{8}{3}\delta\theta^3\Big) F\Big) \\
&S^b_2=- T'\mbox{Str}\int d^2x \sqrt{-\det (\bar{g}_{ab})}\Big[ \xi^4\Big(\dfrac{1}{32}(F_{ab}F^{ab})^2-\dfrac{1}{8}F_{a}{}^bF_{b}{}^cF_c{}^dF_d{}^a\dots \nonumber \\
& +\dfrac{1}{8}D_c\Phi^i D^c\Phi_iF_{ab}F^{ab}-\dfrac{1}{4}D_a\Phi^iD^b\Phi_iD_b\Phi^jD^a\Phi_j +\dfrac{1}{8}(D_a\Phi^iD^a\Phi_i)^2 \nonumber \\
&+\dfrac{i}{2}F_{ab}D^b\Phi^i[\Phi_i,\Phi^j]D_a\Phi_j+\dfrac{1}{8}[\Phi^j,\Phi_k][\Phi^k,\Phi_j]\Big(D_a\Phi^iD^a\Phi_i+\dfrac{1}{2}F_{ab}F^{ab}\Big) \Big) \nonumber \\
&-\xi^2\delta\theta^2 \Big(D_a\Phi^iD^a\Phi_i+\dfrac{1}{2}F_{ab}F^{ab}\Big)+\dfrac{5}{3}\delta\theta^4+\dots 
\Big],\label{SB281}
\ea
where $\bar{g}_{ab}$ is the metric of the 2-dimensional spacetime evaluated at $\theta= \pi/2$ and  $T'=T_5L_1^4\Omega_4=N_c/(3\pi^2\alpha')$. The indices $a,b$ and $i,j$ are raised or lowered in terms of $\bar{g}_{ab}$ and $\bar{g}_{ij}$, respectively. Dots in $S_2^b$ represent higher order terms with $[\Phi^i,\Phi^j]$. Note that $\theta$ already includes the factor of $\xi =2\pi \alpha'$. 

It is shown that the probe D-brane analysis with Abelian symmetry allows a constant solution of transverse scalars. In analogy, we consider the VEV of $U(2)$ adjoint transverse scalars $\theta,\ \Phi^{x_1}$ in non-Abelian symmetry. The VEV of transverse scalars has distance modes in diagonal components. In the matrix form, 
\ba\label{AWA34}
\theta =\dfrac{\pi}{2}\mathbf{1}_{2\times 2}+\xi\begin{pmatrix}\varphi^{1,v} & 0 \\ 0 & L^v \end{pmatrix},\quad \Phi^{x_1}=\begin{pmatrix}
0 & w^{x_1} \\
\bar{w}^{x_1} & 0 \\
 \end{pmatrix}, 
\ea 
These VEV break $U(2)$ symmetry into baryonic $U(1)$. {Note that eigenvalues of $\Phi^{x_1}$ are $\pm \abs{w^{x_1}}$, which give distance modes along a spatial direction.}

Moreover, we describe the general form of $U(2)$ adjoint fields as
\ba\label{AW78}
A_{b}=\begin{pmatrix} a_b^{(1)} & w_b \\
w_b^{\dagger} & a_b^{(2)}
\end{pmatrix},\quad \Phi^i=\begin{pmatrix}\varphi^{1,i} & w^i \\ w^{i,\dagger} & L^v\delta^{iv} \end{pmatrix}, 
\ea 
where $\varphi^{1,i}$ and $L^v$ are chosen to be constant. 
We introduce indices $I,J$ of the worldvolume directions where $I,J\neq v$. 
Moreover, we restrict fields $A_b,\ \Phi^i$ by requiring $w^v=\bar{w}^v=\varphi^{1,I}=0$, and $w_a=\bar{w}_a=0$. The lagrangian is finally written as
\ba\label{bigeq}
&I^r_1=-T'\int d^2x \sqrt{-\bar{g}}\Big[ \xi^2\Big(\dfrac{1}{4}f^{(1)}_{ab}f^{(1)ab}+\dfrac{1}{4}f^{(2)}_{ab}f^{(2)ab}+\dfrac{1}{2} (\partial_a \varphi^v \partial^a\varphi^v+\partial_a L^v \partial^a L^v )\bar{g}_{vv} \nonumber \\
&+\dfrac{1}{2}\partial_a\varphi^I\partial^a\varphi^J\bar{g}_{IJ}+\overline{D_cw^I}D^cw^J\bar{g}_{IJ} +\bar{w}^I((L^v\delta_v{}^k-\varphi^k)(L^v\bar{g}_{vk}-\varphi_k)\bar{g}_{IJ}-\varphi_I\varphi_J)w^J \nonumber \\
&-\dfrac{1}{2}(w_I\bar{w}_J-\bar{w}_Iw_J)(w^I\bar{w}^J-\bar{w}^Iw^J)\Big) -2\,{{(\xi L^v)}}^{2}-2\,{{(\xi \varphi^v)}}^{2}+\dfrac{5}{3}\,{{(\xi \varphi^v)}}^{4}+\dfrac{5}{3}\,{{(\xi L^v
)}}^{4}\Big] \nonumber \\
&+T'\xi \int  \mathcal{L}^{WZ}_0+S_2^b,
\ea
where $f^{(1),(2)}=da^{(1),(2)}$ and the WZ term is expanded as
\ba
&\dfrac{T'\xi \mathcal{L}^{ WZ}_0}{N_c} \sim \dfrac{1}{2}{ f_1}+\dfrac{1}{2}{ f_2} -\dfrac{8}{3\pi} f_2\xi L^v+\dfrac{16}{9\pi}\,{{( \xi L^v)}}^{3}{ f_2}-\dfrac{8}{3\pi} f_1\xi \varphi^v+\dfrac{16}{9\pi}\,{{( \xi \varphi^v)}}^{3}{ f_1}.
\ea
The modes $\varphi^v,\ L^v$ have the negative squared mass. However, these modes are stabilized in the presence of the non-trivial flux $da^{(1),{(2)}}$~\cite{Camino:2001at}. 

Finally, the EOM of the bi-fundamental scalar in the homogeneous phase, i.e. for the Ansatz \eqref{VEVs} with equal gauge fields, becomes
\ba\label{EOMD32}
\dfrac{1}{\sqrt{-\bar{g}}}D_c(\sqrt{-\bar{g}}D^cw^J\bar{g}_{IJ})-(L^v-\phi^v)^2w^J \bar{g}_{IJ}\bar{g}_{vv}-2 (w_Jw^J\bar{w}_I-w_J\bar{w}^Jw_I)=0\,.
\ea

\end{document}